\documentclass{report}

\usepackage[english]{babel}
\usepackage{fancyhdr}
\usepackage{graphicx}
\usepackage{url}
\usepackage{epic}
\usepackage{eepic}
\usepackage{makeidx}
\usepackage{array}
\usepackage{times}
\usepackage{amsmath}
\usepackage{esint}
\usepackage{amssymb}
\usepackage{fancyvrb}
\usepackage[usenames,dvipsnames]{color}
\usepackage{framed}
\usepackage{multirow}
\usepackage{enumerate}
\usepackage[round,sort]{natbib}

\renewcommand{\vec}[1]{\mbox{\boldmath$#1$}}         

\definecolor{shadecolor}{rgb}{0.74,0.74,0.74}

\makeindex

\begin{document}

\begin{titlepage}
\begin{center}
  \hrule \vspace{3mm}
  {\Huge {MITHRA 1.0} \vspace{10mm} } \\
  {\LARGE {\sc A Full-wave Simulation Tool } \vspace{3mm} } \\
  {\LARGE {for Free Electron Lasers \vspace{3mm} } } \\
\end{center}

\vspace{5mm}

\begin{center}
   \includegraphics[width=180mm]{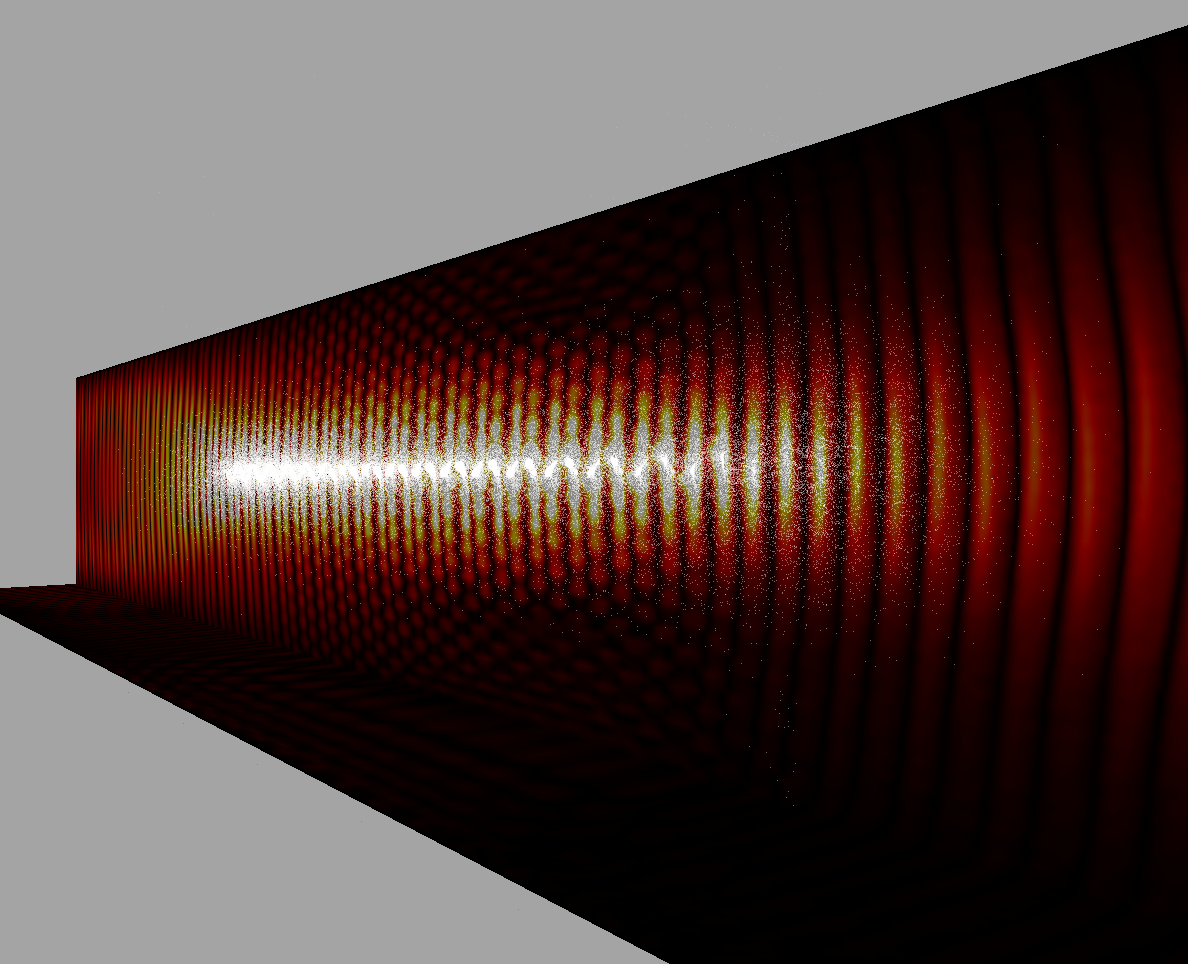}
\end{center}

\vspace{5mm}

\begin{center}
  \Large{ \textbf{Arya Fallahi, Alireza Yahaghi and Franz X. K\"artner} } \\
  \Large{ Group of Ultrafast Optics and X-ray Sources } \\
  \Large{ DESY-Center for Free Electron Laser Science (CFEL) } \\
\end{center}

\end{titlepage}

\newpage

\

\vspace{8cm}

{\large \noindent \textit{Mithra, also spelled Mithras, Sanskrit Mitra, in ancient Indo-Iranian mythology, is the god of light, whose cult spread from India in the east to as far west as Spain, Great Britain, and Germany. The first written mention of the Vedic Mitra dates to 1400 BC. His worship spread to Persia and, after the defeat of the Persians by Alexander the Great, throughout the Hellenic world. In the 3rd and 4th centuries AD, the cult of Mithra, carried and supported by the soldiers of the Roman Empire, was the chief rival to the newly developing religion of Christianity. The Roman emperors Commodus and Julian were initiates of Mithraism, and in 307 Diocletian consecrated a temple on the Danube River to Mithra, ``Protector of the Empire."}

\noindent \textit{According to myth, Mithra was born, bearing a torch and armed with a knife, beside a sacred stream and under a sacred tree, a child of the earth itself. He soon rode, and later killed, the life-giving cosmic bull, whose blood fertilizes all vegetation. Mithra's slaying of the bull was a popular subject of Hellenic art and became the prototype for a bull-slaying ritual of fertility in the Mithraic cult.}

\noindent \textit{As god of light, Mithra was associated with the Greek sun god, Helios, and the Roman Sol Invictus. He is often paired with Anahita, goddess of the fertilizing waters.}

\

\hspace{10cm} \textit{Source: Encyclopaedia Britannica}}

\newpage
\tableofcontents
\listoftables
\listoffigures


\chapter{Introduction}
\label{chapter_introduction}

Free Electron Lasers (FELs) are currently serving as promising and viable solutions for the generation of radiation in the whole electromagnetic spectrum ranging from microwaves to hard X-rays \cite{FEL2,FEL1,freund2012principles}.
Particularly, in portions of spectrum where common solutions like lasers and other electronic sources do not offer efficient schemes, FEL based devices attract considerable attention and interest.
For example, soft and hard X-ray radiation sources as well as THz frequency range are parts of the spectrum where FEL sources are widely used.
In the optical regime, lasers currently serve as the most popular sources, where radiation is generated and amplified based on the stimulated emission.
More accurately, the excited electrons of the gain medium emit coherent photons when changing the energy level to the ground state \cite{siegman1986lasers}.
Since the energy bands of different gain media are fixed curves determined by the material atomic lattice, there are only specific wavelengths obtainable from lasers operating based on stimulated emission in a gain medium.
In contrast, there exist vacuum electronic devices like gyrotrons, klystrons and travelling wave tubes (TWT), in which free electrons travelling along a certain trajectory transform kinetic energy to an electromagnetic wave \cite{gilmour1986microwave}.
Although, these sources are usually not as efficient as medium based lasers, their broadband operations make them promising in portions of the spectrum where no gain media is available.

In a free electron laser, relativistic electrons provided from linear accelerators travel through a static undulator and experience a wiggling motion.
The undulator performance is categorized into two main regimes: (\emph{i}) in a short undulator, each electron radiates as an independent moving charge, which yields an incoherent radiation of electron bunch.
Therefore, the radiation power and intensity is linearly proportional to the number of electrons.
(\emph{ii}) For long interaction lengths, the radiated electromagnetic wave interacts with the bunch and the well-known micro-bunching phenomenon takes place.
Micro-bunching leads to a periodic modulation of charge density inside the bunch with the periodicity equal to the radiation wavelength.
This effect results in a coherent radiation scaling with the square of the bunch numbers.
Coherent X-ray have shown unprecedented promises in enabling biologists, chemists and material scientists to study various evolutions and interactions with nanometer and sub-nanometer resolutions \cite{jaeschke2015synchrotron}.

Owing to the desire of hard X-ray FEL machines for electrons with ultrarelativistic energies (0.5-1 GeV), these sources are usually giant research facilities with high operation costs and energy consumption.
Therefore, it is crucial and additionally very useful to develop sophisticated simulation tools, which are able to capture the important features in a FEL radiation process.
Such tools will be very helpful for designing and optimizing a complete FEL facility and additionally useful for detailed investigation of important effects.
The last decade had witnessed extensive research efforts aiming to develop such simulation tools.
As a result, various softwares like Genesis 1.3 \cite{reiche1999genesis}, MEDUSA \cite{biedron19993d}, TDA3D \cite{tran1989tda,faatz1997tda3d}, GINGER \cite{fawley2002user}, PERSEO \cite{giannessi2006overview}, EURA \cite{bacci2008compact}, RON \cite{dejus1999integral}, FAST \cite{saldin1999fast}, PlaRes \cite{andriyash2015spectral} and Puffin \cite{campbell2012puffin} are developed and introduced to the community.
However, all the currently existing simulation softwares are usually written to tackle special cases and therefore particular assumptions or approximations have been considered in their development \cite{biedron2000multi}.
Some of the common approximations in FEL simulation are tabulated in Table \ref{FELapproximations}.
\begin{table}[h]
\renewcommand{\arraystretch}{1.5}
\caption{Common approximations in modelling free electron laser radiation}
\label{FELapproximations} \centering
\begin{tabular}{|c|c|c|c|c|c|c|c|}
\hline
\multirow{3}{*}{code name} & \multicolumn{7}{|c|}{approximation}  \\
\cline{2-8}
& 1D FEL & steady state & wiggler-average & slow wave & forward & \multirow{2}{*}{no space-charge} & \multirow{2}{*}{slice} \\
& theory & approximation & electron motion & approximation & wave & & \\
\hline
GENESIS 1.3 & \textemdash & \checkmark (optional) & \checkmark & \checkmark & \checkmark & \textemdash &  \checkmark \\
\hline
MEDUSA & \textemdash & \checkmark (optional) & \textemdash & \checkmark & \checkmark & \textemdash & \checkmark \\
\hline
TDA3D & \textemdash & \checkmark & \checkmark & \checkmark & \checkmark & \textemdash & no time-domain \\
\hline
GINGER & \textemdash & \textemdash & \checkmark & \checkmark & \checkmark & \textemdash & \textemdash \\
\hline
PERSEO & \textemdash & \textemdash & \textemdash & \textemdash & \checkmark & \checkmark & \textemdash \\
\hline
EURA & \textemdash & \textemdash & \checkmark & \checkmark & \checkmark & \textemdash & \textemdash \\
\hline
FAST & \textemdash & \textemdash  & \checkmark & \checkmark & \textemdash & \textemdash & \checkmark \\
\hline
Puffin & \textemdash & \textemdash  & \textemdash & \textemdash & \checkmark & \checkmark & \textemdash \\
\hline
\end{tabular}
\end{table}
The main goal in the presented software is the analysis of the FEL interaction without considering any of the above approximation.
The outcome of the research and effort will be a sophisticated software with heavy computation loads.
Nonetheless, it provides a tool for testing the validity of various approximations in different operation regimes and also a reliable approach for preparing the final design of a FEL facility.

Besides the wide investigations and studies on the conventional X-ray FELs, recently research efforts have been devoted to building compact X-ray FELs, where novel schemes for generating X-ray radiations in a so-called table-top setup are examined and assessed.
Various research topics such as laser-plasma wake-field acceleration (LPWA) \cite{mangles2004monoenergetic,faure2004laser,geddes2004high}, laser plasma accelerators (LPA) \cite{tajima1979laser,lundh2011few}, laser dielectric acceleration (LDA) \cite{england2014dielectric} and THz acceleration \cite{nanni2015terahertz,fallahi2016short}, pursue the development of compact accelerators capable of delivering the desired electron bunches to FEL undulators.
Besides such attempts, one promising approach to make a compact undulator is using optical undulators, where the oscillations in an electromagnetic wave realize the wiggling motion of the electrons \cite{kartner2016axsis}.
Many of the approximations in Table \ref{FELapproximations}, which sound reasonable for static undulators are not applicable for studying an optical undulator radiation.
In this regime, due to the various involved length-scales and remarkable impact of the parameter tolerances, having access to a rigorous and robust FEL simulation tool is essential.

One of the difficulties in the X-ray FEL simulation stems from the involvement of dramatically multidimensional electromagnetic effects.
Some of the nominal numbers in a typical FEL simulation are:
\begin{itemize}
  \item Size of the bunch: $\sim 100\,$fs or $300\,\mu$m
	\item Undulator period: $\sim 1\,$cm
	\item Undulator length: $\sim 10-500\,$m
	\item Radiation wavelength: $\sim 1-100\,$nm
\end{itemize}
Comparing the typical undulator lengths with radiation wavelengths immediately communicates the extremely large space for the values.
This in turn predicts very high computation costs to resolve all the physical phenomena, which is not practical even with the existing supercomputer technology.
In order to overcome this problem, we exploit Lorentz boosted coordinate system and implement Finite Difference Time Domain (FDTD) \cite{taflove2000computational} method combined with Particle in Cell (PIC) simulation in the electron rest frame.
This coordinate transformation makes the bunch size and optical wavelengths longer and shortens the undulator period.
Interestingly, these very different length scales transform to values with the same order after the coordinate transformation.
Consequently, the length of the computation domain is reduced to slightly more than the bunch length making the full-wave simulation numerically feasible.
We comment that the simulation of particle interaction with an electromagnetic wave in a Lorentz boosted framework is not a new concept.
The advantage of this technique is already demonstrated and widely used in the simulation of plasma-wakefield acceleration \cite{yu2014modeling,vay2013domain,vay2012novel}.
However, to the best of our knowledge, it has never been used to simulate a FEL mechanism, which is what we are aiming in this study.

The presented manual shows how one can numerically simulate a complete FEL interaction using merely Maxwell equations, equation of motion for a charged particle, and the relativity principles.
In chapter \ref{chapter_methodology}, the whole computational aspects of the software, including the Finite Difference Time Domain (FDTD), Particle In Cell (PIC), current deposition, Lorentz boosting, quantity initialization, and parallelization, are described in detail.
The implementation is explained in a way suitable for a graduate student to start writing the code on his own.
Chapter \ref{chapter_refcard} provides a reference card for a software user to get familiar with MITHRA and the required parameters for performing the simulations.
Finally, in chapter \ref{chapter_examples}, different examples of free electron lasers are analyzed and the results are presented in conjunction with some discussions.
As a new software entering the FEL community, we aim to keep updating this material with new implementations and examples.
In this regard, any assistance and help from the users of this software will be highly appreciated.


\chapter{Methodology}
\label{chapter_methodology}

In this chapter, we present the detailed formalism of Finite Difference Time Domain - Particle In Cell (FDTD/PIC) method in the Lorentz boosted coordinate system.
There are many small still very important considerations in order to obtain reliable results, which converge to the real values.
For example, the method for electron bunch generation, particle pusher algorithm and computational mesh truncation need particular attention.

\section{Finite Difference Time Domain (FDTD)}
\label{section FDTD}

FDTD is perhaps the first choice coming to mind for solving partial differential equations governing the dynamics of a system.
Despite its simple formulation and second order accuracy, there are certain features in this method like explicit time update and zero DC fields, which makes this method a superior choice compared to other algorithms \cite{taflove2000computational}.
FDTD samples the field in space and time at discrete sampling points and represents the partial derivatives with their discrete counterparts.
Subsequently, update equations are derived based on the governing differential equation.
Using these updating equations, a time marching algorithm is acquired which evaluates the unknown functions in the whole computational domain throughout the simulation time.
In the following, we start with the Helmholtz equation which is the governing partial differential equation for our electromagnetic problem.

\subsection{Helmholtz Equation}
The physics of electromagnetic wave and its interaction with charged particles in free space is mathematically formulated through the well-known Maxwell's equations:
\begin{align}
\nabla \times \vec{E} &= -\frac{\displaystyle \partial \vec{B}}{\displaystyle \partial t} \\
\nabla \times \vec{B} &= \mu_0 \vec{J} + \mu_0 \varepsilon_{0} \frac{\displaystyle \partial \vec{E}}{\displaystyle \partial t} \\
\nabla \cdot \vec{E} &= -\frac{\displaystyle \rho}{\displaystyle \varepsilon_{0}} \\
\nabla \cdot \vec{B} &= 0
\end{align}
These equations in conjunction with the electric current equation $\vec{J}=\rho \vec{v}$ ($\vec{v}$ is the charge velocity) and the Lorentz force equation:
\begin{equation}
\label{LorentzForce}
\vec{F} = q(\vec{E} + \vec{v} \times \vec{B})
\end{equation}
are sufficient to describe wave-electron interaction in free space.
Moving free electrons introduce electric current which enters into the Maxwell's equations as the source.
Electric and magnetic fields derived from these equations are subsequently employed in the Lorentz force equation to determine the forces on the electrons, which in turn determine their motions.
As it is evident from the above equations, there are two unknown vectors ($\vec{E}$ and $\vec{B}$) to be evaluated, meaning that six unknown components should be extracted from the equations.
However, since these two vectors are interrelated and specially because there is no magnetic monopole in the nature ($\nabla \cdot \vec{B}=0 $), one can recast Maxwell's equations in a Helmholtz equation for the magnetic vector potential ($\vec{A}$) and a Helmholtz equation for the scalar electric potential ($\varphi$):
\begin{equation}
\label{HelmholtzA}
{\vec{\nabla}}^{2}\vec{A} - \frac{1}{c^2} \frac{\displaystyle \partial^2}{\displaystyle \partial t^2}\vec{A} = -\mu_{0} \vec{J}
\end{equation}
\begin{equation}
\label{HelmholtzF}
{\nabla}^{2}\varphi- \frac{1}{c^2} \frac{\displaystyle \partial^2 \varphi}{\displaystyle \partial t^2} = -\frac{\displaystyle \rho}{\displaystyle \varepsilon_{0}},
\end{equation}
where $c=1/\sqrt{\mu_0 \varepsilon_0}$ is the light velocity in vacuum.
In the derivation of above equations, the Lorentz gauge $\nabla \cdot \vec{A}=-\frac{1}{c^2}\frac{\partial \varphi}{\partial t}$ is used. 
The original $\vec{E}$ and $\vec{B}$ vectors can be obtained from $\vec{A}$ and $\varphi$ as:
\begin{equation}
\label{BvsA}
\vec{B} = \nabla \times \vec{A}
\end{equation}
\begin{equation}
\label{EvsA}
\vec{E} = -\frac{\partial \vec{A}}{\partial t}-\nabla \varphi
\end{equation}
In addition to the above equations, the charge conservation law written as
\begin{equation}
\label{chargeLaw}
\nabla \cdot \vec{J} + \frac{\partial \rho}{\partial t} = 0,
\end{equation}
should not be violated in the employed computational algorithm.
This is the main motivation for seeking proper current deposition algorithms in the FDTD/PIC methods used for plasma simulations.
It is immediately observed that the equations (\ref{HelmholtzA}), (\ref{HelmholtzF}), (\ref{chargeLaw}) and the Lorentz gauge introduce an overdetermined system of equations.
In other words, once a current deposition is implemented that automatically satisfies the charge conservation law, the Lorentz gauge will also hold, provided that the scalar electric potential ($\varphi$) is obtained from (\ref{HelmholtzF}).
However, due to the space-time discretization and the interpolation of quantities to the grids, a suitable algorithm that holds the charge conservation without violating energy and momentum conservation does not exist.
The approach that we follow in MITHRA is using the discretized form of (\ref{HelmholtzA}) and (\ref{HelmholtzF}) with the currents and charges of electrons (i.e. macro-particles) as the source and solving for the vector and scalar potential.
It can be shown that if current density ($\vec{J}$) and charge density ($\rho$) are interpolated similarly to the vertices of each cell, a proper discretization of current density based on positions of the macro-particles holds the charge conservation criterion.
To obtain the fields $\vec{E}$ and $\vec{B}$ at the grid points, we use the momentum conserving interpolation, which will be explained in the upcoming sections.

\subsection{FDTD for Helmholtz Equation}

In cartesian coordinates, a vector Helmholtz equation is written in form of three uncoupled scalar Helmholtz equations.
Therefore, it is sufficient to apply our discretization scheme only on a typical scalar Helmholtz equation: ${\nabla}^{2}\psi- \frac{1}{c^2} \frac{\partial^2 \psi}{ \partial t^2} = \zeta$, where $\psi$ stands for $A_l$ ($l \in \{x,y,z\}$); and $\zeta$ represents the term $-\mu_0 J_l$.
Let us begin with the central-difference discretization scheme for various partial differential terms of the scalar Helmholtz equation at the point $(i\Delta x,j\Delta y,k\Delta z,n\Delta t)$.
In the following equations, $\psi_{i,j,k}^n$ denotes the value of the quantity $\psi$ at the point $(i\Delta x,j\Delta y,k\Delta z)$ and time $n\Delta t$.
The derivatives are written as follows:
\begin{align}
\frac{\partial^{2}}{\partial x^{2}} \psi(x,y,z,t) & \simeq \frac{\psi_{i+1,j,k}^n-2\psi_{i,j,k}^n+\psi_{i-1,j,k}^n}{(\Delta x)^2}
\\
\frac{\partial^{2}}{\partial y^{2}} \psi(x,y,z,t) & \simeq \frac{\psi_{i,j+1,k}^n-2\psi_{i,j,k}^n+\psi_{i,j-1,k}^n}{(\Delta y)^2}
\\
\frac{\partial^{2}}{\partial z^{2}} \psi(x,y,z,t) & \simeq \frac{\psi_{i,j,k+1}^n-2\psi_{i,j,k}^n+\psi_{i,j,k-1}^n}{(\Delta z)^2}
\\
\frac{\partial^{2}}{\partial t^{2}} \psi(x,y,z,t) & \simeq \frac{\psi_{i,j,k}^{n+1}-2\psi_{i,j,k}^n+\psi_{i,j,k}^{n-1}}{(\Delta t)^2}.
\end{align}
Combining these four equations, one obtains the value of $\psi$ at instant $(n+1)\Delta t$ in terms of its value at $n\Delta t$ and $(n-1)\Delta t$:
\begin{equation}
\psi_{i,j,k}^{n+1} = -\psi_{i,j,k}^{n-1}+ \alpha_1 \psi_{i,j,k}^n + \alpha_2 \psi_{i+1,j,k}^n + \alpha_3 \psi_{i-1,j,k}^n + \alpha_4 \psi_{i,j+1,k}^n + \alpha_5 \psi_{i,j-1,k}^n + \alpha_6 \psi_{i,j,k+1}^n + \alpha_7 \psi_{i,j,k-1}^n + \alpha_8 \zeta_{i,j,k}^n \nonumber
\end{equation}
where the coefficients $\alpha_1, \ldots ,\alpha_7$ are obtained from:
\begin{equation}
\begin{array}{l}
\displaystyle
\alpha_1=2 \left[1-\left(\frac{c \Delta t}{\Delta x}\right)^2-\left(\frac{c \Delta t}{\Delta y}\right)^2-\left(\frac{c \Delta t}{\Delta z}\right)^2\right],
\qquad
\alpha_8=\left(c \Delta t\right)^2,
\\ \\ \displaystyle
\alpha_2=\alpha_3=\left(\frac{c \Delta t}{\Delta x}\right)^2,
\qquad
\alpha_4=\alpha_5=\left(\frac{c \Delta t}{\Delta y}\right)^2,
\qquad
\alpha_6=\alpha_7=\left(\frac{c \Delta t}{\Delta z}\right)^2
\end{array}
\end{equation}
The term $\zeta_{i,j,k}^n$ is the magnitude of the source term at the time $n \Delta t$, which is calculated from the particle motions.
Usually, one needs a finer temporal discretization for updating the equation of motion compared to electromagnetic field equations.
If the equation of motion is discretized and updated with $\Delta t_b = \Delta t / N$ time steps, the term $\zeta_{i,j,k}^n$ will be written in terms of the value after each $N$ update:
\begin{equation}
\label{currentIntegral}
\zeta_{i,j,k}^n = -\mu_0 J_l ( n \Delta t ) = -\mu_0 \rho ( n \Delta t ) \frac{\vec{r}^{n+1/2}-\vec{r}^{n-1/2}}{\Delta t}.
\end{equation}
As observed in the above equation, the position of particles are sampled at each $n+1/2$ time step, which later should be considered for updating the scalar potential.
This assumption also results in the calculation of charge density at $n+1/2$ time steps, which should be averaged for obtaining $\rho ( n \Delta t )$.

\subsection{Numerical Dispersion in FDTD}

It is well-known that the FDTD formulation for discretizing the Helmholtz equation suffers from the so-called numerical dispersion.
More accurately, the applied discretization leads to the wave propagation with a speed different from (lower than) the vacuum speed of light.
This may impact the FEL simulation results particularly during the saturation regime, owing to the important role played by the relative phase of electrons with respect to the radiated light.
Therefore, careful scrutiny of this effect and minimizing its impact is essential for the goal pursued by MITHRA.

To derive the equation governing such a dispersion, we assume a plane wave function for $\psi(x,y,z,t) = e^{-j(k_xx+k_yy+k_zz-\omega t)}$ in the discretized Helmholtz equation.
After some mathematical operations, the following equation is obtained for the dispersion properties of central-difference scheme:
\begin{equation}
\label{numericalDispersionCD}
\frac{\sin^2(k_x\Delta x/2)}{(\Delta x)^2} + \frac{\sin^2(k_y\Delta y/2)}{(\Delta y)^2} + \frac{\sin^2(k_z\Delta z/2)}{(\Delta z)^2} = \frac{\sin^2(\omega \Delta t/2)}{(c\Delta t)^2}.
\end{equation}
This equation is evidently different from the vacuum dispersion relation, which reads as
\begin{equation}
\label{vacuumDispersionCD}
k_x^2+k_y^2+k_z^2=\frac{\omega^2}{c^2}.
\end{equation}
Comparison of the two equations shows that the dispersion characteristics are similar, if and only if $\Delta x \rightarrow 0$, $\Delta y \rightarrow 0$, $\Delta z \rightarrow 0$, and $\Delta t \rightarrow 0$.
Another output of the dispersion equation is the stability condition, which is referred to as Courant-Friedrichs-Lewy (CFL) condition \cite{taflove2000computational}.
The spatial and temporal discretization should be related such that the term $\omega$ obtained from equation (\ref{numericalDispersionCD}) has no imaginary part, i.e. $\sin^2(\omega \Delta t/ 2) < 1$.
This implies that
\begin{equation}
c\Delta t < \frac{1}{\sqrt{\frac{\sin^2(k_x\Delta x/2)}{(\Delta x)^2} + \frac{\sin^2(k_y\Delta y/2)}{(\Delta y)^2} + \frac{\sin^2(k_z\Delta z/2)}{(\Delta z)^2}}} .
\end{equation}
The right hand side of the above equation has its minimum when all the sinus functions are equal to one, which leads to the stability condition for the central-difference scheme:
\begin{equation}
\label{CFLcondition}
\Delta t < \frac{1}{c\sqrt{\frac{1}{(\Delta x)^2} + \frac{1}{(\Delta y)^2} + \frac{1}{(\Delta z)^2}}}.
\end{equation}

As mentioned above, for the FEL simulation, it is very important to maintain the vacuum speed of light along the $z$ direction (Throughout this document $z$ is the electron beam and undulator direction).
More accurately, if $k_x=k_y=0$, $k_z=\omega/c$ should be the solution of the dispersion equation.
However, this solution is obtained if and only if $\Delta t = \Delta z /c$, which violates the stability condition.
To resolve this problem, various techniques are developed in the context of compensation of numerical dispersion.
Here, we take advantage from the non-standard finite difference (NSFD) scheme to impose the speed of light propagation along $z$ direction \cite{shlager2003comparison,finkelstein2007finite}.

The trick is to consider a weighted average along $z$ for the derivatives with respect to $x$ and $y$, which is formulated as follows:
\begin{align}
\frac{\partial^{2}}{\partial x^{2}} \psi(x,y,z,t) & \simeq \frac{\bar{\psi}_{i+1,j,k}^n-2\bar{\psi}_{i,j,k}^n+\bar{\psi}_{i-1,j,k}^n}{(\Delta x)^2}
\\
\frac{\partial^{2}}{\partial y^{2}} \psi(x,y,z,t) & \simeq \frac{\bar{\psi}_{i,j+1,k}^n-2\bar{\psi}_{i,j,k}^n+\bar{\psi}_{i,j-1,k}^n}{(\Delta y)^2},
\end{align}
with
\begin{equation}
\label{NSFDtrick}
\bar{\psi}_{i,j,k}^n = \mathcal{A} \psi_{i,j,k-1}^n + (1-2\mathcal{A}) \psi_{i,j,k}^n + \mathcal{A} \psi_{i,j,k+1}^n.
\end{equation}
Such a finite difference scheme leads to the following dispersion equation:
\begin{equation}
\label{NSFDDispersionCD}
\left( 1 - 4 \mathcal{A} \sin^2(k_z\Delta z/2) \right) \left( \frac{\sin^2(k_x\Delta x/2)}{(\Delta x)^2} + \frac{\sin^2(k_y\Delta y/2)}{(\Delta y)^2} \right) + \frac{\sin^2(k_z\Delta z/2)}{(\Delta z)^2} = \frac{\sin^2(\omega \Delta t/2)}{(c\Delta t)^2}.
\end{equation}
It can be shown that if the NSFD coefficient $\mathcal{A}$ is larger than 0.25, and $\sqrt{(\Delta z/\Delta x)^2+ (\Delta z/\Delta y)^2} < 1$, a real $\omega$ satisfies the above dispersion equation for $\Delta t = \Delta z / c$.
This time step additionally yields $k_z=\omega/c$, for $k_x=k_y=0$.

The value we chose for $\mathcal{A}$ in MITHRA is obtained from
\begin{equation}
\label{NSFDCoefficient}
\mathcal{A} = 0.25 \left( 1 + \frac{0.02}{(\Delta z/\Delta x)^2+ (\Delta z/\Delta y)^2} \right).
\end{equation}
The update equation can then be written as
\begin{align}
\label{updateEquation}
\psi_{i,j,k}^{n+1} & = -\psi_{i,j,k}^{n-1}+ \alpha'_1 \psi_{i,j,k}^n \nonumber \\
& + \alpha'_2 ( \mathcal{A} \psi_{i+1,j,k-1}^n + (1-2\mathcal{A}) \psi_{i+1,j,k}^n + \mathcal{A} \psi_{i+1,j,k+1}^n ) + \alpha'_3 ( \mathcal{A} \psi_{i-1,j,k-1}^n + (1-2\mathcal{A}) \psi_{i-1,j,k}^n + \mathcal{A} \psi_{i-1,j,k+1}^n ) \nonumber \\
& + \alpha'_4 ( \mathcal{A} \psi_{i,j+1,k-1}^n + (1-2\mathcal{A}) \psi_{i,j+1,k}^n + \mathcal{A} \psi_{i,j+1,k+1}^n ) + \alpha'_5 ( \mathcal{A} \psi_{i,j-1,k-1}^n + (1-2\mathcal{A}) \psi_{i,j-1,k}^n + \mathcal{A} \psi_{i,j-1,k+1}^n ) \nonumber \\
& + \alpha'_6 \psi_{i,j,k+1}^n + \alpha'_7 \psi_{i,j,k-1}^n + \alpha'_8 \zeta_{i,j,k}^n.
\end{align}
where the coefficients $\alpha'_1, \ldots ,\alpha'_7$ are obtained from:
\begin{equation}
\begin{array}{l}
\displaystyle
\alpha'_1=2 \left[1-(1-2\mathcal{A})\left(\frac{c \Delta t}{\Delta x}\right)^2-(1-2\mathcal{A})\left(\frac{c \Delta t}{\Delta y}\right)^2-\left(\frac{c \Delta t}{\Delta z}\right)^2\right],
\qquad
\alpha'_8=\left(c \Delta t\right)^2,
\\ \\ \displaystyle
\alpha'_2=\alpha'_3=\left(\frac{c \Delta t}{\Delta x}\right)^2,
\qquad
\alpha'_4=\alpha'_5=\left(\frac{c \Delta t}{\Delta y}\right)^2,
\qquad
\alpha'_6=\alpha'_7=\left(\frac{c \Delta t}{\Delta z}\right)^2 - 2\mathcal{A} \left(\frac{c \Delta t}{\Delta x}\right)^2 - 2\mathcal{A} \left(\frac{c \Delta t}{\Delta x}\right)^2.
\end{array}
\end{equation}
To guarantee a dispersion-less propagation along $z$ direction with the speed of light the update time step is automatically calculated from the given longitudinal discretization ($\Delta z$), according to $\Delta t = \Delta z / c$.

\subsection{FDTD for Scalar Potential}

Usually, due to high energy of particles in a FEL process, the FEL simulations neglect the space-charge effects by considering $\varphi \simeq 0$ \cite{andriyash2015spectral}.
However, this is an approximation which we try to avoid in MITHRA.
To account for space-charge forces, one needs to solve the Hemholtz equation for scalar potential, i.e. (\ref{HelmholtzF}).
For this purpose, the same formulation as used for the vector potential is utilized to update the scalar potential.
Nonetheless, since the position of particles are sampled at $t+\Delta t/2$ instants, the obtained value for $\varphi^n$ corresponds to the scalar potential at $(n+1/2)\Delta t$.
This point should be particularly taken into consideration, when electromagnetic fields $\vec{E}$ and $\vec{B}$ are evaluated.

\subsection{Boundary Truncation}

In order to simulate the FEL problem, we consider a cube as our simulation domain.
The absorbing boundary condition is also considered for updating the scalar electric potential $\varphi$ at the boundaries.
Therefore, we introduce the parameter $\xi$, which denotes either $\psi$ or $\varphi$.
The six boundaries of the cube are supposed to be at: $x=\pm l_{x}/2 $, $y=\pm l_{y}/2 $ and $z=\pm l_{z}/2 $.
In the following, the process for implementing Mur absorbing boundary conditions (ABCs) of the first and second order in MITHRA are discussed.
We only present the formulation for the boundary conditions at $z=\pm l_{z}/2 $.
The process to extract the equations for the other four boundaries will be exactly similar.

\subsubsection{First Order ABCs:}

The partial differential equations implying first order ABCs at $ z=\pm l_{z}/2 $ are:
\begin{equation}
\mp \frac{\partial^2 \xi}{\partial z \partial t} - \frac{1}{c}\frac{\partial^2 \xi}{\partial t^2}=0
\end{equation}
The discretized version for different terms appearing in the above equation reads as:
\begin{itemize}
\item At $z=-l_{z}/2 \; \; (k=0)$
\begin{align}
\frac{\partial^2 \xi}{\partial z \partial t} & \simeq \frac{1}{2 \Delta t } \left( \frac{\xi_{i,j,1}^{n+1}-\xi_{i,j,0}^{n+1}}{\Delta z}-\frac{\xi_{i,j,1}^{n-1}-\xi_{i,j,0}^{n-1}}{\Delta z} \right) \\
\frac{1}{c}\frac{\partial^2 \xi}{\partial t^2} & \simeq \frac{1}{2c} \left( \frac{\xi_{i,j,1}^{n+1}-2\xi_{i,j,1}^n+\xi_{i,j,1}^{n-1}}{\Delta t^2}+\frac{\xi_{i,j,0}^{n+1}-2\xi_{i,j,0}^n+\xi_{i,j,0}^{n-1}}{\Delta t^2} \right)
\end{align}
\item At $z=+l_{z}/2 \; \; (k=K=l_{z}/\Delta z)$
\begin{align}
\frac{\partial^2 \xi}{\partial z \partial t} & \simeq \frac{1}{2 \Delta t } \left( \frac{\xi_{i,j,K}^{n+1}-\xi_{i,j,K-1}^{n+1}}{\Delta z}-\frac{\xi_{i,j,K}^{n-1}-\xi_{i,j,K-1}^{n-1}}{\Delta z} \right) \\
\frac{1}{c} \frac{\partial^2 \xi}{\partial t^2} & \simeq \frac{1}{2c} \left( \frac{\xi_{i,j,K}^{n+1}-2\xi_{i,j,K}^n+\xi_{i,j,K}^{n-1}}{\Delta t^2} + \frac{\xi_{i,j,K-1}^{n+1}-2\xi_{i,j,K-1}^n+\xi_{i,j,K-1}^{n-1}}{\Delta t^2} \right)
\end{align}
\end{itemize}
Combining these equations, one obtains the boundary value of $\xi$ at instant $(n+1)\Delta t$ in terms of its values at $n\Delta t$ and $(n-1)\Delta t$: \\
\begin{itemize}
\item At $z=-l_{z}/2 \; \; (k=0)$
\begin{equation}
\xi_{i,j,0}^{n+1} = \beta_0 \xi_{i,j,0}^{n-1} + \beta_1 \xi_{i,j,0}^n+ \beta_2 \xi_{i,j,1}^{n-1} + \beta_3 \xi_{i,j,1}^n + \beta_4 \xi_{i,j,1}^{n+1}
\end{equation}
\item At $z=+l_{z}/2 \; \; (k=K=l_{z}/\Delta z)$
\begin{equation}
\xi_{i,j,K}^{n+1} = \beta_0 \xi_{i,j,K}^{n-1} + \beta_1 \xi_{i,j,K}^n+ \beta_2 \xi_{i,j,K-1}^{n-1} + \beta_3 \xi_{i,j,K-1}^n + \beta_4 \xi_{i,j,K-1}^{n+1}
\end{equation}
\end{itemize}
where:
\begin{equation}
\beta_0 = \beta_4 = \frac{c \Delta t - \Delta z}{c \Delta t + \Delta z},
\qquad
\beta_1 = \beta_3 = \frac{2 \Delta z}{c \Delta t + \Delta z},
\qquad
\beta_2 = -1
\end{equation}

\subsubsection{Second Order ABCs:}

The partial differential equations implying second order ABCs at $ z=\pm l_{z}/2 $ are:
\begin{equation}
\mp \frac{\partial^2 \xi}{\partial z \partial t} - \frac{1}{c}\frac{\partial^2 \xi}{\partial t^2} - \frac{c}{2}\frac{\partial^2 \xi}{\partial x^2} - \frac{c}{2}\frac{\partial^2 \xi}{\partial y^2}=0
\end{equation}
The discretized version for different terms appearing in the above equation reads as:
\begin{itemize}
\item At $z=-l_{z}/2 \; \; (k=0)$
\begin{align}
\frac{\partial^2 \xi}{\partial z \partial t} & \simeq \frac{1}{2 \Delta t } \left( \frac{\xi_{i,j,1}^{n+1}-\xi_{i,j,0}^{n+1}}{\Delta z}-\frac{\xi_{i,j,1}^{n-1}-\xi_{i,j,0}^{n-1}}{\Delta z} \right) \\
\frac{1}{c}\frac{\partial^2 \xi}{\partial t^2} & \simeq \frac{1}{2c} \left( \frac{\xi_{i,j,1}^{n+1}-2\xi_{i,j,1}^n+\xi_{i,j,1}^{n-1}}{\Delta t^2}+\frac{\xi_{i,j,0}^{n+1}-2\xi_{i,j,0}^n+\xi_{i,j,0}^{n-1}}{\Delta t^2} \right) \\
\frac{c}{2}\frac{\partial^2 \xi}{\partial x^2} & \simeq \frac{c}{4} \left( \frac{\xi_{i+1,j,1}^n-2\xi_{i,j,1}^n+\xi_{i-1,j,1}^n}{\Delta x^2}+\frac{\xi_{i+1,j,0}^n-2\xi_{i,j,0}^n+\xi_{i-1,j,0}^n}{\Delta x^2} \right) \\
\frac{c}{2}\frac{\partial^2 \xi}{\partial y^2} & \simeq \frac{c}{4} \left( \frac{\xi_{i,j+1,1}^n-2\xi_{i,j,1}^n+\xi_{i,j-1,1}^n}{\Delta y^2}+\frac{\xi_{i,j+1,0}^n-2\xi_{i,j,0}^n+\xi_{i,j-1,0}^n}{\Delta y^2} \right)
\end{align}
\item At $z=+l_{z}/2 \; \; (k=K=l_{z}/\Delta z)$
\begin{align}
\frac{\partial^2 \xi}{\partial z \partial t} & \simeq \frac{1}{2 \Delta t } \left( \frac{\xi_{i,j,K}^{n+1}-\xi_{i,j,K-1}^{n+1}}{\Delta z}-\frac{\xi_{i,j,K}^{n-1}-\xi_{i,j,K-1}^{n-1}}{\Delta z} \right)
\\
\frac{1}{c} \frac{\partial^2 \xi}{\partial t^2} & \simeq \frac{1}{2c} \left( \frac{\xi_{i,j,K}^{n+1}-2\xi_{i,j,K}^n+\xi_{i,j,K}^{n-1}}{\Delta t^2} + \frac{\xi_{i,j,K-1}^{n+1}-2\xi_{i,j,K-1}^n+\xi_{i,j,K-1}^{n-1}}{\Delta t^2} \right)
\\
\frac{c}{2}\frac{\partial^2 \xi}{\partial x^2} & \simeq \frac{c}{4} \left( \frac{\xi_{i+1,j,K}^n-2\xi_{i,j,K}^n+\xi_{i-1,j,K}^n}{\Delta x^2} + \frac{\xi_{i+1,j,K-1}^n-2\xi_{i,j,K-1}^n+\xi_{i-1,j,K-1}^n}{\Delta x^2} \right)
\\
\frac{c}{2}\frac{\partial^2 \xi}{\partial y^2} & \simeq \frac{c}{4} \left( \frac{\xi_{i,j+1,K}^n-2\xi_{i,j,K}^n+\xi_{i,j-1,K}^n}{\Delta y^2} + \frac{\xi_{i,j+1,K-1}^n-2\xi_{i,j,K-1}^n+\xi_{i,j-1,K-1}^n}{\Delta y^2} \right)
\end{align}
\end{itemize}
Combining these equations, one obtains the boundary value of $\xi$ at instant $(n+1)\Delta t$ in terms of its values at $n\Delta t$ and $(n-1)\Delta t$:
\begin{itemize}
\item At $z=-l_{z}/2 \; \; (k=0)$
\begin{align}
\xi_{i,j,0}^{n+1} = & \gamma_0 \xi_{i,j,0}^{n-1} + \gamma_1 \xi_{i,j,0}^n+ \gamma_2 \xi_{i,j,1}^{n-1} + \gamma_3 \xi_{i,j,1}^n + \gamma_4 \xi_{i,j,1}^{n+1} +
\\ \nonumber
& \gamma_5 \xi_{i+1,j,1}^n + \gamma_6 \xi_{i-1,j,1}^n + \gamma_7 \xi_{i,j+1,1}^n + \gamma_8 \xi_{i,j-1,1}^n +
\\ \nonumber
& \gamma_9 \xi_{i+1,j,0}^n + \gamma_{10} \xi_{i-1,j,0}^n + \gamma_{11} \xi_{i,j+1,0}^n + \gamma_{12} \xi_{i,j-1,0}^n
\end{align}
\item At $z=+l_{z}/2 \; \; (k=K=l_{z}/\Delta z)$
\begin{align}
\xi_{i,j,K}^{n+1} = & \gamma_0 \xi_{i,j,K}^{n-1} + \gamma_1 \xi_{i,j,K}^n+ \gamma_2 \xi_{i,j,K-1}^{n-1} + \gamma_3 \xi_{i,j,K-1}^n + \gamma_4 \xi_{i,j,K-1}^{n+1} +
\\ \nonumber
& \gamma_5 \xi_{i+1,j,K-1}^n + \gamma_6 \xi_{i-1,j,K-1}^n + \gamma_7 \xi_{i,j+1,K-1}^n + \gamma_8 \xi_{i,j-1,K-1}^n +
\\ \nonumber
& \gamma_9 \xi_{i+1,j,K}^n + \gamma_{10} \xi_{i-1,j,K}^n + \gamma_{11} \xi_{i,j+1,K}^n + \gamma_{12} \xi_{i,j-1,K}^n
\end{align}
\end{itemize}
where:
\begin{equation}
\begin{array}{l}
\displaystyle \gamma_0 = \gamma_4 = \frac{c \Delta t - \Delta z}{c \Delta t + \Delta z},
\qquad
\gamma_1 = \gamma_3 = \frac{\Delta z \left( 2 - ( c \Delta t / \Delta y )^2 - (c \Delta t / \Delta x )^2 \right) }{c \Delta t + \Delta z},
\qquad
\gamma_2 = -1 \\
\displaystyle
\gamma_5 = \gamma_6 = \gamma_9 = \gamma_{10} = \frac{(c \Delta t / \Delta x)^2 \Delta z}{2 (c \Delta t + \Delta z) },
\qquad
\gamma_7 = \gamma_8 = \gamma_{11} = \gamma_{12} = \frac{(c \Delta t / \Delta y)^2 \Delta z}{2 (c \Delta t + \Delta z)}
\end{array}
\end{equation}

Particular attention should be devoted to the implementation of Mur second order absorbing boundary condition at edges and corners.
Separate usage of the above equations for second order case encounters problems in the formulation.
On one hand, unknown values at grid points outside the computational domain appears in the equations, and on the other a system of overdetermined equations will be obtained.
The solution to this problem is to discretize all the involved boundary conditions at the center of the cubes (for corners) or squares (for edges).
A simple addition of the obtained equations cancels out the values outside the computational domain and returns the desired value meeting the considered absorbing boundary condition.

\section{Particle In Cell (PIC)}
\label{section PIC}

Particle in cell (PIC) method is the standard algorithm to solve for the motion of particles within an electromagnetic field distribution.
The method takes the time domain data of the fields $\vec{E}$ and $\vec{B}$ and updates the particle position and momentum according to the Lorentz force equation (\ref{LorentzForce}).
We comment that the electromagnetic fields in the motion equation are the total fields in the computational domain, which in a FEL problem is equivalent to the superposition of undulator field, radiated field and the seeded field in case of a seeded FEL problem.
Often considering all the individual particles involved in the problem ($\sim 10^6-10^9$ particles) leads to high computation costs and long simulation times.
The clever solution to this problem is the macro-particle assumption, through which an ensemble of particles ($\sim 10^2-10^4$ particles) are treated as one single entity with charge to mass ratio equal to the particles of interest, which are here electrons.
The relativistic equation of motion for electron macro-particles then reads as
\begin{equation}
\label{equationOfMotion}
\frac{\partial}{\partial t} (\gamma m \vec{v}) = -e(\vec{E}+ \vec{v} \times \vec{B}), \qquad \mathrm{and} \qquad \frac{\partial \vec{r}}{\partial t} = \vec{v},
\end{equation}
where $\vec{r}$ and $\vec{v}$ are the position and velocity vectors of the electron, $e$ is the electron charge and $m$ is its rest mass.
$\gamma$ stands for the Lortenz factor of the moving particle.

\subsection{Update Algorithm}

There are numerous update algorithms proposed for the time domain solution of (\ref{equationOfMotion}), including various Runge-Kutta and finite difference algorithms.
Among these methods, Boris scheme has garnered specific attention owing to its interesting peculiarity which is being simplectic.
Simplectic update algorithms are update procedures which maintain the conservation of any parameter in the equation which obey a physical conservation law.
Since in a FEL problem effect of the magnetic field on a particle motion plays the most important role, using a simplectic algorithm is essential to obtain reliable results.
This was the main motivation to choose the Boris scheme for updating the particle motion in MITHRA.

We sample the particle position at times $m\Delta t_b$, which is represented by $\vec{r}^m$ and the particle normalized momentum at times $(m-\frac{1}{2})\Delta t_b$ which is written as $\gamma \vec{\beta}^{m-1/2}$.
Then, by having $\vec{r}^m$ and $\gamma \vec{\beta}^{m-1/2}$ as the known parameters and $\vec{E}^m_t$ and $\vec{B}^m_t$ as the \emph{total} field values imposed on the particle at instant $m\Delta t$, the values $\vec{r}^{m+1}$ and $\gamma \vec{\beta}^{m+1/2}$ are obtained as follows:
\begin{equation}
\arraycolsep=1.0pt\def\arraystretch{2.0}
\label{BorisScheme}
\begin{array}{rcl}
\vec{t}_1 & = & \displaystyle \gamma \vec{\beta}^{m-1/2} - \frac{e\Delta t_b \vec{E}^m_t}{2mc} \\
\vec{t}_2 & = & \displaystyle \vec{t}_1 + \alpha \vec{t}_1 \times \vec{B}^m_t \\
\vec{t}_3 & = & \displaystyle \vec{t}_1 + \vec{t}_2 \times \frac{2 \alpha \vec{B}^m_t}{1+\alpha^2 \vec{B}^m_t \cdot \vec{B}^m_t} \\
\gamma \vec{\beta}^{m+1/2} & = & \displaystyle \vec{t}_3 - \frac{e\Delta t_b \vec{E}^m_t}{2mc} \\
\displaystyle \vec{r}^{m+1} & = & \displaystyle \vec{r}^l + \frac{c \Delta t_b \gamma \vec{\beta}^{m+1/2} }{\sqrt{1+\gamma \vec{\beta}^{m+1/2} \cdot \gamma \vec{\beta}^{m+1/2} } }
\end{array}
\end{equation}
with $\alpha = -e\Delta t_b / (2 m \sqrt{1+\vec{t}_1 \cdot \vec{t}_1})$.
$\vec{E}^m_t = \vec{E}^m_{ext} + \vec{E}^m$ and $\vec{B}^m_t = \vec{B}^m_{ext} + \vec{B}^m$ are total fields imposed on the particle, which are equal to the superposition of the radiated field with the external fields, i.e. the undulator or the seed fields.
In order to figure out the derivation of the equations (\ref{BorisScheme}), the reader is referred to \cite{boris1,boris2}.
As seen from the above equations, the electric and magnetic fields at time $m \Delta t_b$ and the position $\vec{r}$ of the particle are needed to update the motion.
In the next section, the equations to extract these values from the computed values of the magnetic and scalar potential are presented.
Note that to achieve a certain precision level, the required time step in updating the bunch properties ($\Delta t_b$) is usually much smaller than the time step for field update ($\Delta t$).
In MITHRA, there exists the possibility for setting different time steps for PIC and FDTD algorithms.

\subsection{Field Evaluation}

As described in section \ref{section FDTD}, the propagating fields in the computational domain are evaluated by solving the Helmholtz equation for the magnetic vector potential, i.e. (\ref{HelmholtzA}).
To update the particle position and momentum, one needs to obtain the field values $\vec{E}^m$ and $\vec{B}^m$ from the potentials $\vec{A}$ and $\varphi$.
For this purpose, the equations (\ref{BvsA}) and (\ref{EvsA}) need to be discretized in a consistent manner to provide the accelerating field with lowest amount of dispersion and instability error.
\begin{figure}
\centering
\includegraphics[height=2.0in]{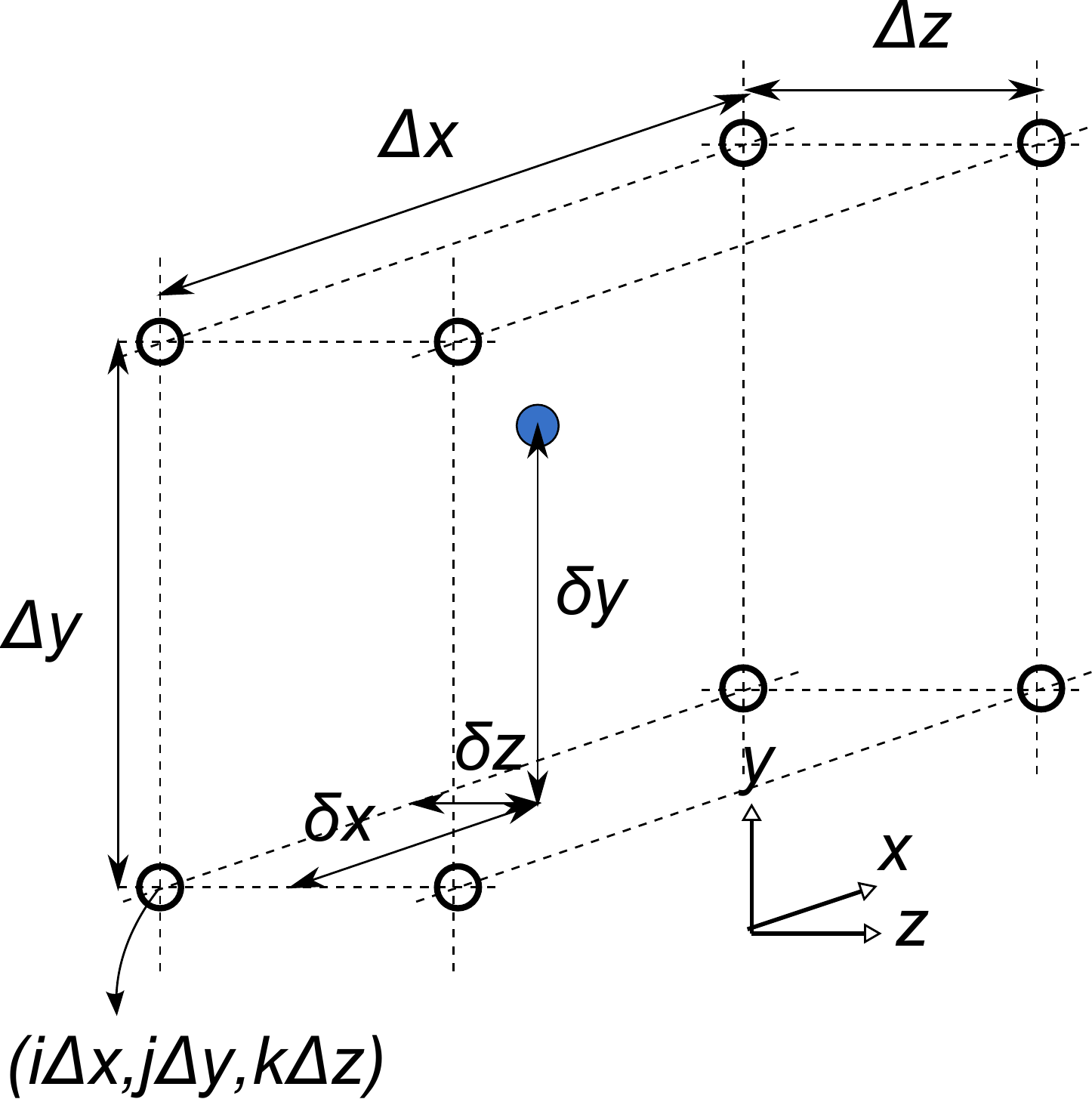}
\caption{Schematic illustration of the parameters used to locate a particle within the computational domain.}
\label{FDTDPICFig1}
\end{figure}
First, the values of magnetic and scalar potentials at $t+\Delta t/2$ are used to evaluate the electromagnetic fields at the cell vertices.
Subsequently, the field values are interpolated to the particle location for updating the equation of motion.
An important consideration at this stage is compatible interpolation of fields from the cell vertices with the interpolations used for current and charge densities.
Similar interpolation algorithms should be followed to cancel the effect of self-forces on particle motion.

Using the equation (\ref{BvsA}), the magnetic field $\vec{B}^n_{i,j,k}$ at cell vertex $(i,j,k)$ is calculated as follows:
\begin{align}
{B_x}^n_{i,j,k} = & \frac{1}{2} \left( \frac{{A_z}_{i,j+1,k}^n-{A_z}_{i,j-1,k}^n}{2\Delta y} - \frac{{A_y}_{i,j,k+1}^n-{A_y}_{i,j,k-1}^n}{2\Delta z} + \frac{{A_z}_{i,j+1,k}^{n+1}-{A_z}_{i,j-1,k}^{n+1}}{2\Delta y} - \frac{{A_y}_{i,j,k+1}^{n+1}-{A_y}_{i,j,k-1}^{n+1}}{2\Delta z} \right), \\
{B_y}^n_{i,j,k} = & \frac{1}{2} \left( \frac{{A_x}_{i,j,k+1}^n-{A_x}_{i,j,k-1}^n}{2\Delta z} - \frac{{A_z}_{i+1,j,k}^n-{A_z}_{i-1,j,k}^n}{2\Delta x} + \frac{{A_x}_{i,j,k+1}^{n+1}-{A_x}_{i,j,k-1}^{n+1}}{2\Delta z} - \frac{{A_z}_{i+1,j,k}^{n+1}-{A_z}_{i-1,j,k}^{n+1}}{2\Delta x} \right), \\
{B_z}^n_{i,j,k} = & \frac{1}{2} \left( \frac{{A_y}_{i+1,j,k}^n-{A_y}_{i-1,j,k}^n}{2\Delta x} - \frac{{A_x}_{i,j+1,k}^n-{A_x}_{i,j-1,k}^n}{2\Delta y} + \frac{{A_y}_{i+1,j,k}^{n+1}-{A_x}_{i-1,j,k}^{n+1}}{2\Delta x} - \frac{{A_x}_{i,j+1,k}^{n+1}-{A_x}_{i,j-1,k}^{n+1}}{2\Delta y} \right).
\end{align}
Similarly, equation (\ref{EvsA}) is employed to evaluate the electric field at the cell vertices.
The electric field $\vec{E}^n_{i,j,k}$ is obtained from the following equations:
\begin{align}
{E_x}^n_{i,j,k} = & \left( -\frac{{A_x}_{i,j,k}^{n+1}-{A_x}_{i,j,k}^n}{\Delta t} - \frac{\varphi_{i+1,j,k}^n-\varphi_{i-1,j,k}^n}{2\Delta x} \right), \\
{E_y}^n_{i,j,k} = & \left( -\frac{{A_y}_{i,j,k}^{n+1}-{A_y}_{i,j,k}^n}{\Delta t} - \frac{\varphi_{i,j+1,k}^n-\varphi_{i,j-1,k}^n}{2\Delta y} \right), \\
{E_z}^n_{i,j,k} = & \left( -\frac{{A_z}_{i,j,k}^{n+1}-{A_z}_{i,j,k}^n}{\Delta t} - \frac{\varphi_{i,j,k+1}^n-\varphi_{i,j,k-1}^n}{2\Delta z} \right).
\end{align}

Suppose that a particle resides at the cell $ijk$ with the grid point indices shown in Fig.\,\ref{FDTDPICFig1}. %
As illustrated in Fig.\ref{FDTDPICFig1}, the distance to the corner $(i\Delta x, j\Delta y, k \Delta z)$ is assumed to be $(\delta x, \delta y, \delta z)$.
We use a linear interpolation of the fields from the vertices to the particle position to calculate the imposed field.
If $\varsigma$ denotes for a component of the electric or magnetic field, i.e. $\varsigma \in \{ E_x, E_y, E_z, B_x, B_y, B_z \}$, one can write
\begin{equation}
\label{fieldInterpolation}
\varsigma^p = \sum\limits_{I,J,K} \left(\frac{1}{2} + (-1)^I \left| \frac{1}{2} - \frac{\delta x}{\Delta x} \right| \right)
\left(\frac{1}{2} + (-1)^J \left| \frac{1}{2} - \frac{\delta y}{\Delta y} \right| \right)
\left(\frac{1}{2} + (-1)^K \left| \frac{1}{2} - \frac{\delta z}{\Delta z} \right| \right) \varsigma_{i+I,j+J,k+K},
\end{equation}
where $I$, $J$, and $K$ are equal to either 0 or 1, producing the eight indices corresponding to the eight corners of the mesh cell.

\subsection{Current Deposition}

Once the position and momentum of all the particles over the time interval $\Delta t $ is known, one needs to couple the pertinent currents into the Helmholtz equation (\ref{HelmholtzA}).
As described before, this coupling over time is implemented through the equation (\ref{currentIntegral}).
The remaining question is how to evaluate the related currents on the grid points, i.e. the method for performing an spatial interpolation.
To maintain consistency, we should use a similar interpolation scheme as used for the field evaluation.
This assumption leads to the following equation for spatial interpolation.
\begin{equation}
{\rho^p}_{i+I,j+J,k+K} = \rho \left(\frac{1}{2} + (-1)^I \left| \frac{1}{2} - \frac{\delta x}{\Delta x} \right| \right)
\left(\frac{1}{2} + (-1)^J \left| \frac{1}{2} - \frac{\delta y}{\Delta y} \right| \right)
\left(\frac{1}{2} + (-1)^K \left| \frac{1}{2} - \frac{\delta z}{\Delta z} \right| \right)
\end{equation}
where $\rho$ is the charge density attributed to each macro-particle, namely $q/(\Delta x \Delta y \Delta z)$.
${\rho^p}_{i,j,k}$ is the charge density at the grid point $(i,j,k)$ due to the moving particle $p$ in the computational mesh cell (Fig.\,\ref{FDTDPICFig1}a).
$I$, $J$, and $K$ are equal to either 0 or 1, which produce the eight indices corresponding to the eight corners of the mesh cell.
The total charge density $\rho_{i,j,k}$ will be a superposition of all the charge densities due to the moving particles of the bunch.
We have removed the superscripts corresponding to the time instant, to avoid the confusion due to different time marching steps $\Delta t$ and $\Delta t_b$.
The above interpolation is carried out at each update step of the field values.
One can consider the above interpolation equations as a rooftop charge distribution centered at the particle position and expanding in the regions $(-\Delta x < x < \Delta x, -\Delta y < y < \Delta y, -\Delta z < z < \Delta z)$.
Eventually, the equation (\ref{currentIntegral}) is used to calculate the corresponding current densities.

\section{Quantity Initialization}

The previous two sections on FDTD and PIC algorithms present a suitable and efficient framework for the computation of interaction between charged particles and propagating waves.
However, the initial conditions are always required for a complete determination of the problem of interest.
For a FEL simulation, the initial conditions corresponding to the FEL input are given to the FDTD/PIC solver.
For example, in case of a SASE (Self Amplified Spontaneous Emission) FEL, the initial fields are zero and there is no excitation entering the computational domain, whereas for a seeded FEL, an outside excitation should be considered entering the computational domain.
The explanation of how such initializations are implemented in MITHRA is the goal in this section.

One novel feature of the method, followed here, is the solution of Maxwell's equations in the bunch rest frame.
It can be shown that a proper coordinate transformation yields the matching of all the major parameters in a FEL simulation, namely bunch length, undulator period, undulator length, and radiation wavelength.
\begin{figure}
\centering
\includegraphics[width=7.0in]{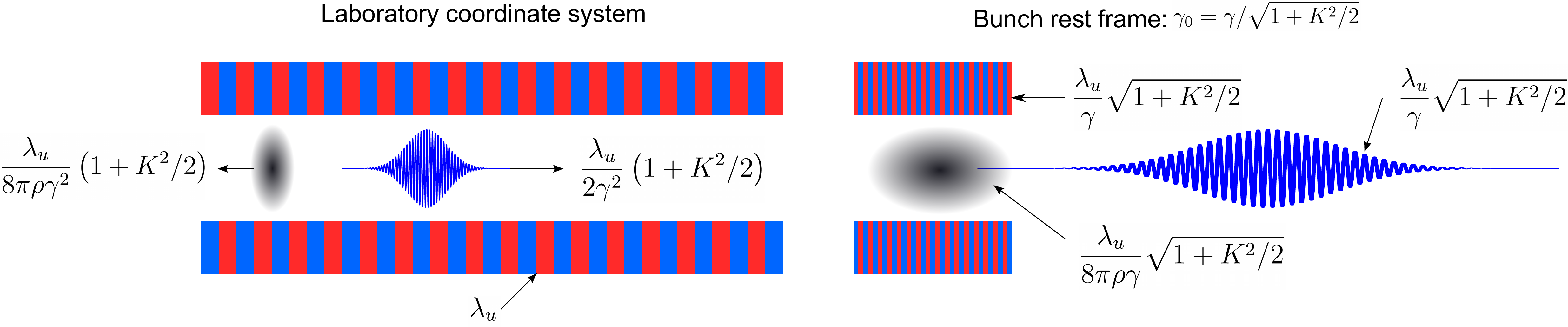}
\caption{Schematic illustration of the Lorentz boosting to transform the problem from the laboratory frame to the bunch rest frame.}
\label{FDTDPICFig2}
\end{figure}
Fig.\,\ref{FDTDPICFig2} schematically describes the advantage of moving into the bunch rest frame.
In a typical FEL problem, the FEL parameter $\rho_{FEL}$ is about $10^{-3}$.
Therefore, simulation of FEL interaction with a bunch equal to the cooperation length of the FEL ($L_c=\lambda_l/(4 \pi \rho_{FEL})$, with $\lambda_l$ being the radiation wavelength) requires a simulation domain only 100 times larger than the wavelength.
This becomes completely possible with the today computer technology and constitutes the main goal of MITHRA.
In this section, the main basis for Lorentz boosting the simulation coordinate is described first.
Afterwards, the relations for evaluating the undulator fields in the Lorentz boosted framework are presented.
Finally, the electron bunch initialization in the Lorentz-boosted framework is discussed.

\subsection{Lorentz Transformation}

It is known from the FEL thoery that a bunch with central Lorentz factor equal to $\gamma$ moves in an undulator with an average Lorentz factor equal to $\gamma_0=\gamma/\sqrt{(1+K^2/2)}$, where $K=eB\lambda_u/(2\pi m c)$ is the undulator parameter determining the amplitude of the wiggling motion.
Consequently, a frame moving with normalized velocity $\beta_0 = \sqrt{1-1/\gamma_0^2}$ is indeed the bunch rest frame, where the volume of the computational domain stays minimal.
Transforming into this coordinate system necessitates tailoring the bunch and undulator properties.
For this purpose, the Lorentz length contraction, time dilation and relativistic velocity addition need to be employed.

In MITHRA, the input parameters are all taken in the laboratory frame and the required Lorentz transformations are carried out based on the bunch energy.
The required transformations for the computational mesh are as the following:
\begin{align}
\Delta z & = \Delta z' \gamma_0,  \label{LorentzTransformB}\\
\Delta t & = \Delta t' / \gamma_0,  \\
\Delta t_b & = \Delta t'_b / \gamma_0,
\end{align}
where the prime sign stands for the quantities in the laboratory frame.
The quantities without prime are values in the bunch rest frame, which are used in the FDTD/PIC simulation.
With the consideration of the above transformations, the length of the total computational domain along the undulator period and the total simulation time is also transformed similarly.

In addition to the data for the computational mesh, the properties of the electron bunch also changes after the Lorentz boosting.
This certainly affects the bunch initialization process which is thoroughly explained in the next section.
An electron bunch in MITHRA is initialized and characterized by the following parameters:
\begin{enumerate}[(i)]
\item Mean electron position: $(\bar{x}_b, \bar{y}_b, \bar{z}_b)$,
\item Mean electron normalized momentum: $(\overline{\gamma \beta}_x, \overline{\gamma \beta}_y, \overline{\gamma \beta}_z)$,
\item RMS value of the electron position distribution: $(\sigma_x, \sigma_y, \sigma_z)$,
\item RMS value of the electron normalized momentum distribution: $(\sigma_{\gamma \beta_x}, \sigma_{\gamma \beta_y}, \sigma_{\gamma \beta_z})$.
\end{enumerate}
As mentioned previously, the above parameters are entered by the user in the laboratory frame.
To transform the given values to the bunch rest frame the position related parameters are changed as
\begin{equation}
\begin{array}{cccccc}
\bar{x}_b = \bar{x}'_b, & \bar{y}_b = \bar{y}'_b, & \bar{z}_b = \displaystyle \frac{\bar{z}'_b}{\gamma_0 (1 - \bar{\beta'}_z \beta_0)} , & \sigma_x = \sigma'_x, & \sigma_y = \sigma'_y, & \sigma_z = \displaystyle \frac{\sigma'_z}{{\gamma_0 (1 - \bar{\beta'}_z \beta_0)}}.
\end{array}
\end{equation}
To transfer the momentum related quantities, we assume that the main contribution to the Lorentz factor is the momentum along $z$ direction or the undulator period.
In other words, $(\overline{\gamma \beta}_x, \overline{\gamma \beta}_y, \overline{\gamma \beta}_z) = \gamma ( \bar{\beta}_x, \bar{\beta}_y, \bar{\beta}_z )$, with $\gamma = 1/\sqrt{1-\bar{\beta}_z^2}$.
Similarly, the RMS values can also be written as $(\sigma_{\gamma \beta_x}, \sigma_{\gamma \beta_y}, \sigma_{\gamma \beta_z}) = \gamma (\sigma_{\beta_x}, \sigma_{\beta_y}, \sigma_{\beta_z})$.
Using the relativistic velocity transformation \cite{JacksonClassical}, the transformation equations for the above values are found as follows:
\begin{align}
\gamma & = \gamma' \gamma_0 (1 - \bar{\beta'}_z \beta_0), \\
( \bar{\beta}_x, \bar{\beta}_y, \bar{\beta}_z ) & = ( \bar{\beta'}_x, \bar{\beta'}_y, \sqrt{1 - 1/\gamma^2} ), \\
(\sigma_{\gamma \beta_x}, \sigma_{\gamma \beta_y}, \sigma_{\gamma \beta_z}) & = (\sigma'_{\gamma \beta_x}, \sigma'_{\gamma \beta_y}, \sigma'_{\gamma \beta_z} ) \gamma_0 (1 - \bar{\beta'}_z \beta_0). \label{LorentzTransformE}
\end{align}
Equations (\ref{LorentzTransformB})-(\ref{LorentzTransformE}) provide a sufficient set of equations to perform the Lorentz boost to the bunch rest frame.

\subsection{Field Initialization}
\label{fieldInitialization}

The utilized FDTD/PIC algorithm solves the Maxwell's equation coupled with the motion equation of an ensemble of particles.
Therefore, in addition to the field values, particle initial conditions should also be initialized.
For a SASE FEL problem, the initial field profile is zero everywhere, whereas for a seeded FEL the initial seed should enter the computational domain through the boundaries.
In both cases, the external field which is the undulator field should separately be initialized.

\subsubsection{Undulator Field:}

By solving the Laplace equation for the magnetic field, the undulator field in the laboratory frame is found to be as the following (Fig.\,\ref{FDTDPICFig3}) \cite{FEL2}:
\begin{figure}
\centering
\includegraphics[width=3.2in]{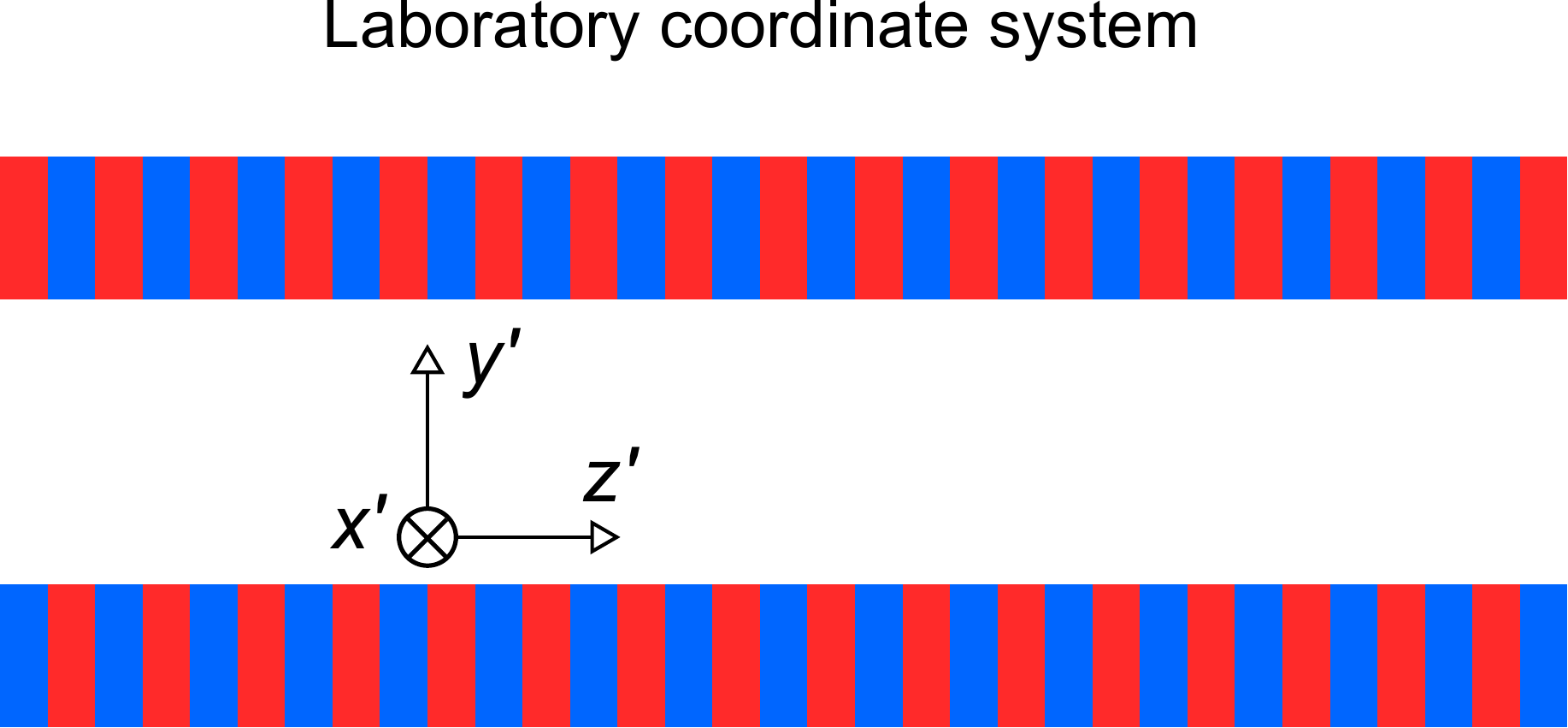}
\caption{Schematic illustration of the undulator in the lab frame and the definition of the coordinates.}
\label{FDTDPICFig3}
\end{figure}
\begin{align}
\label{undulatorField}
B'_x & = 0, \nonumber \\
B'_y & = B_0 \cosh(k_uy')\sin(k_uz'), \\
B'_z & = B_0 \sinh(k_uy')\cos(k_uz'), \nonumber
\end{align}
where $B_0$ is the maximum transverse field of the undulator.
To calculate the undulator field in the bunch rest frame, first the position is transformed to laboratory frame $(x',y',z')$ through the Lorentz boost equations.
Afterwards, the field is evaluated using the equation (\ref{undulatorField}).
Eventually, these fields are transformed back into the bunch rest frame.
Following this procedure, the undulator field in the bunch rest frame is obtained as
\begin{align}
\label{undulatorFieldRest}
B_x & = 0, \nonumber & E_x & = 0 \nonumber \\
B_y & = \gamma_0 B_0 \cosh(k_u y)\sin(k_u l_z), & E_y & = 0 \\
B_z & = B_0 \sinh(k_uy)\cos(k_u l_z), & E_z & = \gamma_0 \beta_0 c B_0 \cosh(k_u y)\sin(k_u l_z), \nonumber
\end{align}
with $l_z = \gamma_0 ( z' + \beta_0 c t - z_0 )$ and $z_0$ being the initial distance between the bunch and undulator in the bunch rest frame.

An important consideration in the initialization of undulator field is the entrance region of the undulator.
A direct usage of the equation (\ref{undulatorFieldRest}) with zero field for $z'<0$ causes an abrupt variation in the particles motion, which results in a spurious coherent radiation.
In fact, in a real undulator, there exists fringing fields at the undulator entrance, which remove any abrupt transition in the undulator field and consequently the particle radiations  \cite{sagan2003magnetic}.
To the best of our knowledge, the fringing fields are always modeled numerically and there exists no analytical solution for the problem.
Here, we approximate the fringing fields by a gradually decreasing magnetic field in form of a Neumann function.
The coefficients in the function are set such that the particles do not gain any net transverse momentum and stay in the computational domain as presumed.
The undulator field for $z'<0$ in the laboratory frame is obtained as the following:
\begin{align}
B'_x & = 0, \nonumber \\
B'_y & = B_0 \cosh(k_uy')k_uz'e^{-(k_uz')^2/2}, \\
B'_z & = B_0 \sinh(k_uy')e^{-(k_uz')^2/2}, \nonumber
\label{fringingField}
\end{align}
The same transformations as in (\ref{undulatorFieldRest}) can be used to approximate the fringing field values in the bunch rest frame.

\subsubsection{Seed Field:}

External excitation of free electron laser process using a seed mechanism has proved to be advantageous in terms of output spectrum, photon flux and the required undulator length  \cite{pellegrini2016physics,FEL2}.
Such benefits has propelled the proposal of seeded FEL schemes.
To simulate such a mechanism, MITHRA uses the TF/SF (total-field/scattered-field) technique to introduce an external excitation into the computational domain.
When seeding is enabled by having a non-zero seed amplitude, the second and third points (after the boundary points) constitute the scattered and total field boundaries, respectively.
Therefore, during the time marching process, after each update according to equation (\ref{updateEquation}) the excitation terms are added to the fields at TF/SF boundaries.
For example for the TF/SF boundaries close to $z=z_{min}$ plane, the field values to be used in the next time steps are obtained as the following:
\begin{align}
\text{ SF boundary:  } & \psi_{i,j,k}^{'n+1} = \psi_{i,j,k}^{n+1} + \mathcal{A} ( \alpha'_2 f_{i+1,j,k+1}^n  + \alpha'_3 f_{i-1,j,k+1}^n + \alpha'_4 f_{i,j+1,k+1}^n + \alpha'_5 f_{i,j-1,k+1}^n ) + \alpha'_6 f_{i,j,k+1}^n, \nonumber \\
\text{ TF boundary:  } & \psi_{i,j,k}^{'n+1} = \psi_{i,j,k}^{n+1} - \mathcal{A} ( \alpha'_2 f_{i+1,j,k-1}^n  + \alpha'_3 f_{i-1,j,k-1}^n + \alpha'_4 f_{i,j+1,k-1}^n + \alpha'_5 f_{i,j-1,k-1}^n ) - \alpha'_7 f_{i,j,k-1}^n,
\end{align}
where $f_{i,j,k}^n$ is the excitation value at time $n\Delta t$ and position $(i\Delta x,j\Delta y,k\Delta z)$.
The excitation value is calculated based on the imposed seed fields, which are usually either a plane wave or a Gaussian beam radiation.

\subsection{Electron Bunch Generation}

\subsubsection{Position and momentum initialization:}

As described previously, the evolution of the electron bunch is always simulated by following the macro-particle approach, where an ensemble of particles are represented by one sample particle.
This typically reduces the amount of computation cost for updating the bunch properties by three or four orders of magnitude.
Due to the high sensitivity of a FEL problem to the initial conditions, correct and proper initialization of these macro-particles play a critical role in obtaining reliable results.
In computational accelerator physics, different approaches are introduced and developed for bunch generation.
Some examples are random generation of particles, mirroring macro-particles at different phases to prevent initial average bunching factors, and independent random filling of different coordinates to prevent unrealistic correlations \cite{reiche2000numerical}.
Among all the different methods, using the sophisticated methods to load the bunch in a "quasi-random" manner seem to be the most appropriate solutions.
The Halton or Hammersley sequences, as generalizations of the bit-reverse techniques, are implemented in MITHRA for particle generation.
These sequences compared to random based filling of the phase space avoid the appearance of local clusters in the bunch distribution.
In addition, the uniform filling of the phase space prevents initial bunching factor of the generated electron bunch, making it well-suited for FEL simulations.

For details on the nature of Halton sequences, the reader is referred to the specialized documents.
Here, we only present the implemented algorithm to generate the required sequence of numbers filling the interval $[0,1]$.
The following C++ code is integrated into MITHRA which produces $N<20$ uncorrelated sequences of $M$ numbers in the interval $[0,1]$:
\begin{Verbatim}[frame=single, fontsize=\small, tabsize=4]
std::vector<std::vector<Double> > haltonSequence (unsigned int N, unsigned int M)
{
	std::vector<std::vector<Double> > s (M, std::vector<Double> (N, 0.0) );
	unsigned int prime [20] = {2, 3, 5, 7, 11, 13, 17, 19, 23, 29, 31, 37, 41, 43, 47, 53, 59,
	                           61, 67, 71};
	int p0, p, k, k0, a;
	double x = 0.0;
	for (unsigned int i = 0; i < size; i++)
	{
		k0 = i;
		for (unsigned int j = 0; j < dimension; j++)
		{
			p  = prime[j];
			p0 = p;
			k  = k0;
			x  = 0.0;
			while (k > 0)
			{
				a   = k % p;
				x  += a / (double) p0;
				k   = int (k/p);
				p0 *= p;
			}
			s[i][j] = x;
		}
	}
	return s;
}
\end{Verbatim}
By having the above uniform distributions, the 6D phase space of the initial bunch can be filled according to the desired bunch properties.

In MITHRA, different schemes for the user is implemented to generate the initial electron bunch, which are described in chapter \ref{chapter_refcard}.
The main requirements for initializing the bunches is to generate 1D and 2D set of numbers with either uniform or Gaussian distributions.
Suppose $x_1$ and $x_2$ are two uncorrelated number sequences produced by the Halton algorithm.
A 1D uniform distribution $y_1$ with average $y_{m1}$ and total width $y_{s1}$ is found by the following transformation:
\begin{equation}
\label{uniform1D}
y_1 = y_{s1} (x_1 - \frac{1}{2}) + y_{m1}.
\end{equation}
Such a distribution is used when a bunch with uniform current profile ($z$ distribution of particles) is to be initialized.
On the other hand, a 1D Gaussian distribution is needed when radiation of a bunch with Gaussian current profile is modelled.
To generate bunches with Gaussian distribution, we employ Box-muller's theory to extract a sequence of numbers with Gaussian distribution from two uncorrelated uniform distributions.
Based on this theory, a 1D Gaussian distribution $y_2$ with average $y_{m2}$ and deviation width $y_{s2}$ is found by the following transformation:
\begin{equation}
\label{gaussian1D}
y_2 = y_{s2} \sqrt{-2 \ln x_1} \cos(2\pi x_2) + y_{m2}.
\end{equation}

Similar to the undulator fields, an abrupt variation in the bunch profile results in an unrealistic coherent scattering emission (CSE), which happens if the uniform bunch distribution is directly initialized from equation (\ref{uniform1D}).
CSE is avoided by imposing smooth variations in the particle distribution.
For this purpose, we follow the procedure proposed in \cite{penman199282} and \cite{reiche2000numerical}.
A small Gaussian bunch with the same density as the real bunch and a width equal to an undulator wavelength is produced.
The lower half of the bunch (particles with smaller $z$) is transferred to the tail and the other half is placed at the head of the uniform bunch.
Hence, a uniform current profile with smooth variations at its head and tail is created.

The transverse coordinates of the bunches are initialized using 2D distributions.
In MITHRA, a 2D Gaussian distribution is assumed for transverse coordinates.
To generate such a distribution, two independent sets of numbers $x_1$ and $x_2$ are generated based on Halton sequence.
The desired 2D Gaussian distribution with average position $(y_{m3},y_{m4})$ and total deviation $(y_{s3},y_{s4})$ is produced as the following: 
\begin{equation}
\label{gaussian2D}
\displaystyle y_3 = y_{s3} \sqrt{-2 \ln x_1} \cos(2\pi x_2) + y_{m3}, \qquad \mathrm{and} \qquad
\displaystyle y_4 = y_{s4} \sqrt{-2 \ln x_1} \sin(2\pi x_2) + y_{m4}.
\end{equation}
Such algorithms are similarly used to generate the distribution in particle momenta.
The only difference is that for initializing a distribution in momentum merely Gaussian profiles are considered in transverse and longitudinal coordinates.
The method to introduce these bunch types are described in the next chapter.

\subsubsection{Bunching factor:}

Free electron laser radiation should start from a nonzero initial radiation.
This radiation can be in form of an initial seed field, initial modulation in the bunch, or the radiation from bunch shot noise.
In the current version of MITHRA, we have implemented the first two types.
The implementation of shot noise is postponed to the next versions of the software.
The implementation of seeding through an external excitation using TF/SF boundaries was described in \ref{fieldInitialization}.
Here, we explain how an initial bunching factor, $<e^{jk_uz}>$, is introduced to the electron bunch profile.

For this purpose, a small variation $\delta z$ is applied to a particle distribution, which is generated using the above formulations.
$\delta z$ for each particle is obtained from
\begin{equation}
\label{bunchingFactor}
\delta z = \xi \gamma_0 k_u b_f \sin (2 \xi \gamma_0 k_u z),
\end{equation}
where $b_f$ is the given bunching factor of the distribution, and $\xi=1+\bar{\beta}_z/\beta_0$ accounts for the change in the bunch longitudinal velocity after entering the undulator.
The introduced variation to the bunch coordinates, i.e. $z \rightarrow z+\delta z$, yields a bunch with all the given particle and momentum distributions and the desired bunching factor, $b_f$.

\section{Parallelization}

\begin{figure}
\centering
$\begin{array}{ccc}
\includegraphics[width=2.5in]{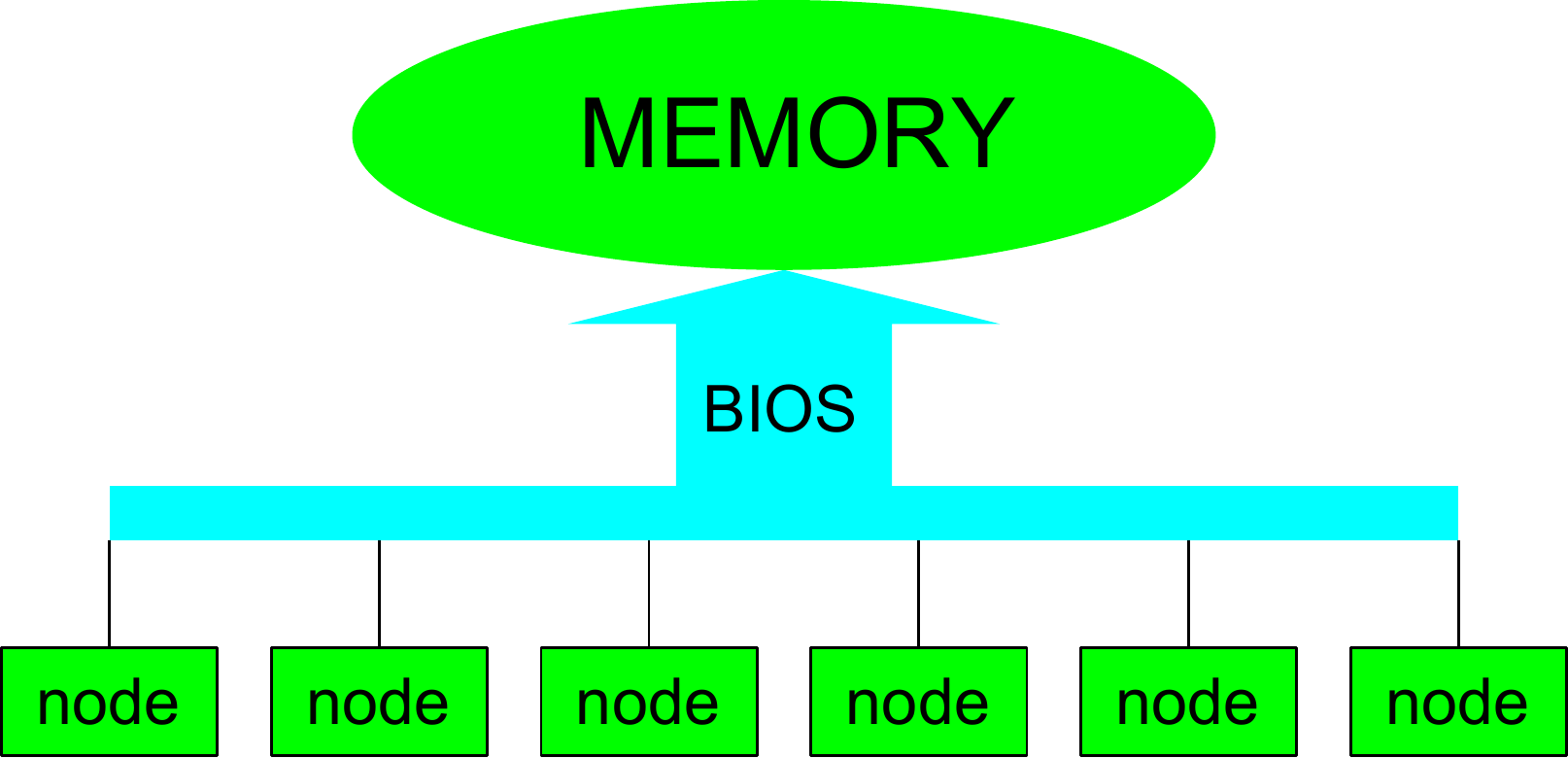} & &
\includegraphics[width=2.5in]{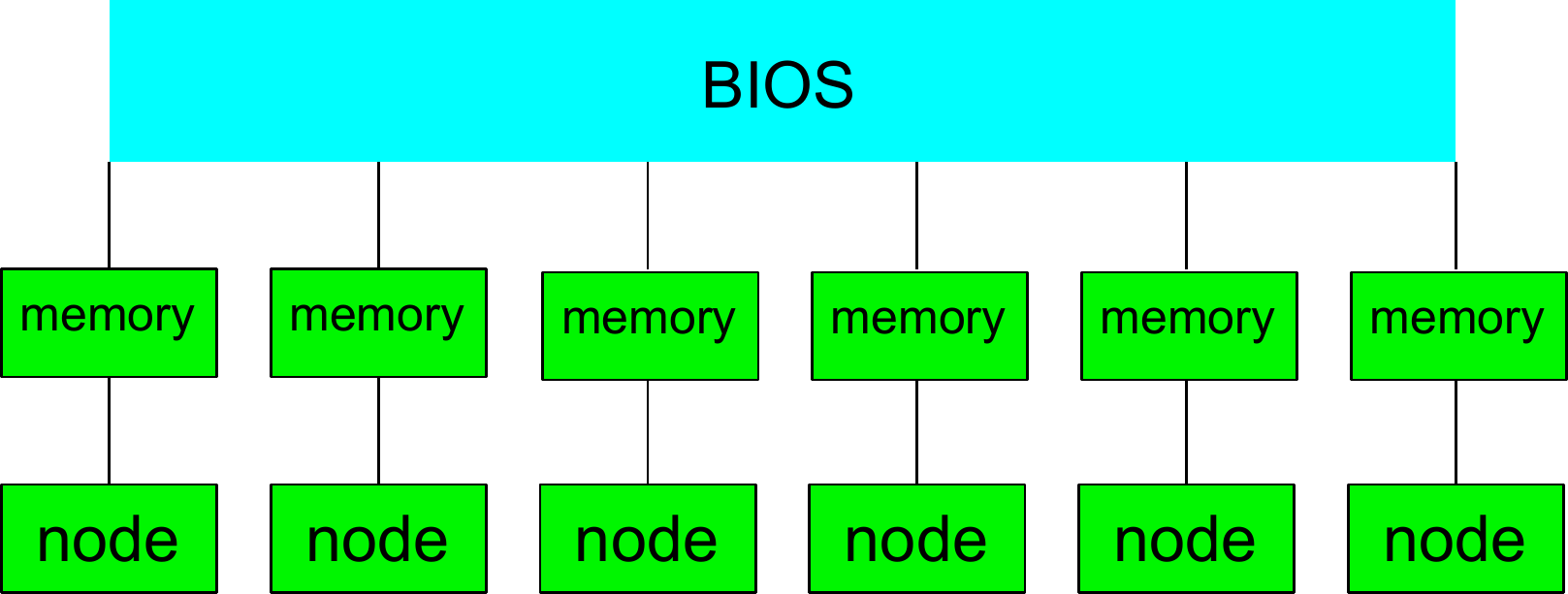} \\
(a) & &  (b)
\end{array}$
\caption{Schematic illustration of the (a) shared and (b)  distributed memory parallelization schemes.}
\label{FDTDPICFig4}
\end{figure}

The large and demanding computation cost needed for the simulation of the FEL process even in the Lorentz boosted coordinate frame necessitates solving the problem on multiple processors to achieve reasonable computation times.
Therefore, efficient parallelization techniques should be implemented in the FDTD/PIC algorithm to develop an efficient software.
Traditionally, there are two widely used techniques to run a computation in parallel on several processors: (1) \textit{shared} memory, and (2) \textit{distributed} memory parallelization.
In the shared memory parallelization or the so-called multi-threading technique, several processors run a code using the variables saved in one shared memory (Fig.\,\ref{FDTDPICFig4}a).
This technique is very suitable for PIC algorithms because it avoids the additional costs of communicating the particle position and momenta between the processors.
On the other hand, distributed memory technique distributes the involved variables among several processors, solves the problem in each processor independently and communicates the required variables whenever they are called (Fig.\,\ref{FDTDPICFig4}b).
The distributed memory technique is very suitable for FDTD algorithm due to the ease of problem decomposition beyond various machines.
The advantage is fast reading and writing of the data and the possibility to share the computational load between different machines.

\begin{figure}
\centering
\includegraphics[width=5.0in]{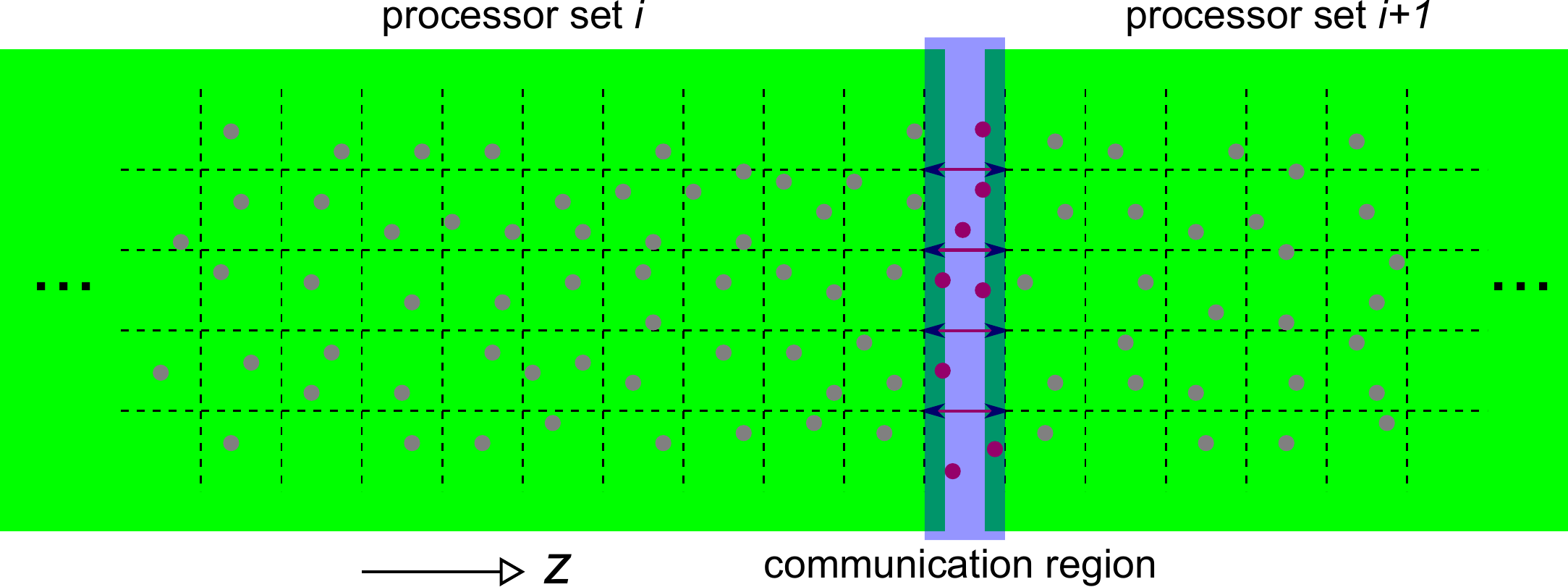}
\caption{Schematic illustration of the domain decomposition used for distributed memory parallelization in MITHRA}
\label{FDTDPICFig5}
\end{figure}

Choosing a suitable parallelization scheme for the hybrid FDTD/PIC algorithm depends on both problem size and machine implementations.
Therefore, both techniques are implemented in the software and finding an efficient distribution of the computational task among the available processors is offered to the user.
More accurately, the user has the possibility to determine how many processors are threaded (shared memory) and how far the multi-threaded algorithm is parallelized among separate processors.
The multi-threading parallelization in MITHRA is implemented using OpenMP package.
The implementation simply requires distributing the loops among different threads along with certain considerations to avoid race conditions.
The total computational domain is then decomposed to several separate regions, each of them solved by one set of multi-threaded processors.
These sets of processors communicate the required variables based on the technique visualized in Fig.\,\ref{FDTDPICFig5}.

To parallelize the computation among $N$ sets of multi-threaded processors, the whole computational domain is divided into $N$ domains along $z$ (undulator period) axis.
In each time update of the field, the field values at the boundaries of each domain are communicated with the corresponding processor set.
To parallelize the PIC solver, we define a communication domain which as shown in Fig.\,\ref{FDTDPICFig5}, is the region between the boundaries of each processor.
After each update of the particles position, it is checked if the particle has entered a communication domain.
In case of residing in the communication region, the master processor, which is the processor containing the particle in the previous time step, communicates the new coordinates to the slave processor, which is the processor sharing the communication region with the master one.
Through this simple algorithm, both parallelization schemes function simultaneously to achieve the fastest computation feasible and compatible with an available computing machine.


\chapter{User Interface}
\label{chapter_refcard}

This chapter, as apparent from its name, is considered as a reference card for the MITHRA code.
We aim to present the functions and variables which can be delivered to the MITHRA software and can be handled for a FEL simulation problem.
In what follows in this chapter, we introduce the defined language of MITHRA to write a compatible job file.
This chapter can also be considered as a reference for the current capabilities of MITHRA and with time will be updated with the further improvement of the software capabilities.
\begin{description}
\item[\textbf{Iron Rule:}] parameters that are used for the solution of a specific electromagnetic problem are delivered to the code at only one single location, \emph{the job file}. This is indeed the only thing that the solver takes as an input parameter.
\end{description}
It should be noted that all the parameters in the job file is given in the laboratory frame.
The Lorentz boost into the bunch rest frame will be done by the software automatically.

To run a job file using the software MITHRA, the following command should be written in the linux command line:

\begin{itemize}
	\item  mpirun -np "number of distributed processors" "MITHRA object file name" "job file name"
\end{itemize}

The transferred job file to the solver contains five main sections, each one defining an essential part of the electromagnetic problem.
These sections include:
\begin{enumerate}
\item \textbf{\texttt{MESH:}} The parameters of the FDTD solver like the computational domain, cell sizes and time steps are set in this section.
\item \textbf{\texttt{BUNCH:}} The required data to initialize the electron bunch in the computational domain is set in this section. In addition, the desired type of recording the bunch evolution is entered in this section by the user.
\item \textbf{\texttt{FIELD:}} This section fulfills the same task as the previous section for the electromagnetic fields. The field initialization in case of a seeded FEL and the desired output type for the field evolution is given in this section to the software.
\item \textbf{\texttt{UNDULATOR:}} This section introduced the different parameters of the undulator.
\item \textbf{\texttt{FEL-OUTPUT:}} The desired data concerning the FEL radiation and how to record this data is set in this section.
\end{enumerate}
In the next subsections, we explain each part and the supported parameters, respectively.
To write comments in your job file use the sign "\#" at the beginning of the comment and the text will be commented to the end of the line.

\section{\texttt{MESH}}

\begin{figure}
\centering
\includegraphics[width=5.0in]{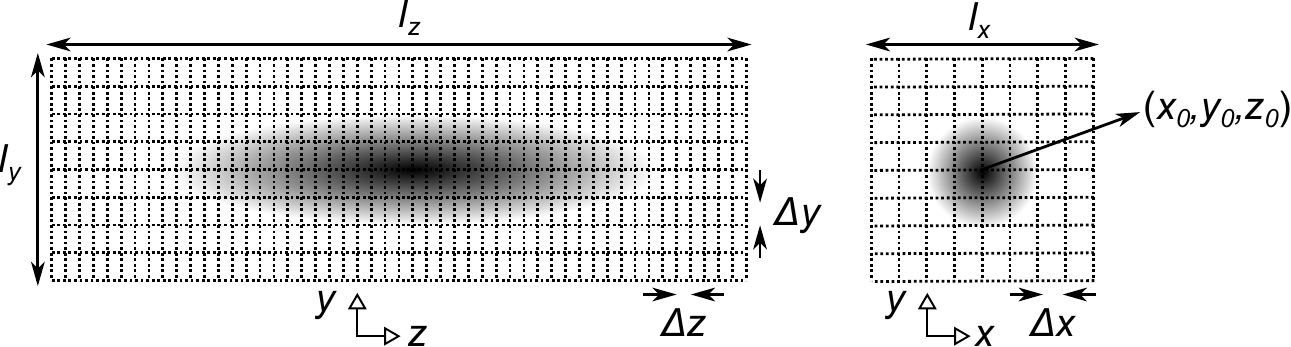}
\caption{The definition of the spatial mesh parameters in MITHRA}
\label{RefcardFig1}
\end{figure}

As mentioned above, this part is dedicated to the determination of the FDTD/PIC parameters.
In Fig\,\ref{RefcardFig1}, a typical computation domain assumed in MITHRA is depicted.
The mesh and update parameters of the solver are defined through the following eleven parameters:
\begin{itemize}
\item \textbf{\texttt{length-scale}} is the scaling of the length and all the spatial parameters in the job file. The capability to play with length scales is crucial to avoid working with very large or very small numbers.
\item \textbf{\texttt{time-scale}} is the scaling of the time and all the temporal parameters in the job file. Similar to above, through this capability working with very large or very small numbers is avoided.
\item \textbf{\texttt{mesh-lengths}} is a three dimensional vector equal to the lengths $(l_x,l_y,l_z)$ of the computational domain (\ref{RefcardFig1}) along the three Cartesian axes.
\item \textbf{\texttt{mesh-resolution}} defines the length of one single grid cell or in other words the spatial discretization resolution of the FDTD mesh in the laboratory coordinate system $(\Delta x',\Delta y',\Delta z')$.
\item \textbf{\texttt{mesh-center}} is the position of the central point of the computational rectangle, i.e. $(x_0',y_0',z_0')$ in Fig.\,\ref{RefcardFig1}.
\item \textbf{\texttt{total-time}} is the total computation time in the scale given by the time scale. This is indeed the time it takes for the electron bunch to travel through the considered undulator length.
\item \textbf{\texttt{bunch-time-step}} is the time step for updating the macro-particles' coordinates in the PIC solver.
\item \textbf{\texttt{bunch-time-start}} is the time point when the bunch of macro-particles emerges in the computational domain.
\item \textbf{\texttt{mesh-truncation-order}} is the truncation order of the absorbing boundary condition in the computational domain. This parameter can be either 1 or 2, representing the first order and second order absorbing boundary condition.
\item \textbf{\texttt{number-of-threads}} is the level of multi-threading in the solver. This corresponds to the number of processors using a shared memory in one processor set.
\item \textbf{\texttt{space-charge}} is a boolean flag determining if the space-charge effect should be considered or not. If this flag is false, the scalar potential $\phi$ is zero throughout the calculation. Otherwise, the scalar potential is calculated using the corresponding Helmholtz equation.
\end{itemize}

The format of the \texttt{MESH} group is:
\begin{Verbatim}[frame=single,fontsize=\small,tabsize=4]
MESH
{
	length-scale                    = < real value | METER | DECIMETER | CENTIMETER | MILLIMETER
									  | MICROMETER | NANOMETER | ANGSTROM >
	time-scale                      = < real value | SECOND | MILLISECOND | MICROSECOND |
									  NANOSECOND | PICOSECOND | FEMTOSECOND | ATTOSECOND >
	mesh-lengths                    = < ( real value,  real value, real value) >
	mesh-resolution                 = < ( real value,  real value, real value) >
	mesh-center                     = < ( real value,  real value, real value) >
	total-time                      = < real value >
	bunch-time-step                 = < real value >
	bunch-time-start                = < real value >
	mesh-truncation-order           = < 1 | 2 >
    number-of-threads               = < integer value >
    space-charge           	     = < true | false >
}
\end{Verbatim}
An example of the computational mesh definition looks as the following:
\begin{snugshade}
\begin{Verbatim}[fontsize=\small, tabsize = 4]
MESH
{
  length-scale                      = MICROMETER
  time-scale                        = PICOSECOND
  mesh-lengths                      = ( 3200,  3200.0,    280.0)
  mesh-resolution                   = ( 50.0,    50.0,      0.1)
  mesh-center                       = ( 0.0,      0.0,      0.0)
  total-time                        = 30000
  bunch-time-step                   = 1.6
  bunch-time-start                  = 0.0
  number-of-threads                 = 2
  mesh-truncation-order             = 2
  space-charge                      = false
}
\end{Verbatim}
\end{snugshade}
Note that there are some conditions, which should be fulfilled for the numerical integrator to obtain reliable dispersion-less results.
The software checks for these conditions before starting to solve the problem, if the conditions are violated the closest value to the given number meeting the violated conditions will be used.
Regarding the above parameters. the software checks for the stability condition $\sqrt{(\Delta z/\Delta x)^2+ (\Delta z/\Delta y)^2} < 1$, adapts the values of $\Delta x$ and $\Delta y$ accordingly, and finally sets the time step for field update equal to $\Delta z / c$.
In addition, the bunch update time step should be an integer fraction of the field time step to avoid redundant dispersion in the calculated values.
Therefore, the closest value to the given bunch time step, which satisfies the above criterion, will be chosen.

\section{\texttt{BUNCH}}

The section \texttt{BUNCH} is the main part of the job file to establish the required data for the bunch input and output framework.
This section consists of four groups: (1) \texttt{bunch-initialization}, (2) \texttt{bunch-sampling}, (3) \texttt{bunch-visualization}, and (4) \texttt{bunch-profile}.
As apparent from the name the first group determines the set of parameters to initialize the bunch and the other three groups are dedicated to reporting the bunch evolution in different formats.
In what follows, the parameters in each group are introduced:

\begin{enumerate}
\item \textbf{\texttt{bunch-initialization:}} This group mainly determines the parameters whose values are needed for initializing a bunch of electrons with different types. If several bunches are present in a simulation, this group should simply be repeated in the \texttt{BUNCH} section. The set of values accepted in this group include:
\begin{itemize}
\item \textbf{\texttt{type}} is the type of the bunch to be initialized in the computational domain. There are five bunch types supported by MITHRA:
\begin{enumerate}
	\item \texttt{manual} initializes charges at the points specified by the position vector. At each appearance of this type of bunch only one single macro-particle will be initialized. Therefore, to have multiple manual initialization, the \texttt{bunch-initialization} group should be repeated. Using the \texttt{file} type is a better solution for high number of manual inputs.
	\item \texttt{ellipsoid} initializes charges with a
	\item \texttt{3D-crystal} initializes multiple bunches on the points of a 3D crystal centered at the coordinate specified by the position vector and extends over the space by the number vector and the considered lattice constant. Each single bunch has a ellipsoid Gaussian property with the values read from the deviation parameters.
	\item \texttt{file} reads a list of 6D position and momentum coordinates from a file and initializes the macro-particles correspondingly in the solver.
\end{enumerate}
\item \textbf{\texttt{distribution}} determines if the initialized particle distribution should have a uniform or Gaussian current profile.
\item \textbf{\texttt{number-of-particles}} is the total number of particles (or macro-particles) considered in the bunch.
\item \textbf{\texttt{charge}} is the total charge of the bunch in one electron charge unit.
\item \textbf{\texttt{gamma}} is the initial mean Lorentz factor of the .
\item \textbf{\texttt{beta}} is the initial mean normalized velocity of the particles, if it is not determined here the value will be calculated from the \texttt{gamma} parameter, otherwise the same \texttt{beta} will be used.
\item \textbf{\texttt{direction}} is the average momentum direction of the bunch, i.e. $(\beta_x, \beta_y, \beta_z)/\beta$.
\item \textbf{\texttt{position}} is the central position of the bunch.
\item \textbf{\texttt{sigma-position}} is the deviation in position of the bunch, i.e. $(\sigma_x, \sigma_y, \sigma_z)$ for Gaussian distributions. $\sigma_z$ half the bunch length for the uniform distribution.
\item \textbf{\texttt{sigma-momentum}} is the deviation in energy of the bunch, i.e. $(\sigma_{\gamma \beta_x}, \sigma_{\gamma \beta_y}, \sigma_{\gamma \beta_z})$.
\item \textbf{\texttt{numbers}} is a parameter read only when the bunch type is a \texttt{3d-crystal} type. It is the number of bunch replication in the three directions.
\item \textbf{\texttt{lattice-constants}} is a parameter read only when the bunch type is a \texttt{3d-crystal} type. It is the length of lattice constants of the crystal in the three directions.
\item \textbf{\texttt{transverse-truncation}} determines a limit to transversely truncate the bunches. This factor brings the possibility to control particle initialization and prevents them from escaping out of the computational domain. The bunch initializer truncates the bunch at the given distance from the bunch center.
\item \textbf{\texttt{longitudinal-truncation}} determines a limit to longitudinally truncate the bunches. This factor brings the possibility to control particle initialization and prevents them from escaping out of the computational domain. The bunch initializer truncates the bunch at the given distance from the bunch center.
\item \textbf{\texttt{bunching-factor}} is a value larger than zero and less than one, which determines the bunching factor, i.e. $<e^{jk_uz}>$, of the initialized bunch.
\end{itemize}
\item \textbf{\texttt{bunch-sampling:}} This group defines the required parameters for saving the bunch properties with time. The bunch mean position, mean momentum, position spread, and momentum spread along the three Cartesian coordinates are saved respectively with time. There are different parameters required for this definition which include:
\begin{itemize}
	\item \textbf{\texttt{sample}} is a boolean value determining if the bunch sampling should be activated.
	\item \textbf{\texttt{base-name}} is the file name (no suffix) with the required address of the file to save the output data.
	\item \textbf{\texttt{directory}} is the address where the above file should be saved. The file name with the address can also be given in the base-name section. The software eventually considers the combination of directory and base-name as the final complete file name.
	\item \textbf{\texttt{rhythm}} is the rhythm of bunch sampling, i.e. the time interval between two consecutive visualization times.
\end{itemize}
\item \textbf{\texttt{bunch-visualization:}} This group defines the required parameters for visualizing the charge distribution in the whole computational domain. The output will be a set of \texttt{.vtu} files at each time for each processor which are connected with a set of \texttt{.pvtu} files. They can be very nicely visualized using the open source ParaView package. There are different parameters required for this definition which include:
\begin{itemize}
	\item \textbf{\texttt{sample}} is a boolean value determining if the charge visualization should be activated.
	\item \textbf{\texttt{base-name}} is the file name (no suffix) with the required address of the file to save the output data.
	\item \textbf{\texttt{directory}} is the address where the above file should be saved. The file name with the address can also be given in the base-name section. The software eventually considers the combination of directory and base-name as the final complete file name.
	\item \textbf{\texttt{rhythm}} is the rhythm of charge illustration, i.e. the time interval between two consecutive visualization times.
\end{itemize}
\item \textbf{\texttt{bunch-profile:}} This group defines the required parameters for saving a histogram of the charges. It means that at a specific time instant the charge values, positions and momenta of all the particles (or macro-particles) will be written and saved in a file. The parameters entered by the user for saving the histogram include:
\begin{itemize}
	\item \textbf{\texttt{sample}} is a boolean value determining if the writing of the histogram during the PIC simulations should be activated.
	\item \textbf{\texttt{base-name}} is the file name (no suffix) with the required address of the file to save the output data.
	\item \textbf{\texttt{directory}} is the address where the above file should be saved. The file name with the address can also be given in the base-name section. The software eventually considers the combination of directory and base-name as the final complete file name.
	\item \textbf{\texttt{time}} is the time instant for saving the histogram. If this needs to be done in several time instants, simply this line should be repeated with different time values.
    \item \textbf{\texttt{rhythm}} is the rhythm of writing the bunch profile, i.e. the time interval between two consecutive profiling times. If this value is nonzero, the sequence of times will be considered in addition to the specific time points given by the time variable.
\end{itemize}
\end{enumerate}

The format of the \texttt{BUNCH} group is:
\begin{Verbatim}[frame=single,fontsize=\small,tabsize=4]
BUNCH
{
	bunch-initialization
	{
		type                        = < manual | ellipsoid | 3D-crystal | file >
		distribution                = < uniform | gaussian >
		charge                      = < real value >
		number-of-particles         = < integer value >
		gamma                       = < real value >
		beta                        = < real value >
		direction                   = < ( real value, real value, real value ) >
		position                    = < ( real value, real value, real value ) >
		sigma-position              = < ( real value, real value, real value ) >
		sigma-momentum              = < ( real value, real value, real value ) >
		transverse-truncation       = < real value >
		longitudinal-truncation     = < real value >
		bunching-factor             = < real value between zero and one >
	}
	
	bunch-sampling
	{
		sample                      = < true | false >
		directory                   = < address according to UNIX convention >
		base-name                   = < name of the file >
		rhythm                      = < real value >
	}
	
	bunch-visualization
	{
		sample                      = < true | false >
		directory                   = < address according to UNIX convention >
		base-name                   = < name of the file >
		rhythm                      = < real value >
	}

	bunch-profile
	{
		sample                      = < true | false >
		directory                   = < address according to UNIX convention >
		base-name                   = < name of the file >
		time                        = < real value >
        rhythm                      = < real value >
	}
}
\end{Verbatim}
An example of the bunch category definition looks as the following:
\begin{snugshade}
\begin{Verbatim}[fontsize=\small, tabsize = 4]
BUNCH
{
  bunch-initialization
  {
    type                            = ellipsoid
    distribution                    = uniform
    charge                          = 1.846e8
    number-of-particles             = 131072
    gamma                           = 100.41
    direction                       = ( 0.0, 0.0, 1.0)
    position                        = ( 0.0, 0.0, 0.0)
    sigma-position                  = ( 260.0, 260.0, 50.25)
    sigma-momentum                  = ( 1.0e-8, 1.0e-8, 100.41e-4)
    transverse-truncation           = 1040.0
    longitudinal-truncation         = 90.0
    bunching-factor                 = 0.01
  }

  bunch-sampling
  {
    sample                          = false
    directory                       = ./
    base-name                       = bunch-sampling/bunch
    rhythm                          = 3.2
  }

  bunch-visualization
  {
    sample                          = true
    directory                       = ./
    base-name                       = bunch-visualization/bunch
    rhythm                          = 32
  }

  bunch-profile
  {
    sample                          = false
    directory                       = ./
    base-name                       = bunch-profile/bunch
    time                            = 5000
    time                            = 10000
    time                            = 15000
    time                            = 20000
    time                            = 25000
    time                            = 30000
  }
}
\end{Verbatim}
\end{snugshade}

\section{\texttt{FIELD}}

In section \texttt{FIELD}, the required data for the input and output framework of the field in the FDTD algorithm is produced. 
This section consists of four groups: (1) \texttt{field-initialization}, (2) \texttt{field-sampling}, (3) \texttt{field-visualization}, and (4) \texttt{field-profile}. 
As apparent from the name, the first group determines the set of parameters to initialize the field and the other three groups are dedicated to reporting the field propagation in different formats. 
In what follows, the parameters in each group are introduced:

\begin{enumerate}
\item \textbf{\texttt{field-initialization:}} This group mainly determines the parameters whose values are needed for initializing a field excitation entering the computational domain. The excitation may have different types. This group is where a seed can be added to the simulations to simulate a seeded-FEL problem. The set of values accepted in this group include:
\begin{itemize}
	\item \textbf{\texttt{type}} is the type of the excitation. The accepted excitation types in MITHRA include plane wave and Gaussian beam.
	\item \textbf{\texttt{position}} is the reference position of the excitation. It is the reference position of the plane wave propagation in the plane wave excitation and the focusing point in the Gaussian beam excitation.
	\item \textbf{\texttt{direction}} is the propagation direction of the excitation in the plane wave and Gaussian beam types.
	\item \textbf{\texttt{polarization}} is the polarization of the incoming excitation and is used by both plane wave and Gaussian beam types.
	\item \textbf{\texttt{rayleigh-radius-parallel}} is the Rayleigh radius of the Gaussian beam in the direction parallel to the polarization and is only meaningful for the Gaussian beam type.
	\item \textbf{\texttt{rayleigh-radius-perpendicular}} is the Rayleigh radius of the Gaussian beam in the direction perpendicular to the polarization and is only meaningful for the Gaussian beam type.
	\item \textbf{\texttt{signal-type}} determines the time signature of the signal exciting the fields according to the particular type. The accepted signal types in MITHRA include modulated Neumann, modulated Gaussian, modulated secant hyperbolic and the sinusoidal pulse. The equation representing the time domain variation of each pulse is as follows:
	\begin{equation}
	\renewcommand{\arraystretch}{1.5}
	\begin{array}{lrcl}
	\mbox{modulated Neumann:}  \qquad & f(t) & = & \displaystyle - A_0 4 \ln 2 \: \cos( 2 \pi f (t - t_0) + \phi_\mathrm{CEP} ) \frac{t - t_0}{\tau^2} e^{-2 \ln 2 \: (t - t_0)^2/\tau^2 } \\
	\mbox{modulated Gaussian:} \qquad & f(t) & = & \displaystyle A_0 \cos( 2 \pi f (t - t_0) + \phi_\mathrm{CEP} ) e^{-2 \ln 2 \: (t - t_0)^2/\tau^2 } \\
	\mbox{modulated hyperbolic secant:} \qquad & f(t) & = & \displaystyle A_0 \cos( 2 \pi f (t - t_0) + \phi_\mathrm{CEP} ) \frac{1}{\cosh ( (t - t_0)/\tau ) } \\
	\mbox{sinusoidal pulse:}   \qquad & f(t) & = & \displaystyle \left\{ \begin{array}{ll} A_0 \cos( 2 \pi f (t - t_0) + \phi_\mathrm{CEP} ) e^{-2 \ln 2 \: (t - t_0)^2/\tau^2 } & t \leq t_0 \\ A_0 \cos( 2 \pi f (t - t_0) + \phi_\mathrm{CEP} ) & t > t_0 \end{array} \right.
	\end{array}
	\end{equation}
	\item \textbf{\texttt{strength-parameter}} is the normalized amplitude $a_0 = e A_0 / m_ec $ of the beam.
	\item \textbf{\texttt{offset}} is the distance offset of the signal $ct_0$.
	\item \textbf{\texttt{variance}} is the variance of the signal in length units $c\tau$.
	\item \textbf{\texttt{wavelength}} is the modulation wavelength $\lambda_0$ of the modulated signal.
	\item \textbf{\texttt{CEP}} is the carrier envelope phase $\phi_{\mathrm{CEP}}$ of the modulated signal.
\end{itemize}
\item \textbf{\texttt{field-sampling:}} This group defines the required parameters for saving the field value at specific points with time. There are different parameters required for this definition which include:
\begin{itemize}
	\item \textbf{\texttt{sample}} is a boolean value determining if the field sampling should be activated.
	\item \textbf{\texttt{type}} determines if the field should be sampled at the given points (at-point) or the field should be sampled at the points over a line (over-line).
	\item \textbf{\texttt{field}} determines which electromagnetic field is to be sampled. The available options are the electric field, magnetic field, magnetic vector potential, scalar electric potential, charge and current. This item can be repeated to assign several fields for the sampling. In the text file, the fields appear with the same order.
	\item \textbf{\texttt{base-name}} is the file name (no suffix) with the required address of the file to save the output data.
	\item \textbf{\texttt{directory}} is the address where the above file should be saved. The file name with the address can also be given in the base-name section. The software eventually considers the combination of directory and base-name as the final complete file name.
	\item \textbf{\texttt{rhythm}} is the rhythm of field sampling, i.e. the time interval between two consecutive visualization times.
	\item \textbf{\texttt{position}} is the coordinate of the points where the fields should be sampled. By repeating this line any number of points can be added to the set of sampling locations. This option is merely used when the sampling type is set to "at-point" option.
	\item \textbf{\texttt{line-begin}} defines the position of the line begin over which the fields should be sampled and is used when the sampling type is set to "over-line" option.
	\item \textbf{\texttt{line-end}} defines the position of the line end over which the fields should be sampled and is used when the sampling type is set to "over-line" option.
	\item \textbf{\texttt{resolution}} defines the resolution of the line discretization for sampling. In other words, the value is the distance between two adjacent sampling points. This value is used when the sampling type is set to "over-line" option.
\end{itemize}
\item \textbf{\texttt{field-visualization:}} This group defines the required parameters for visualizing the fields in the whole computational domain. The output will be a set of \texttt{.vtu} files at each time for each processor which are connected with a set of \texttt{.pvtu} files. They can be very nicely visualized using the open source ParaView package. There are different parameters required for this definition which include:
\begin{itemize}
	\item \textbf{\texttt{sample}} is a boolean value determining if the field visualization should be activated.
	\item \textbf{\texttt{base-name}} is the file name (no suffix) with the required address of the file to save the output data.
	\item \textbf{\texttt{directory}} is the address where the above file should be saved. The file name with the address can also be given in the base-name section. The software eventually considers the combination of directory and base-name as the final complete file name.
	\item \textbf{\texttt{rhythm}} is the rhythm of field visualization, i.e. the time interval between two consecutive visualization times.
	\item \textbf{\texttt{field}} determines which electromagnetic field is to be saved. The available options are the electric field, magnetic field, magnetic vector potential, scalar electric potential, charge and current. This item can be repeated to assign several fields for the sampling. In the text file, the fields appear with the same order.
\end{itemize}
\item \textbf{\texttt{field-profile:}} This group defines the required parameters for saving a histogram of the field over the whole computational domain. It means that at a specific time instant the field values and the corresponding positions at all the grid points will be written and saved in a text file. The parameters entered by the user for saving the histogram include:
\begin{itemize}
	\item \textbf{\texttt{sample}} is a boolean value determining if the writing of the histogram during the FDTD simulations should be activated.
	\item \textbf{\texttt{base-name}} is the file name (no suffix) with the required address of the file to save the output data.
	\item \textbf{\texttt{directory}} is the address where the above file should be saved. The file name with the address can also be given in the base-name section. The software eventually considers the combination of directory and base-name as the final complete file name.
	\item \textbf{\texttt{time}} is the time instant for saving the histogram. If this needs to be done in several time instants, simply this line should be repeated with different time values.
	\item \textbf{\texttt{rhythm}} is the rhythm of field profiling, i.e. the time interval between two consecutive visualization times. Both rhythmic profiling and saving the fields at specific times can be given to the software.
	\item \textbf{\texttt{field}} determines which electromagnetic field is to be profiled. The available options are the electric field, magnetic field, magnetic vector potential, scalar electric potential, charge and current. This item can be repeated to assign several fields for the sampling. In the text file, the fields appear with the same order.
\end{itemize}
\end{enumerate}

The format of the \texttt{FIELD} group is:
\begin{Verbatim}[frame=single,fontsize=\small,tabsize=4]
FIELD
{
	field-initialization
	{
		type                            = < hertzian-dipole | plane-wave | gaussian-beam >
		position                        = < ( real value , real value , real value ) >
		direction                       = < ( real value , real value , real value ) >
		polarization                    = < ( real value , real value , real value ) >
		rayleigh-radius-parallel        = < real value >
		rayleigh-radius-perpendicular   = < real value >
		signal-type                     = < neumann | gaussian | secant-hyperbolic | flat-top >
		strength-parameter              = < real value >
		offset                          = < real value >
		variance                        = < real value >
		wavelength                      = < real value >
		CEP                             = < real value >
	}
	
	field-sampling
	{
		sample                          = < true | false >
		type							= < over-line | at-point >
		field                           = < Ex | Ey | Ez | Bx | By | Bz | Ax | Ay | Az | Jx | Jy
                                          | Jz | F | Q >
		directory                       = < address according to UNIX convention >
		base-name                       = < name of the file >
		rhythm                          = < real value >
		position                        = < ( real value , real value , real value ) >
		line-begin                      = < ( real value , real value , real value ) >
		line-end                        = < ( real value , real value , real value ) >
		resolution                      = < real value >
	}

	field-visualization
	{
		sample                          = < true | false >
		field                           = < Ex | Ey | Ez | Bx | By | Bz | Ax | Ay | Az | Jx | Jy
                                          | Jz | F | Q >
		directory                       = < address according to UNIX convention >
		base-name                       = < name of the file >
		rhythm                          = < real value >
	}
	
	field-profile
	{
		sample                          = < true | false >
		field                           = < Ex | Ey | Ez | Bx | By | Bz | Ax | Ay | Az | Jx | Jy
                                          | Jz | F | Q >
		directory                       = < address according to UNIX convention >
		base-name                       = < name of the file >
		rhythm                          = < real value >
		time                            = < real value >
	}
}
\end{Verbatim}
An example of the field category definition looks as the following:
\begin{snugshade}
\begin{Verbatim}[fontsize=\small, tabsize = 4]
FIELD
{
  field-initialization
  {
    type                                = gaussian-beam
    position                            = ( 0.0, 0.0, -2500.0)
    direction                           = ( 0.0, 0.0, 1.0)
    polarization                        = ( 0.0, 1.0, 0.0)
    rayleigh-radius-parallel            = 0.5
    rayleigh-radius-perpendicular       = 0.5
    strength-parameter                  = 0.0
    signal-type                         = gaussian
    offset                              = 0.00
    variance                            = 1.00
    wavelength                          = 0.0
    CEP                                 = 0.0
  }

  field-sampling
  {
    sample                              = true
    type                                = at-point
    field                               = Ex
    field                               = Ey
    field                               = Ez
    directory                           = ./
    base-name                           = field-sampling/field
    rhythm                              = 3.2
    position                            = (0.0, 0.0, 110.0)
  }

  field-visualization
  {
    sample                              = true
    field                               = Ex
    field                               = Ey
    field                               = Ez
    field                               = Q
    directory                           = ./
    base-name                           = field-visualization/field
    rhythm                              = 32
  }

  field-profile
  {
    sample                              = false
    field                               = Ex
    field                               = Ey
    field                               = Ez
    directory                           = ./
    base-name                           = field-profile/field
    rhythm                              = 100
  }
}
\end{Verbatim}
\end{snugshade}

\section{\texttt{UNDULATOR}}

In section \texttt{UNDULATOR}, the properties of the undulator considered in the FEL problem are introduced.
The parameters, through which the complete data for establishing undulator fields are obtained, include:

\begin{itemize}
	\item \textbf{\texttt{undulator-type}} is the type of the undulator considered for the FEL interaction. It can be either an optical or static undulator.
    \item \textbf{\texttt{undulator-parameter}} is the undulator parameter of the undulator, i.e. the so-called K parameter.
	\item \textbf{\texttt{period}} is the period of the undulator in the given length-scale determined in the mesh class.
	\item \textbf{\texttt{length}} is the total length of the undulator.
    \item \textbf{\texttt{polarization-angle}} is the angle between the magnetic field polarization and the $x$-axis in degrees.
    \item \textbf{\texttt{beam-type}} is the type of the pulse for an optical undulator. The accepted undulator beam types in MITHRA include plane wave and Gaussian beam.
	\item \textbf{\texttt{position}} is the reference position of the undulator. It is the reference position of the plane wave propagation in the plane wave undulator and the focusing point in the Gaussian beam undulator.
	\item \textbf{\texttt{direction}} is the propagation direction of the optical undulator in the plane wave and Gaussian beam types.
	\item \textbf{\texttt{polarization}} is the polarization of the undulator and is used by both plane wave and Gaussian beam types.
	\item \textbf{\texttt{rayleigh-radius-parallel}} is the Rayleigh radius of the Gaussian beam in the direction parallel to the polarization and is only meaningful for the Gaussian beam type.
	\item \textbf{\texttt{rayleigh-radius-perpendicular}} is the Rayleigh radius of the Gaussian beam in the direction perpendicular to the polarization and is only meaningful for the Gaussian beam type.
	\item \textbf{\texttt{signal-type}} determines the time signature of the undulator. The accepted signal types in MITHRA are listed in the field section. Here, the same set of signal can be given as an undulator envelope.
	\item \textbf{\texttt{strength-parameter}} is the normalized amplitude $a_0 = e A_0 / m_ec $ of the undulator, which is equivalent to the undulator-parameter in the static case.
	\item \textbf{\texttt{offset}} is the distance offset of the signal $ct_0$.
	\item \textbf{\texttt{variance}} is the time variance of the signal in length unit $c\tau$.
	\item \textbf{\texttt{wavelength}} is the modulation wavelength $\lambda_0$ of the modulated signal.
	\item \textbf{\texttt{CEP}} is the carrier envelope phase $\phi_{\mathrm{CEP}}$ of the modulated signal.
\end{itemize}

The format of the \texttt{UNDULATOR} group is:
\begin{Verbatim}[frame=single,fontsize=\small,tabsize=4]
UNDULATOR
{
	undulator-type                  = < static | optical >
    undulator-parameter             = < real value >
	period                          = < real value >
	length                          = < real value >
    polarization-angle              = < real value >
    beam-type                       = < hertzian-dipole | plane-wave | gaussian-beam >
	position                        = < ( real value , real value , real value ) >
	direction                       = < ( real value , real value , real value ) >
	polarization                    = < ( real value , real value , real value ) >
	rayleigh-radius-parallel        = < real value >
	rayleigh-radius-perpendicular   = < real value >
	signal-type                     = < neumann | gaussian | secant-hyperbolic | flat-top >
	strength-parameter              = < real value >
	offset                          = < real value >
	variance                        = < real value >
	wavelength                      = < real value >
	CEP                             = < real value >
}
\end{Verbatim}
As explained before, MITHRA always initializes the bunch outside the undulator.
It may be already noticed that there exists no option to determine the beginning of the static undulator with respect to the bunch.
This is automatically set by the solver, to avoid particle initialization inside the undulator.
For the optical undulator type, the user should control this effect through the parameter offset.
An example of the undulator category definition looks as the following:
\begin{snugshade}
\begin{Verbatim}[fontsize=\small, tabsize = 4]
UNDULATOR
{
  undulator-type                    = static
  undulator-parameter               = 1.417
  period                            = 3.0e4
  length                            = 300
  polarization-angle                = 0.0
}
\end{Verbatim}
\end{snugshade}
An instance of optical undulator definition reads as follows:
\begin{snugshade}
\begin{Verbatim}[fontsize=\small, tabsize = 4]
UNDULATOR
{
  undulator-type                    = optical
  beam-type                         = plane-wave
  position                          = ( 0.0, 0.0, 0.0 )
  direction                         = ( 0.0, 0.0,-1.0 )
  polarization                      = ( 0.0, 1.0, 0.0 )
  strength-parameter                = 0.5
  signal-type                       = flat-top
  wavelength                        = 1.0e3
  variance                          = 1200.0e3
  offset                            = 600118.0
  CEP                               = 0.0
}
\end{Verbatim}
\end{snugshade}

\section{\texttt{FEL-OUTPUT}}

The typical parameters for a free electron laser instrument are calculated from the radiated fields using the parameter definitions at this section. 
Currently, there is only one group implemented in MITHRA, which calculates the total radiated power in the $+z$ direction and saves the power versus the undulator position to a text file. 
In what follows, the parameters in each group are introduced:

\begin{enumerate}
\item \textbf{\texttt{radiation-power:}} This group mainly determines the parameters whose values are needed for calculating the radiation power from the field distribution. The output is a .txt file with the first column the time point and the second column the radiated power at the given point. For multiple points the column is repeated. If multiple wavelengths are given there will be several rows with similar time value listing the radiated power with different wavelengths. This group can be repeated to obtain various files for different output definitions. 

The set of values accepted in this group include:
\begin{itemize}
	\item \textbf{\texttt{sample}} is a boolean parameter which activates the computation of the total radiated power at each instant.
	\item \textbf{\texttt{base-name}} is the file name (no suffix) with the required address of the file to save the output data.
	\item \textbf{\texttt{directory}} is the address where the above file should be saved. The file name with the address can also be given in the base-name section. The software eventually considers the combination of directory and base-name as the final complete file name.
	\item \textbf{\texttt{type}} determines if the radiated power should be sampled at given distances from the bunch (at-point) or should be sampled at the points over a line (over-line).
	\item \textbf{\texttt{distance-from-bunch}} gives the total distance from the bunch where a sampling plate to capture the whole radiated power will be placed.  One can enter several sampling positions by repeating this line. This option is only considered if the type variable is set to "at-point".
	\item \textbf{\texttt{line-begin}} defines the distance from the bunch center for the line begin over which the fields should be sampled and is used when the sampling type is set to "over-line" option. In case of multiple bunches, the center of the first bunch is considered as the reference position.
	\item \textbf{\texttt{line-end}} defines the distance from the bunch center for the line end over which the fields should be sampled and is used when the sampling type is set to "over-line" option. In case of multiple bunches, the center of the first bunch is considered as the reference position.
	\item \textbf{\texttt{resolution}} defines the resolution of the line discretization for sampling. In other words, the value is the distance between two adjacent sampling points. This value is used when the type variable is set to "over-line" option.
	\item \textbf{\texttt{normalized-frequency}} is the central wavelength normalized to the radiation wavelength of the radiation spectrum.
	\item \textbf{\texttt{minimum-normalized-frequency}} is the minimum wavelength normalized to the radiation wavelength. This parameters and the next two parameters are used to sweep over the normalized wavelength and save the radiation power spectrum.
	\item \textbf{\texttt{maximum-normalized-frequency}} is the maximum wavelength normalized to the radiation wavelength.
	\item \textbf{\texttt{normalized-frequency-resolution}} is the sweep resolution for the wavelength normalized to the radiation wavelength.
\end{itemize}
\end{enumerate}

The format of the \texttt{FEL-OUTPUT} group is:
\begin{Verbatim}[frame=single,fontsize=\small,tabsize=4]
FEL-OUTPUT
{
	radiation-power
	{
		sample                           = < false | true >
		type                             = < at-point | over-line >
		directory                        = < address according to UNIX convention >
		base-name                        = < name of the file >
		distance-from-bunch			     = < real value >
		line-begin                       = < real value >
		line-end                         = < real value >
		resolution                       = < real value >
		normalized-frequency            = < real value >
		minimum-normalized-frequency    = < real value >
		maximum-normalized-frequency    = < real value >
		normalized-frequency-resolution = < real value >
	}
}
\end{Verbatim}
An example of the FEL output category definition looks as the following:
\begin{snugshade}
\begin{Verbatim}[fontsize=\small, tabsize = 4]
FEL-OUTPUT
{
	radiation-power
	{
		sample                           = true
		type                             = at-point
		directory                        = ./
		base-name                        = power-sampling/power
		distance-from-bunch              = 110.0
		normalized-frequency            = 1.00
	}
}
\end{Verbatim}
\end{snugshade} 


\chapter{Examples}
\label{chapter_examples}

The goal in this chapter is to present several examples for the MITHRA users to more easily get familiar with the interface of the software.
In addition, through the presented examples the pros and cons of using the developed FDTD/PIC algorithm are more accurately evaluated.
For example, the computation time, numerical stability and numerical convergence and more importantly the reliability of the results are studied based on some standard examples.
The software developers aim to update this chapter with the most recent examples where MITHRA is used for the FEL simulation.

\section{Example 1: Infrared FEL}

\subsection{Problem Definition}

\begin{table}[h]
\label{example1}
\caption{Parameters of the Infrared FEL configuration considered as the first example.}
\centering
\begin{tabular}{|c||c|}
\hline
FEL parameter & Value \\ \hline \hline
Current profile & Uniform \\ \hline
Bunch size & $(260\times260\times100)\,\mu$m \\ \hline
Bunch charge & 29.5\,pC \\ \hline
Bunch energy & 51.4\,MeV \\	\hline
Bunch current & 88.5\,A \\ \hline
Longitudinal momentum spread & 0.01\% \\ \hline
Transverse momentum spread & 0.0 \\	\hline
Undulator period & 3.0\,cm \\ \hline
Magnetic field & 0.5\,T \\ \hline
Undulator parameter & 1.4 \\ \hline
Undulator length & 5\,m \\ \hline
Radiation wavelength & 3\,$\mu$ m \\ \hline
Electron density & $2.72\times10^{13} 1/\text{cm}^3$ \\ \hline
Gain length (1D) & 22.4\,cm \\ \hline
FEL parameter & 0.006 \\ \hline
Cooperation length & $39.7\,\mu$m \\ \hline
Initial bunching factor & $0.01$ \\ \hline
\end{tabular}
\end{table}
As the first example, we consider an infrared FEL with the parameters tabulated in table \ref{example1}, which is inspired by the numerical analysis presented in \cite{tran1989tda}.
The bunch distribution is assumed to be uniform in order to compare the results with one-dimensional FEL theory.
For the same purpose, the transverse energy spread is considered to be zero and a minimal longitudinal energy spread is assumed.

To simulate the considered FEL configuration, the following job file is written and given to the software to analyze the interaction.
As observed in the mesh definition, the transverse size of the computational domain is almost 10 times larger than the bunch transverse size.
In the contrary, the longitudinal size of the mesh is only three times larger than the bunch length.
This needs to be considered due to the failure of absorbing boundary conditions for the oblique incidence of the field.
Furthermore, the bunch and undulator both have some tapering sections to avoid abrupt transitions which produce coherent scattering emission (CSE).
It is recommended that the undulator begin is initialized at least ten radiation wavelengths apart from the bunch head to reduce the CSE.
This also introduces corresponding limitations on the mesh size, meaning that the minimum distance from the bunch tail and the mesh boundary should be at least ten radiation wavelengths.
In the illustrated job file, some of the output formats are turned off which can always be activated to obtain the required data.
\begin{snugshade}
\begin{Verbatim}[fontsize=\small, tabsize = 4]
MESH
{
  length-scale                     = MICROMETER
  time-scale                       = PICOSECOND
  mesh-lengths                     = ( 3200,  3200.0,    280.0)
  mesh-resolution                  = ( 50.0,    50.0,      0.1)
  mesh-center                      = ( 0.0,      0.0,      0.0)
  total-time                       = 30000
  bunch-time-step                  = 1.6
  bunch-time-start                 = 0.0
  number-of-threads                = 2
  mesh-truncation-order            = 2
  space-charge                     = false
}

BUNCH
{
  bunch-initialization
  {
    type                           = ellipsoid
    distribution                   = uniform
    charge                         = 1.846e8
    number-of-particles            = 131072
    gamma                          = 100.41
    direction                      = (    0.0,     0.0,       1.0)
    position                       = (    0.0,     0.0,       0.0)
    sigma-position                 = (  260.0,   260.0,     50.25)
    sigma-momentum                 = ( 1.0e-8,  1.0e-8, 100.41e-4)
    transverse-truncation          = 1040.0
    longitudinal-truncation        = 90.0
    bunching-factor                = 0.01
  }

  bunch-sampling
  {
    sample                         = false
    directory                      = ./
    base-name                      = bunch-sampling/bunch
    rhythm                         = 0.01
  }

  bunch-visualization
  {
    sample                         = true
    directory                      = ./
    base-name                      = bunch-visualization/bunch
    rhythm                         = 32
  }

  bunch-profile
  {
    sample                         = false
    directory                      = ./
    base-name                      = bunch-profile/bunch
    time                           = 15.0
  }
}

FIELD
{
  field-initialization
  {
    type                           = gaussian-beam
    position                       = ( 0.0, 0.0, -2500.0)
    direction                      = ( 0.0, 0.0, 1.0)
    polarization                   = ( 0.0, 1.0, 0.0)
    rayleigh-radius-parallel       = 0.5
    rayleigh-radius-perpendicular  = 0.5
    strength-parameter             = 0.0
    signal-type                    = gaussian
    offset                         = 0.00e-9
    variance                       = 1.00e-100
    wavelength                     = 0.0
    CEP                            = 0.0
  }

  field-sampling
  {
    sample                         = true
    type                           = at-point
    field                          = Ex
    field                          = Ey
    field                          = Ez
    directory                      = ./
    base-name                      = field-sampling/field
    rhythm                         = 3.2
    position                       = (0.0, 0.0, 110.0)
  }

  field-visualization
  {
    sample                         = true
    field                          = Ex
    field                          = Ey
    field                          = Ez
    field                          = Q
    directory                      = ./
    base-name                      = field-visualization/field
    rhythm                         = 32
  }

  field-profile
  {
    sample                         = false
    field                          = Ex
    field                          = Ey
    field                          = Ez
    directory                      = ./
    base-name                      = field-profile/field
    rhythm                         = 100
  }
}

UNDULATOR
{
  undulator-parameter              = 1.417
  period                           = 3.0e4
  length                           = 300
  polarization-angle               = 0.0
}

FEL-OUTPUT
{
  radiation-power
  {
    sample                         = false
    type                           = at-point
    directory                      = ./
    base-name                      = power-sampling/power
    distance-from-bunch            = 110.0
    normalized-frequency          = 1.00
  }
}
\end{Verbatim}
\end{snugshade}

\subsection{Simulation Results}

\begin{figure}
\centering
$\begin{array}{cc}
\includegraphics[height=2.5in]{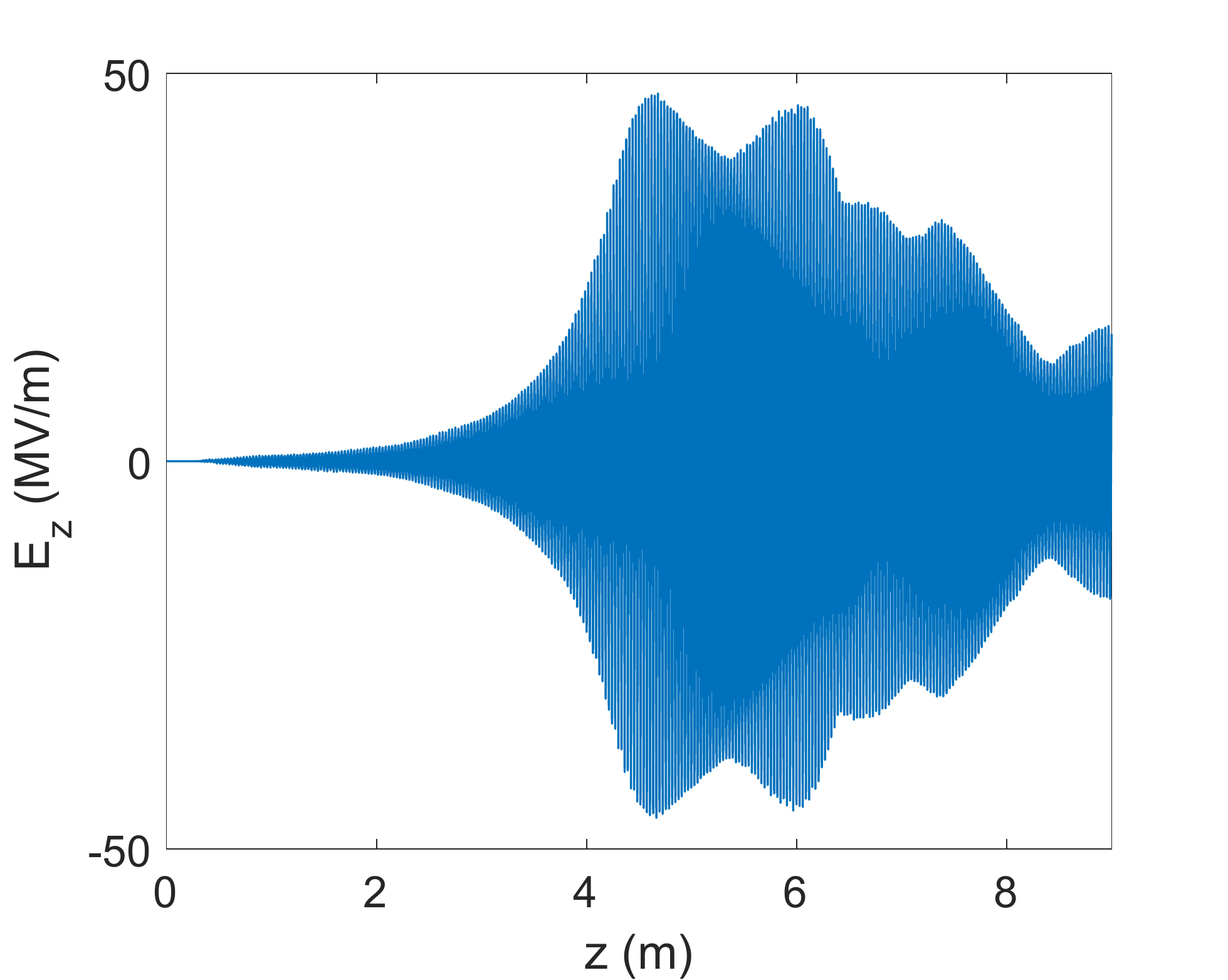} & \includegraphics[height=2.5in]{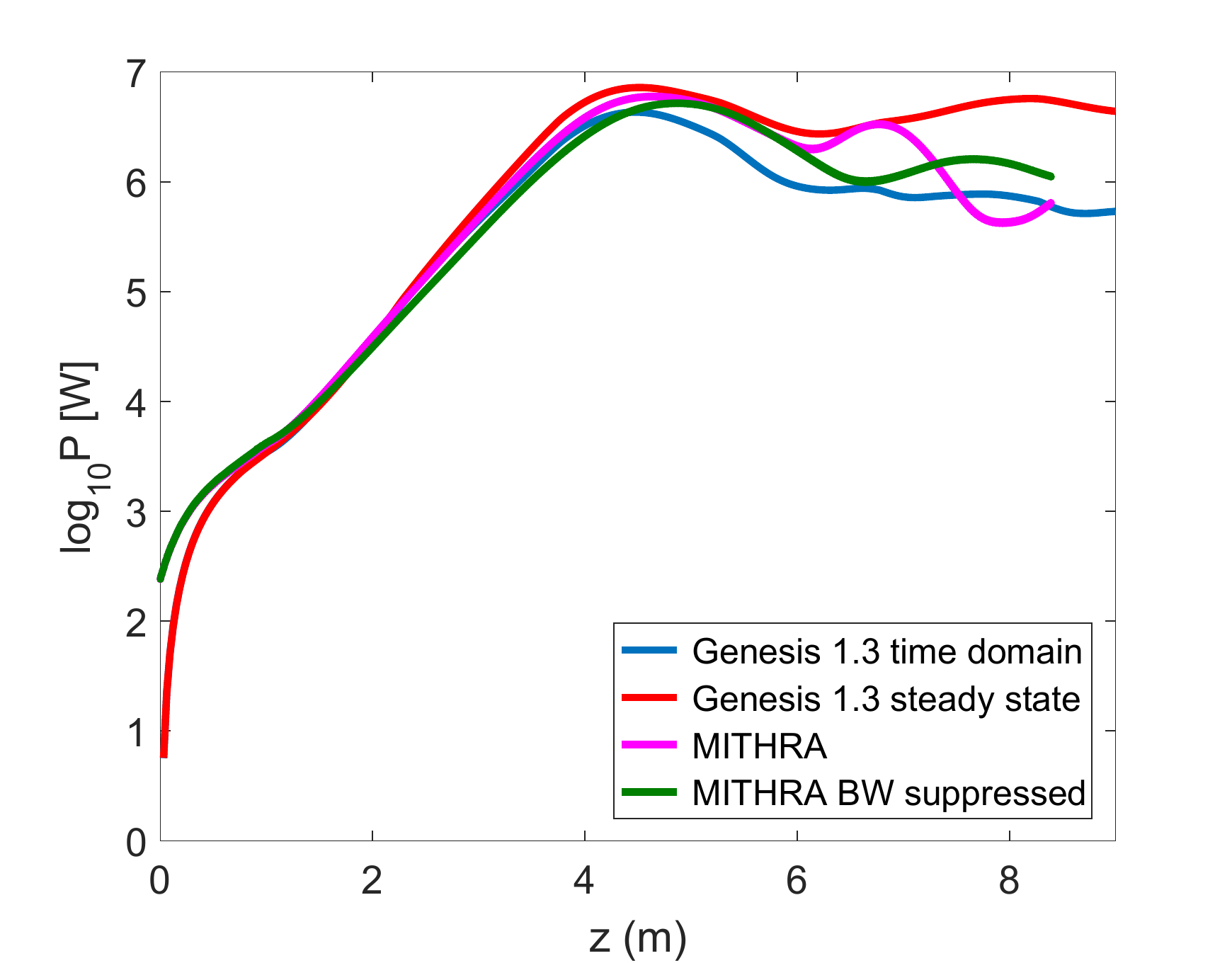} \\
(a) & (b)
\end{array}$
\caption{(a) The transverse field $E_y$ at 110\,$\mu$m distance from the bunch center and (b) the total radiated power calculated at 110\,$\mu$m distance from the bunch center in terms of the traveled undulator length.}
\label{power-example1}
\end{figure}

In the beginning, we neglect the space-charge effect only to achieve a good assessment of MITHRA simulation results.
The investigation of space-charge effect will be performed in the second step.
Fig.\,\ref{power-example1}a shows the transverse electric field sampled at 110\,$\mu$m in front of the bunch center.
The logarithmic plot of the radiated power for different propagation lengths ($z$) is also depicted in Fig.\,\ref{power-example1}b.
According to the 1D FEL theory the gain length of the considered SASE FEL configuration is $L_G=22.4\,\mathrm{cm}$.
The gain length calculated from the slope of the power curve is $L_G=22\,\mathrm{cm}$.
There exists also a good agreement in the computed saturation power.
The beam energy according to the data in table \ref{example1} is 1.52\,mJ which for the bunch length of 100\,$\mu$m corresponds to $P_{beam}=4.55\,\mathrm{GW}$ beam power.
The estimated saturation power according to the 1D theory is equal to $P_{sat} = \rho P_{beam} = 2.7\,\mathrm{GW}$.
The saturation power computed by MITHRA is $2.6\,\mathrm{GW}$.

We have also performed a comparison study between the obtained results from MITHRA and the code Genesis 1.3, which is presented in Fig.\,\ref{power-example1}b.
As observed, both codes produce similar results in the initial state and the gain regime.
Nonetheless, there exists a considerable discrepancy between the calculated radiated power in the saturation regime.
The illustrated results in Fig.\,\ref{power-example1}b show that the steady state and time domain analyses using Genesis do not produce similar results.
This shows that the bunch is not long enough to justify the steady state approximation, and dictates a time domain analysis for accurate simulation.
However, the results obtained by MITHRA at saturation do not match with the Genesis results even in the time domain.

The origin of such a discrepancy is described as follows:
As explained in chapter \ref{chapter_introduction}, Genesis 1.3 and all the existing softwares for FEL simulation neglect the backward radiation of the electrons.
Such an approximation is motivated by the inherent interest in forward radiation in the FEL process.
However, the backward radiation although is seldom used due to its long wavelength, it influences the motion of electrons, the charge distribution and in turn the FEL output.
The influence of low-frequency backward radiation on the performance of free electron lasers has been already studied in a 1D regime \cite{maroli2000effects}.
The effect becomes stronger in the saturation regime, when the electrons are microbunched and the FEL forward radiation is a strong function of the particles distribution.
To demonstrate this effect, we used a mathematical trick in MITHRA through the domain decomposition algorithm to suppress the propagation of backward radiation.
The results of such an analysis is also shown in Fig.\,\ref{power-example1}b, which shows a relatively better agreement with time domain simulation results returned by Genesis 1.3.
The still existing discrepancy is attributed to the different formulations of FDTD and TDA algorithms as well as the introduced tapers in bunch current and undulator fields.

\begin{figure}
\centering
\includegraphics[width=6.0in]{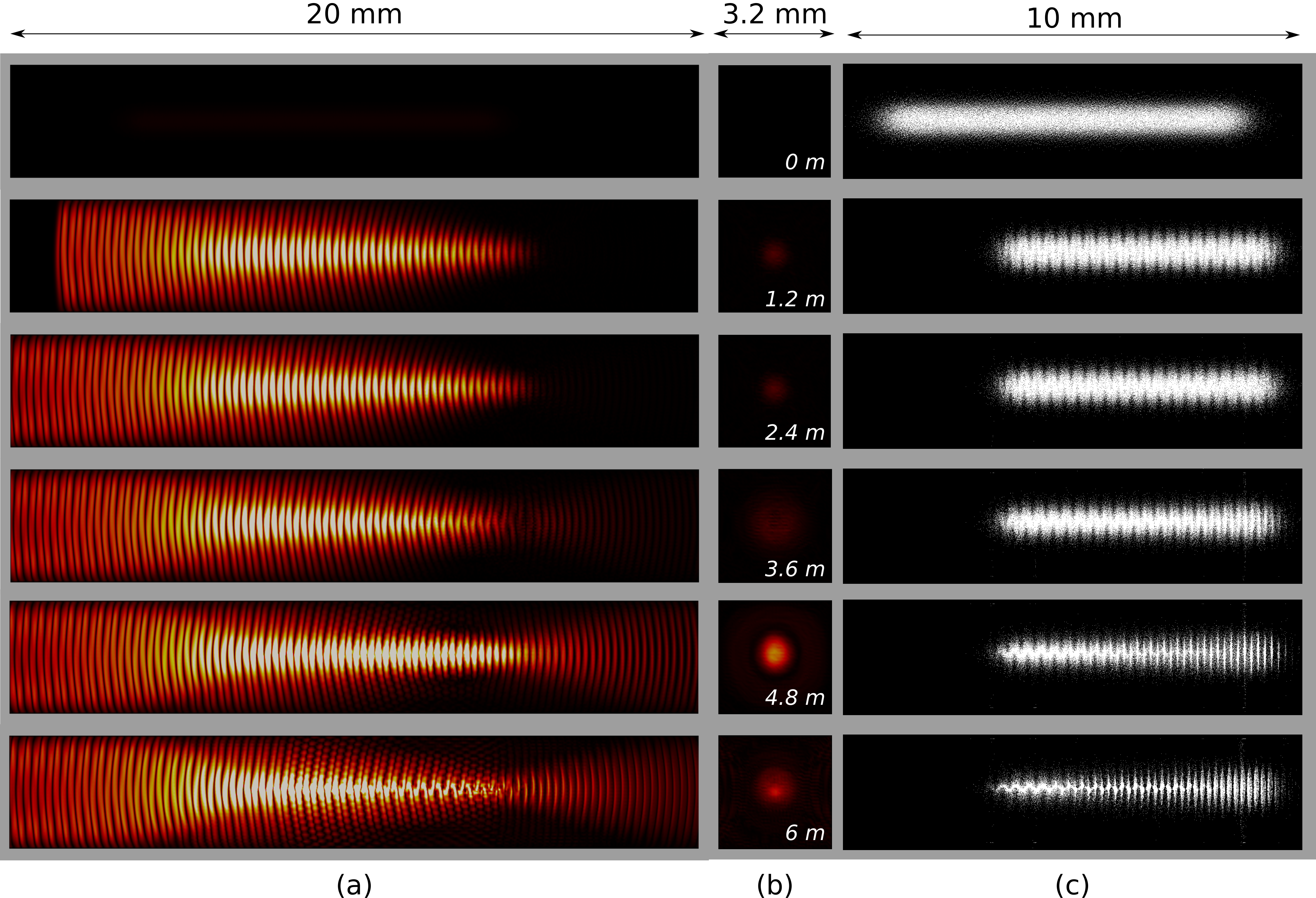}
\caption{Snapshots of the radiated field profile taken at (a) $x=0$ and (b) $z=60\,\mu m$ plane and (c) the bunch profile viewed from the $x$ axis.}
\label{profile-example1}
\end{figure}

As a 3D electromagnetic simulation, it is always beneficial to investigate the electromagnetic field profile in the computational domain.
Using the field visualization capability in MITHRA, snapshots of the field profile at different instants and from various view points are provided.
In Fig.\,\ref{profile-example1}, snapshots of the radiated field profile at different time instants are illustrated.
The emergence of lasing radiation at the end of the undulator motion is clearly observed in the field profile.

Furthermore, snapshots of the bunch profile are also presented beside the field profile.
The main FEL principle which is the lasing due to micro-bunching of the electron bunch is observed from the field and bunch profiles.

\subsection{Convergence Analysis}

The convergence rate of the results is the main factor used to assess a numerical algorithm. 
In our FEL analysis, there are several parameters introduced by the numerical method which may affect the final result. 
These parameters include (1) number of macro-particles ($n$), (2) time step for updating equation of motion ($\Delta t_b$), (3) longitudinal mesh size ($l_z$), (4) transverse mesh size ($l_x=l_y$), (5) longitudinal discretization ($\Delta z$) and (6) transverse discretization ($\Delta x = \Delta y$). 
Studying the convergence of the results is crucial to acquire an estimate for the uncertainty in the reported values. 
Here, this task is accomplished by sweeping over the above parameters and plotting the error function defined as the following:
\begin{equation}
\label{errorDefinition}
\mathrm{error} = \frac{\int_{z_i}^{z_f} | P(z)-P_0(z) | \mathrm{dz}}{\int_{z_i}^{z_f} P_0(z) \mathrm{dz}},
\end{equation}
where $z_i$ and $z_f$ are the beginning and end of the undulator, respectively and $P_0$ is the reference simulation result which is chosen as the results with the highest resolution. 

\begin{figure}
\centering
\includegraphics[width=7.0in]{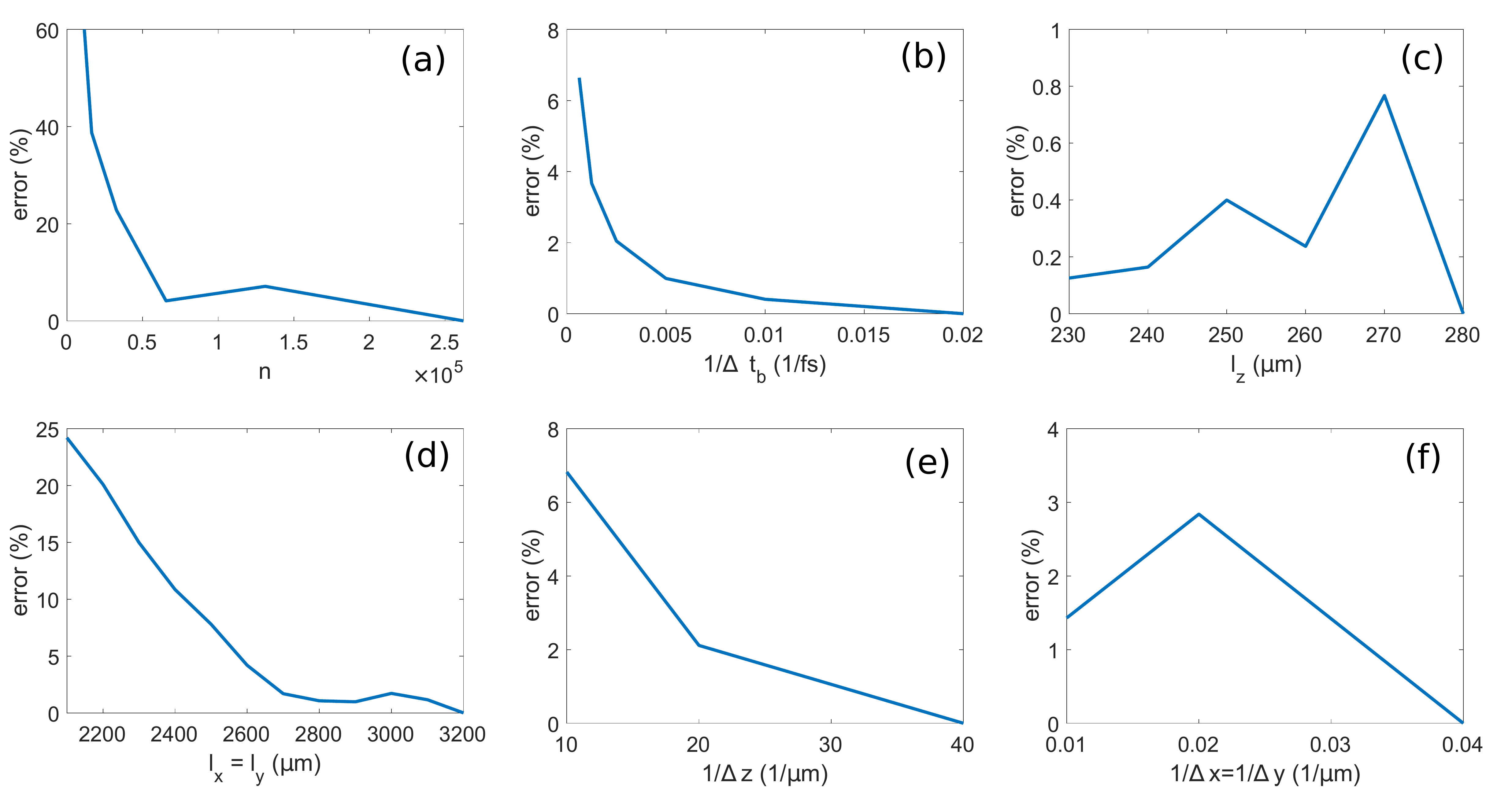}
\caption{Convergence study for the different involved parameters in the considered FEL simulation: (a) $n$, (b) $\Delta t_b$, (c) $l_z$, (d) $l_x=l_y$, (e) $\Delta z$ and (f) $\Delta x = \Delta y$}
\label{convergenceStudy}
\end{figure}

In Fig.\,\ref{convergenceStudy} the convergence study is shown for the aforementioned parameters. 
Generally, accuracy of less than 3\% is achieved by using the initially suggested values.

\subsection{Space-charge effect}

\begin{figure}
\centering
$\begin{array}{cc}
\includegraphics[height=2.5in]{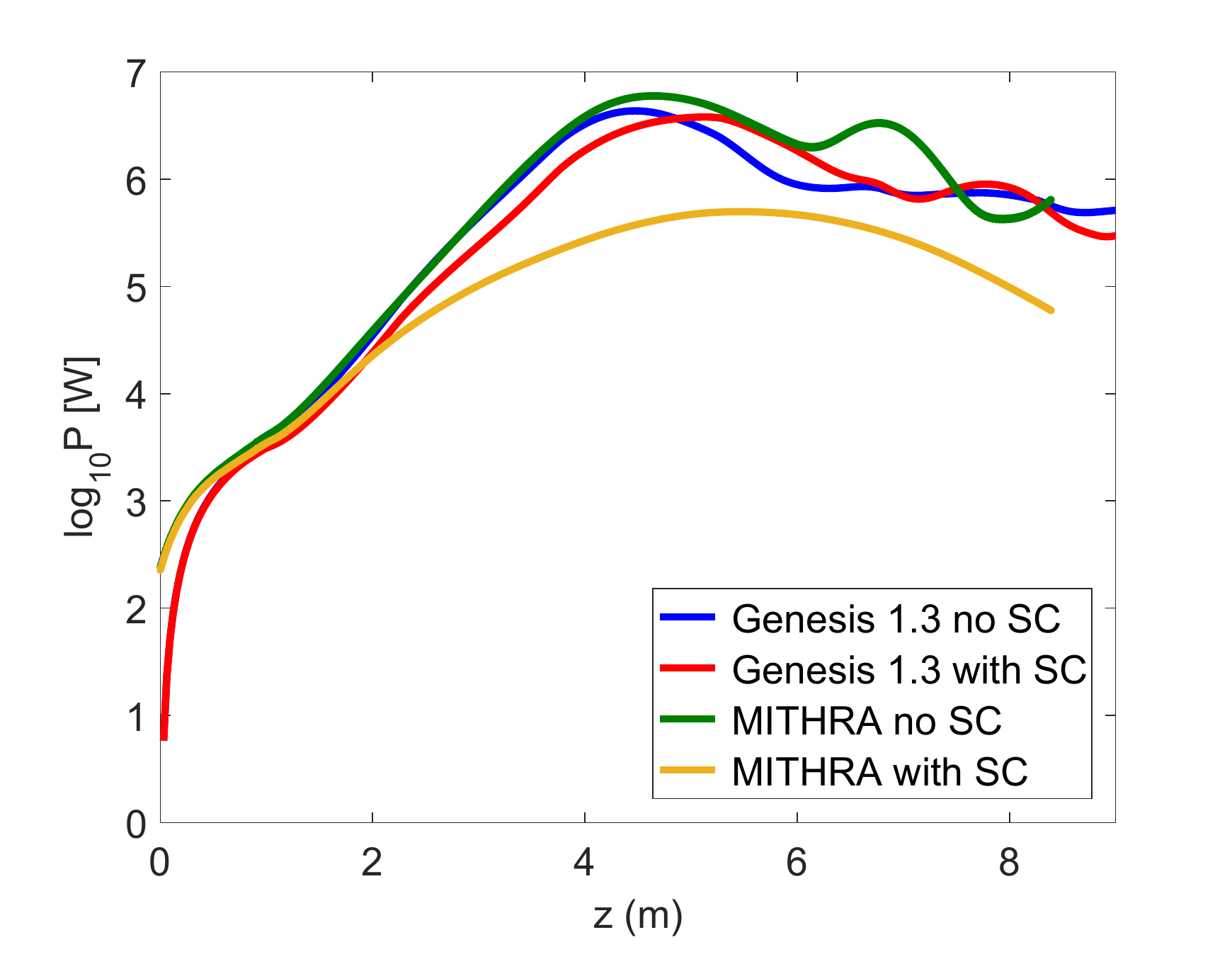} & \includegraphics[height=2.5in]{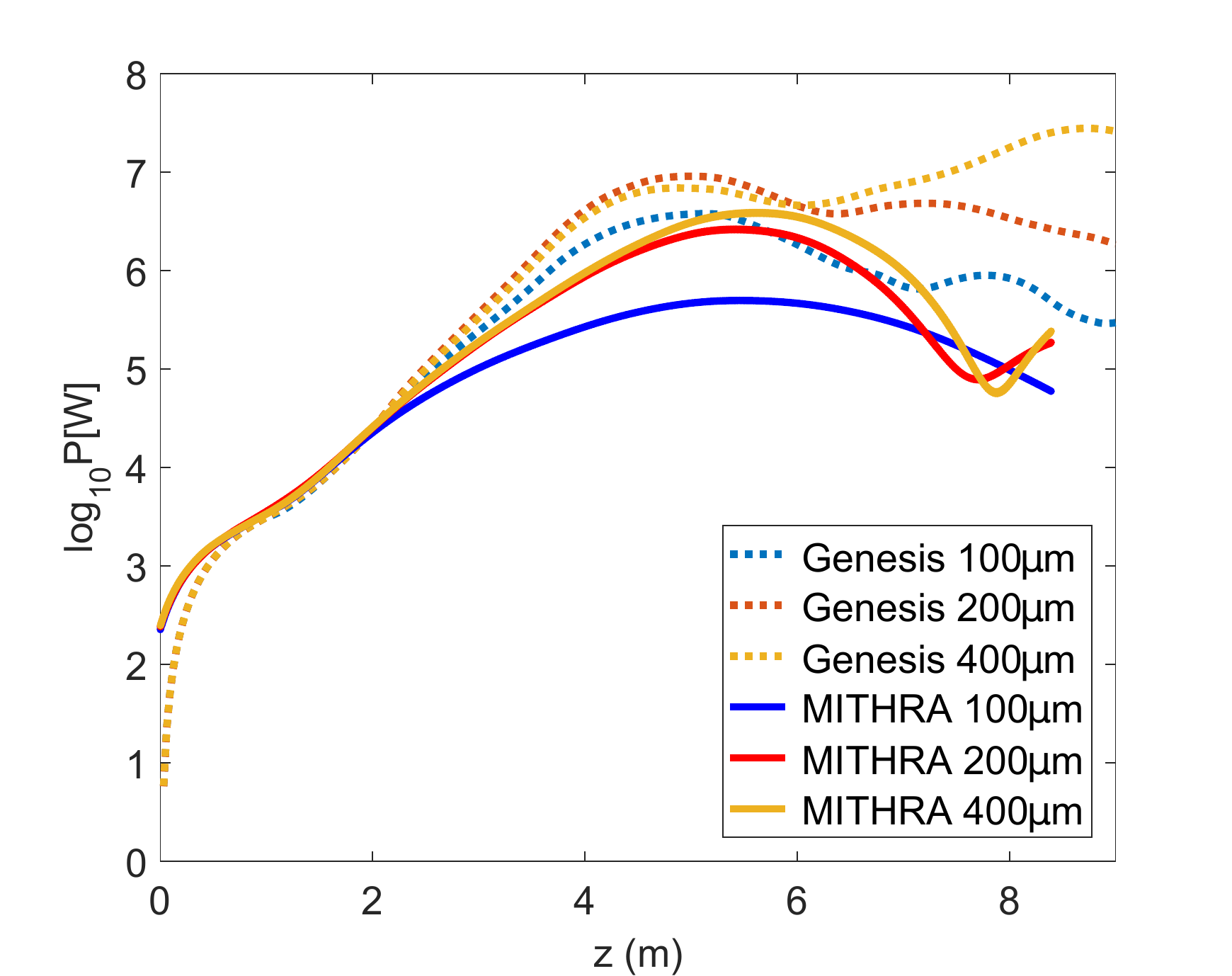} \\
(a) & (b)
\end{array}$
\caption{The total radiated power calculated at 110\,$\mu$m distance from the bunch center in terms of the traveled undulator length (a) with and without space-charge consideration and (b) various lengths of the bunch with space-charge assumption.}
\label{spaceChargeEffect}
\end{figure}

A promising benefit offered by MITHRA is the assessment of various approximations used in the previously developed FEL codes.
As an example, the algorithm used in the TDA method to evaluate the space-charge effect can be examined and verified using this code.
The TDA method implemented in Genesis 1.3 software considers a periodic variation of space-charge force throughout the electron bunch \cite{tranFEL,reiche2000numerical}.
However, a simple investigation of bunch profiles shown in Fig.\,\ref{profile-example1}c shows that a periodic assumption for the electron distribution may be a crude approximation.
In addition, this assumption is favored by the FEL gain process and potentially decreases any detrimental influence of the space-charge fields on the FEL radiation.
In Fig.\,\ref{spaceChargeEffect}a, we are comparing the solution of the FEL problem using MITHRA and Genesis 1.3 with and without considering the space-charge effect.
As observed in the results, the effect of space-charge on the radiation gain predicted by MITHRA is much stronger than the same effect predicted by Genesis.
This is attributed to the assumption of periodic variations in the space-charge force made in TDA algorithm.
If such a hypothesis is correct, the observed discrepancy should reduce once the radiation from a longer bunch is simulated, because the accuracy of periodicity assumption increases for longer bunches.
Indeed, this is observed after repeating the simulation for longer electron bunches with similar charge and current densities.
The results of such a study is illustrated in Fig.\,\ref{spaceChargeEffect}b.

\section{Example 2: Seeded UV FEL}

\subsection{Problem Definition}

\begin{table}[h]
\label{example2}
\caption{Parameters of the UV seeded FEL configuration considered as the second example.}
\centering
\begin{tabular}{|c||c|}
\hline
FEL parameter & Value \\ \hline \hline
Current profile & Uniform \\ \hline
Bunch size & $(95.3\times95.3\times150)\,\mu$m \\ \hline
Bunch charge & 54.9\,pC \\ \hline
Bunch energy & 200\,MeV \\	\hline
Bunch current & 110\,A \\ \hline
Longitudinal momentum spread & 0.01\% \\ \hline
Transverse momentum spread & 0.0026\% \\	\hline
Undulator period & 2.8\,cm \\ \hline
Magnetic field & 0.7\,T \\ \hline
Undulator parameter & 1.95 \\ \hline
Undulator length & 15\,m \\ \hline
Radiation wavelength & 0.265\,$\mu$ m \\ \hline
Electron density & $2.52\times10^{14} 1/\text{cm}^3$ \\ \hline
Gain length (1D) & 8.9\,cm \\ \hline
FEL parameter & 0.015 \\ \hline
Cooperation length & $3.34\,\mu$m \\ \hline
Initial bunching factor & $0.0$ \\ \hline
Seed type & Gaussian beam \\ \hline
Seed focal point & 70\,cm \\ \hline
Seed beam radius & 183.74\,$\mu$m \\ \hline
Seed pulse length & infinite \\ \hline
Seed power & 10\,kW \\ \hline
\end{tabular}
\end{table}
As the second example, we consider a seeded FEL in the UV regime to verify the implemented features for simulating a seeded FEL.
The parameters of the considered case are taken from \cite{giannessi2006overview}, which are tabulated in table \ref{example2}.
The bunch distribution is again assumed to be uniform with a long current profile ($\sim$1000 times the radiation wavelength) in order to compare the results with the steady state simulations.
For the same reason, the seed pulse length is considered to be infinitely long, i.e. a continuous wave pulse.
The transverse energy spread is calculated for a bunch with normalized transverse emittance equal to 1\,mm\,mrad.
Because of the very long bunch compared to the previous example, the number of required micro-particles to obtain convergent results is 8 times larger.
Furthermore, the stronger undulator parameter dictates a smaller time step for the simulation of bunch dynamics.
Note that MITHRA, takes the bunch step value as an initial guess, it automatically adjusts the value based on the calculated time step for mesh update.
To simulate the considered FEL configuration, the following job file is written and given to the software to analyze the interaction.
\begin{snugshade}
\begin{Verbatim}[fontsize=\small, tabsize = 4]
MESH
{
  length-scale                     = MICROMETER
  time-scale                       = PICOSECOND
  mesh-lengths                     = ( 1600.0, 1600.0, 165.0)
  mesh-resolution                  = ( 10.0,   10.0,   0.01 )
  mesh-center                      = ( 0.0,    0.0,    0.0  )
  total-time                       = 50000
  bunch-time-step                  = 0.8
  bunch-time-start                 = 0.0
  number-of-threads                = 2
  mesh-truncation-order            = 2
  space-charge                     = true
}

BUNCH
{
  bunch-initialization
  {
    type                           = ellipsoid
    distribution                   = uniform
    charge                         = 3.4332e8
    number-of-particles            = 1048576
    gamma                          = 391.36
    direction                      = ( 0.0,  0.0,  1.0 )
    position                       = ( 0.0,  0.0,  0.0 )
    sigma-position                 = ( 95.3, 95.3, 75.0)
    sigma-momentum                 = ( 0.0105, 0.0105, 391.36e-4)
    transverse-truncation          = 400.0
    longitudinal-truncation        = 78.0
    bunching-factor                = 0.0
  }

  bunch-sampling
  {
    sample                         = false
    directory                      = ./
    base-name                      = bunch-sampling/bunch
    rhythm                         = 0.01
  }

  bunch-visualization
  {
    sample                         = false
    directory                      = ./
    base-name                      = bunch-visualization/bunch
    rhythm                         = 10
  }

  bunch-profile
  {
    sample                         = true
    directory                      = ./
    base-name                      = bunch-profile/bunch
    time                           = 40000
  }
}

FIELD
{
  field-initialization
  {
    type                           = gaussian-beam
    position                       = ( 0.0, 0.0, 700080)
    direction                      = ( 0.0, 0.0, 1.0)
    polarization                   = ( 0.0, 1.0, 0.0)
    rayleigh-radius-parallel       = 183.74
    rayleigh-radius-perpendicular  = 183.74
    strength-parameter             = 9.857e-7
    signal-type                    = gaussian
    offset                         = 700005.0
    variance                       = 1e12
    wavelength                     = 0.265187
    CEP                            = 0.0
  }

  field-sampling
  {
    sample                         = false
    type                           = at-point
    field                          = Ex
    directory                      = ./
    base-name                      = field-sampling/field
    rhythm                         = 0.5
    position                       = (0.0, 0.0, 55.0)
  }

  field-visualization
  {
    sample                         = false
    field                          = Ay
    directory                      = ./
    base-name                      = field-visualization/field
    rhythm                         = 100
  }

  field-profile
  {
    sample                         = false
    field                          = Ex
    directory                      = ./
    base-name                      = field-profile/field
    rhythm                         = 100
  }
}

UNDULATOR
{
  undulator-parameter              = 1.95
  period                           = 2.8e4
  length                           = 535
  polarization-angle               = 0.0
}

FEL-OUTPUT
{
  radiation-power
  {
    sample                         = true
    type                           = at-point
    directory                      = ./
    base-name                      = power-sampling/power
    distance-from-bunch            = 80.0
    normalized-frequency          = 1.00
  }
}
\end{Verbatim}
\end{snugshade}

\subsection{Simulation Results}

\begin{figure}
\centering
$\begin{array}{cc}
\includegraphics[height=2.5in]{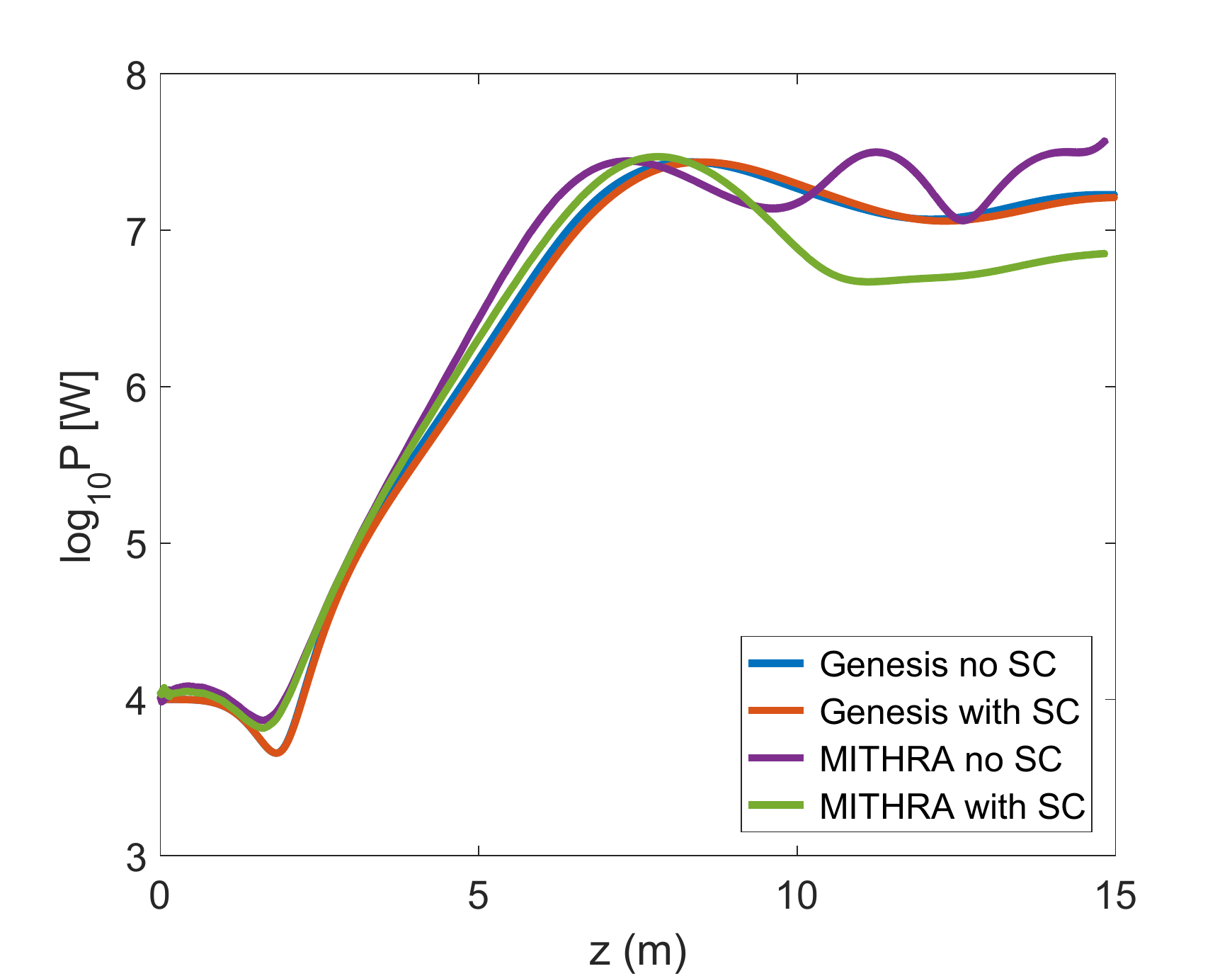} & \includegraphics[height=2.5in]{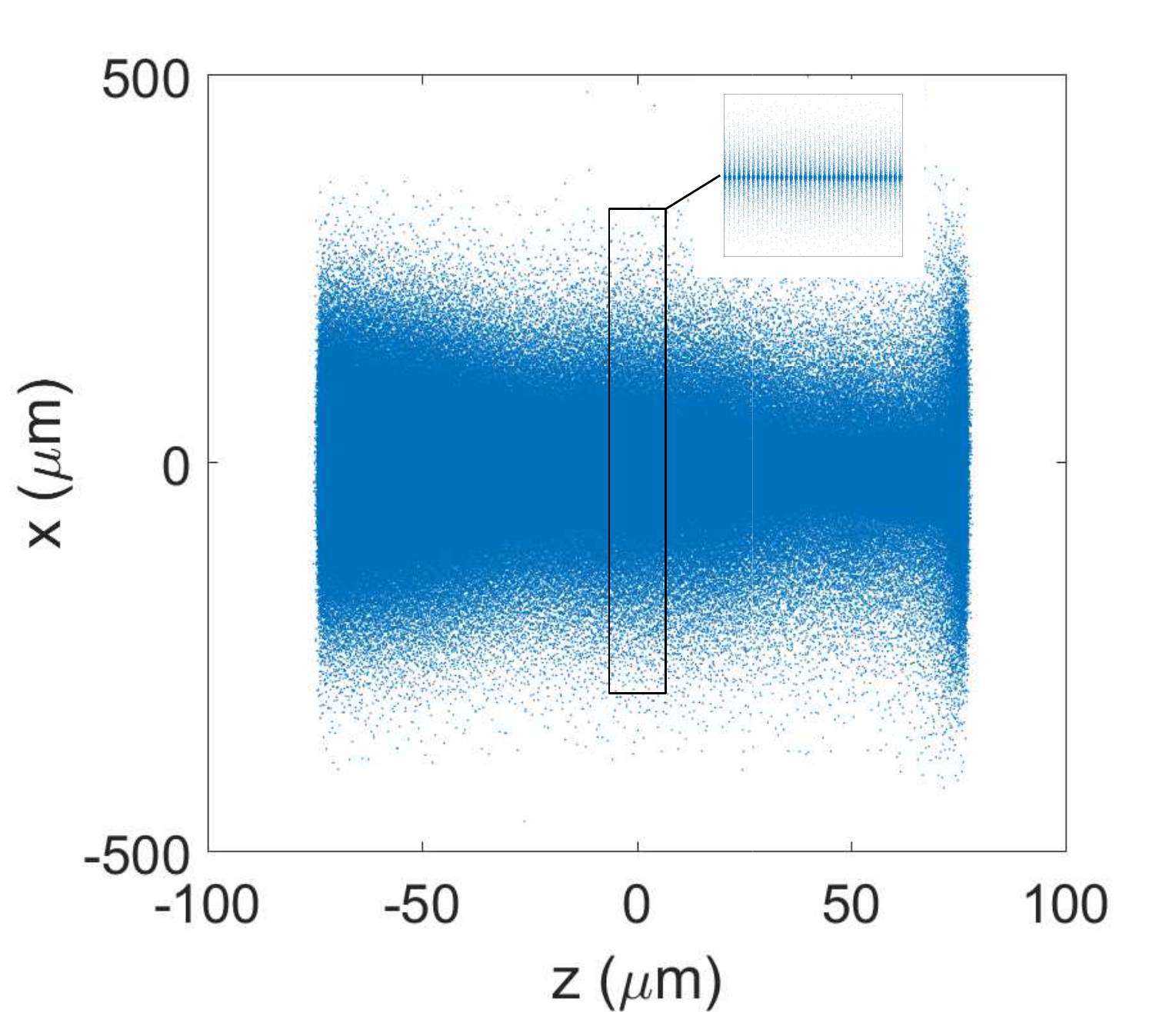} \\
(a) & (b)
\end{array}$
\caption{(a) The total radiated power measured at 80\,$\mu$m distance from the bunch center in terms of the traveled undulator length and (b) the bunch profile at 12\,m from the undulator begin.}
\label{power-example2}
\end{figure}

Fig.\,\ref{power-example2}a shows the radiated power in terms of travelled undulator distance computed using MITHRA and Genesis.
As observed again in this example, the results agree very well in the seeded and gain regime, with notable discrepancies in the saturation regime.
In Fig.\,\ref{power-example2}b, the bunch profile after 12\,m propagation in the undulator is also depicted.
The micro-bunching of the large bunch is only visible once a zoom into a part of the bunch is considered.
The investigation of the results with and without considering space-charge effect shows that in the seeded and gain intervals, space charge plays a negligible role.
However, in the saturation regime the effect of space-charge predicted by MITHRA is stronger than the effect predicted by Genesis.

\section{Example 3: Optical Undulator}

As explained in the introduction of this manual, one of the milestones considered for the development of MITHRA is full-wave simulation of inverse Compton scattering (ICS) or the so-called optical undulator.
The possibility of lasing or the so-called micro-bunching in an electron beam due to an interaction with a counter-propagating laser beam has been under debate for several years.
A full-wave analysis of such an interaction definitely gives valuable physical insight to this process.
Note that the classical treatment of this interaction within MITHRA does not allow for any consideration of quantum mechanical effects.
It is known that the radiation of photons results in a backward force on electrons which leads to a change in their momenta.
This effect is considered in the simulations presented by MITHRA within a classical framework.
Nonetheless, the fact that this momentum change is a quantized value and equal to $n\hbar\omega$ can not be considered by MITHRA.

\begin{figure}
\centering
\includegraphics[height=2.5in]{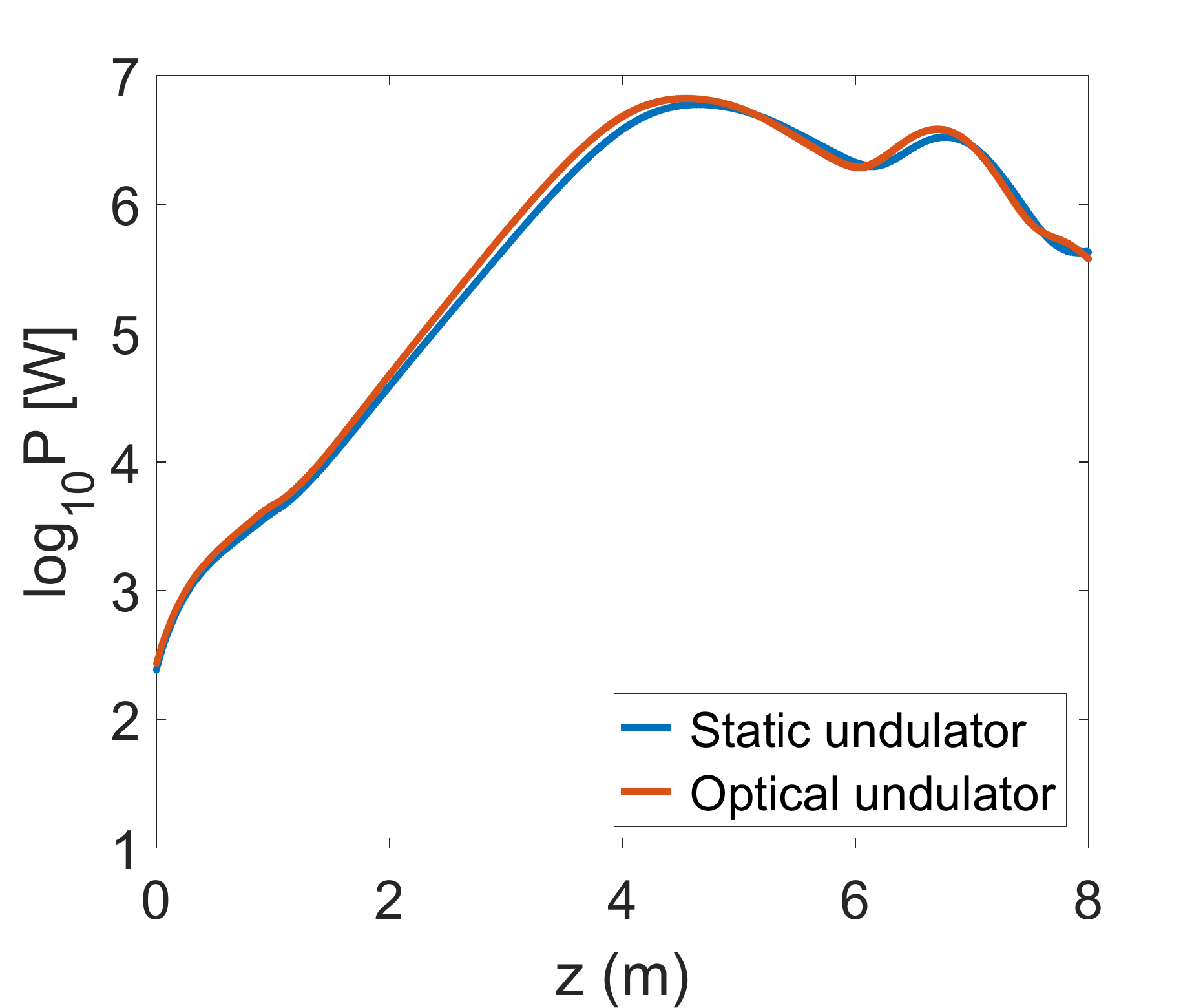}
\caption{The total radiated power calculated at 110\,$\mu$m distance from the bunch center in terms of the traveled undulator length compared for two cases of an optical and static undulator.}
\label{ICS-benchmark}
\end{figure}

Before embarking on the analysis and interpretation of results for a typical ICS experiment, a benchmark to validate the analysis of optical undulators using FDTD/PIC is presented.
It is known that electron trajectories in a static undulator with undulator parameter $K$ and periodicity $\lambda_u$ are similar to the trajectories in an electromagnetic undulator setup with normalized vector potential $a_0=K$ and wavelength $\lambda_l=2\lambda_u$ \cite{esarey1993nonlinear}.
We take the first SASE FEL example in table \ref{example1} into account and analyze the same configuration but with an equivalent optical undulator.
For this purpose, the undulator definition of example 1 is entered according to the following:
\begin{snugshade}
\begin{Verbatim}[fontsize=\small, tabsize = 4]
UNDULATOR
{
  undulator-type                   = optical
  beam-type                        = plane-wave
  position                         = ( 0.0, 0.0, 0.0 )
  direction                        = ( 0.0, 0.0,-1.0 )
  polarization                     = ( 0.0, 1.0, 0.0 )
  strength-parameter               = 1.417
  signal-type                      = flat-top
  wavelength                       = 6.0e4
  variance                         = 1800.0e4
  offset                           = 9000150
  CEP                              = 0.0
}
\end{Verbatim}
\end{snugshade}
Fig.\,\ref{ICS-benchmark} illustrates a comparison between the radiated infrared light for the static and optical undulator cases.
The very close agreement between the two results validates the implementation of optical undulators in MITHRA.

The parameters of FEL interaction in an optical undulator, considered as the third example, are tabulated in table \ref{example3}.
Since we observe drastic deviation from the predictions of one-dimensional FEL theory in our simulations, we have not listed the FEL parameters calculated using the 1D theory.
We believe the discrepancies are originated from the small number of electrons in each 3D wave bucket, i.e. only 2 electrons.
This strongly intensifies the 3D effects, dramatically reduces the transverse coherence of the radiation, and indeed makes analysis using 1D FEL theory completely invalid.
Detailed verification of the above claims will be presented in a publication.
Here, the optical undulator is merely treated as an example analyzed with FDTD/PIC.
\begin{table}[h]
\label{example3}
\caption{Parameters of the FEL configuration with optical undulator considered as the third example.}
\centering
\begin{tabular}{|c||c|}
\hline
FEL parameter & Value \\ \hline \hline
Current profile & Uniform \\ \hline
Bunch size & $(60\times60\times144)$\,nm \\ \hline
Bunch charge & 0.45\,fC \\ \hline
Bunch energy & 15\,MeV \\	\hline
Bunch current & 0.93\,A \\ \hline
Longitudinal momentum spread & 0.1\% \\ \hline
Transverse momentum spread & 0.1\% \\	\hline
Laser wavelength & 1\,$\mu$m \\ \hline
Laser strength parameter & 1.0 \\ \hline
Pulse duration & 4\,ps \\ \hline
Laser pulse type & flat-top \\ \hline
Radiation wavelength & 0.41\,nm \\ \hline
Electron density & $5.4\times10^{18} 1/\text{cm}^3$ \\ \hline
Initial bunching factor & $0.0$ \\ \hline
\end{tabular}
\end{table}

To simulate the considered FEL configuration, the following job file is written and given to the software to analyze the interaction.
\begin{snugshade}
\begin{Verbatim}[fontsize=\small, tabsize = 4]
MESH
{
  length-scale                     = NANOMETER
  time-scale                       = ATTOSECOND
  mesh-lengths                     = ( 2000.0, 2000.0, 165.0 )
  mesh-resolution                  = (    5.0,    5.0,   0.05)
  mesh-center                      = (    0.0,    0.0,   0.0 )
  total-time                       = 2000000
  bunch-time-step                  = 20.0
  bunch-time-start                 = 0.0
  number-of-threads                = 4
  mesh-truncation-order            = 2
  space-charge                     = true
}

BUNCH
{
  bunch-initialization
  {
    type                           = ellipsoid
    distribution                   = uniform
    charge                         = 2800
    number-of-particles            = 2800
    gamma                          = 30.0
    direction                      = (  0.0,   0.0,   1.0)
    position                       = (  0.0,   0.0,   0.0)
    sigma-position                 = ( 60.0,  60.0,  72.0)
    sigma-momentum                 = ( 0.03,  0.03,  0.03)
    transverse-truncation          = 240.0
    longitudinal-truncation        = 77.0
    bunching-factor                = 0.0
  }

  bunch-sampling
  {
    sample                         = false
    directory                      = ./
    base-name                      = bunch-sampling/bunch
    rhythm                         = 1000
  }

  bunch-visualization
  {
    sample                         = false
    directory                      = ./
    base-name                      = bunch-visualization/bunch
    rhythm                         = 1600
  }

  bunch-profile
  {
    sample                         = false
    directory                      = ./
    base-name                      = bunch-profile/bunch
    rhythm                         = 1600
  }
}

FIELD
{
  field-initialization
  {
    type                           = gaussian-beam
    position                       = ( 0.0, 0.0, -2500.0)
    direction                      = ( 0.0, 0.0, 1.0)
    polarization                   = ( 0.0, 1.0, 0.0)
    rayleigh-radius-parallel       = 0.5
    rayleigh-radius-perpendicular  = 0.5
    strength-parameter             = 0.0
    signal-type                    = gaussian
    offset                         = 0.00e-9
    variance                       = 1.00
    wavelength                     = 0.0
    CEP                            = 0.0
  }

  field-sampling
  {
    sample                         = false
    type                           = at-point
    field                          = Ex
    field                          = Ey
    field                          = Ez
    directory                      = ./
    base-name                      = field-sampling/field
    rhythm                         = 3.2
    position                       = (0.0, 0.0, 110.0)
  }

  field-visualization
  {
    sample                         = false
    field                          = Ex
    field                          = Ey
    field                          = Ez
    directory                      = ./
    base-name                      = field-visualization/field
    rhythm                         = 8.0
  }

  field-profile
  {
    sample                         = false
    field                          = Q
    directory                      = ./
    base-name                      = field-profile/field
    rhythm                         = 100
  }
}

UNDULATOR
{
  undulator-type                   = optical
  beam-type                        = plane-wave
  position                         = ( 0.0, 0.0, 0.0 )
  direction                        = ( 0.0, 0.0,-1.0 )
  polarization                     = ( 0.0, 1.0, 0.0 )
  strength-parameter               = 0.5
  signal-type                      = flat-top
  wavelength                       = 1.0e3
  variance                         = 1200.0e3
  offset                           = 600082.0
  CEP                              = 0.0
}

FEL-OUTPUT
{
  radiation-power
  {
    sample                         = true
    type                           = at-point
    directory                      = ./
    base-name                      = power-sampling/power
    distance-from-bunch            = 82
    normalized-frequency           = 1.000
    normalized-frequency           = 2.000
    normalized-frequency           = 3.000
  }
}
\end{Verbatim}
\end{snugshade}

\begin{figure}
\centering
$\begin{array}{cc}
\includegraphics[height=2.5in]{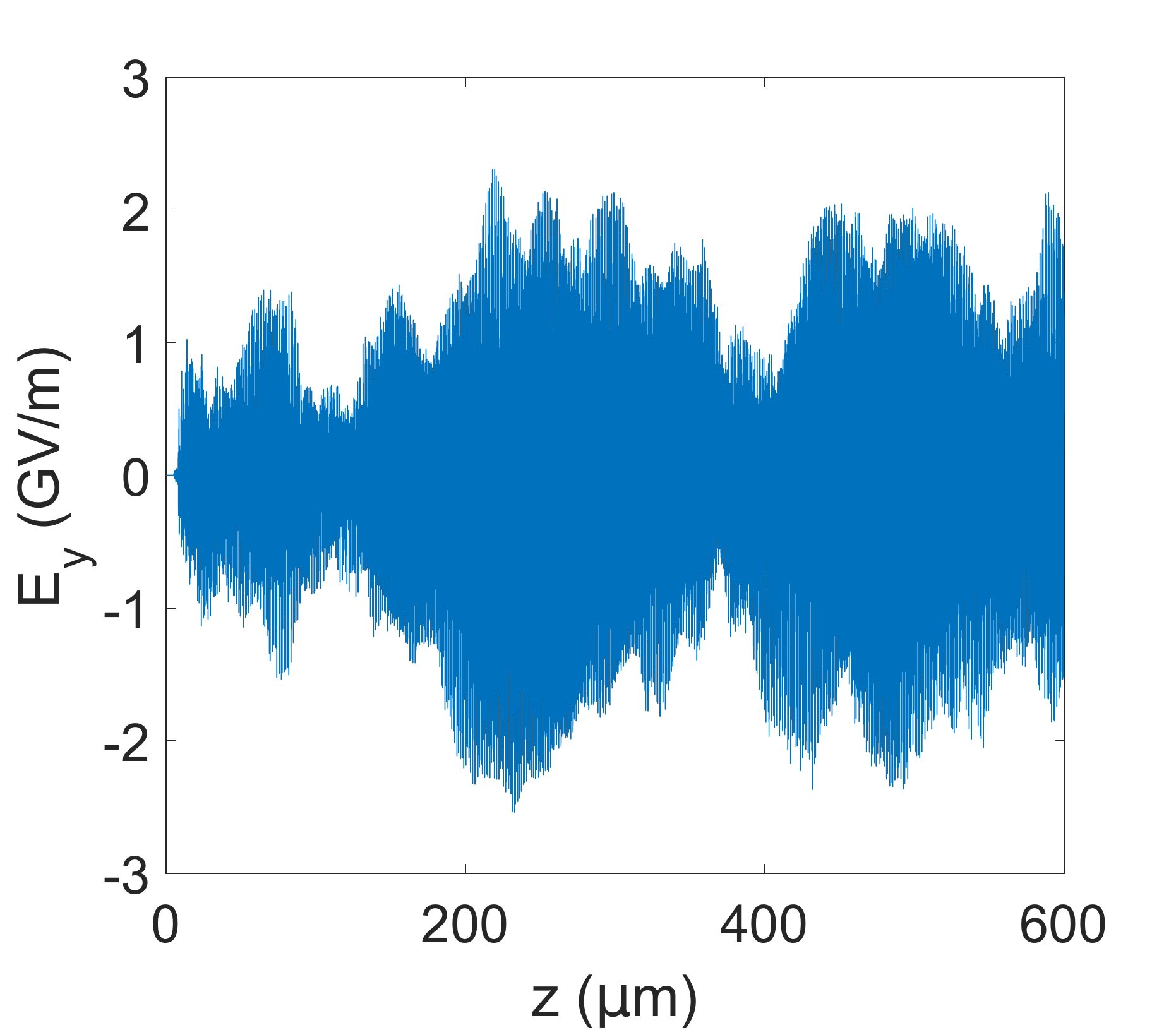} & \includegraphics[height=2.5in]{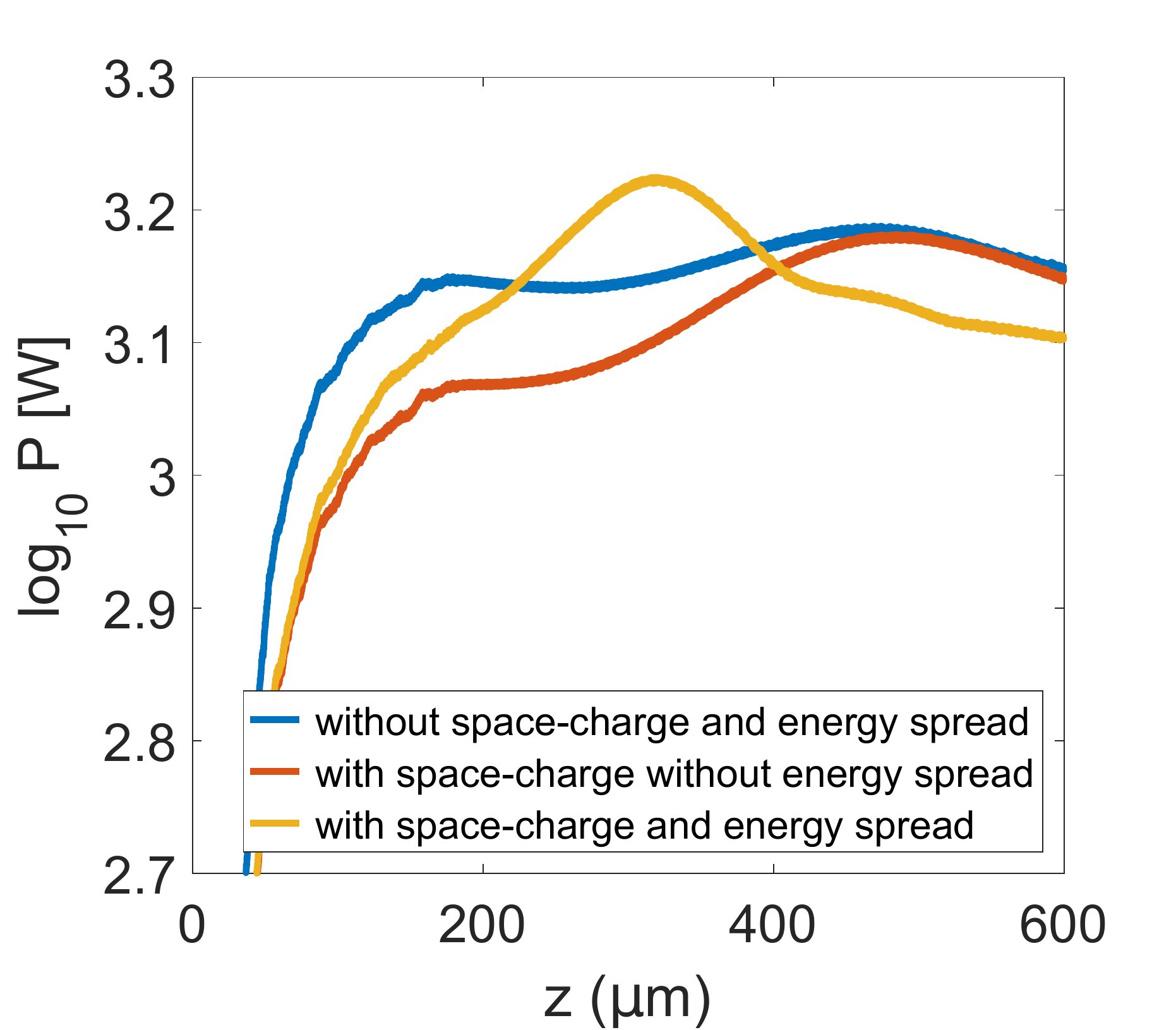} \\
(a) & (b) \\
\includegraphics[height=2.5in]{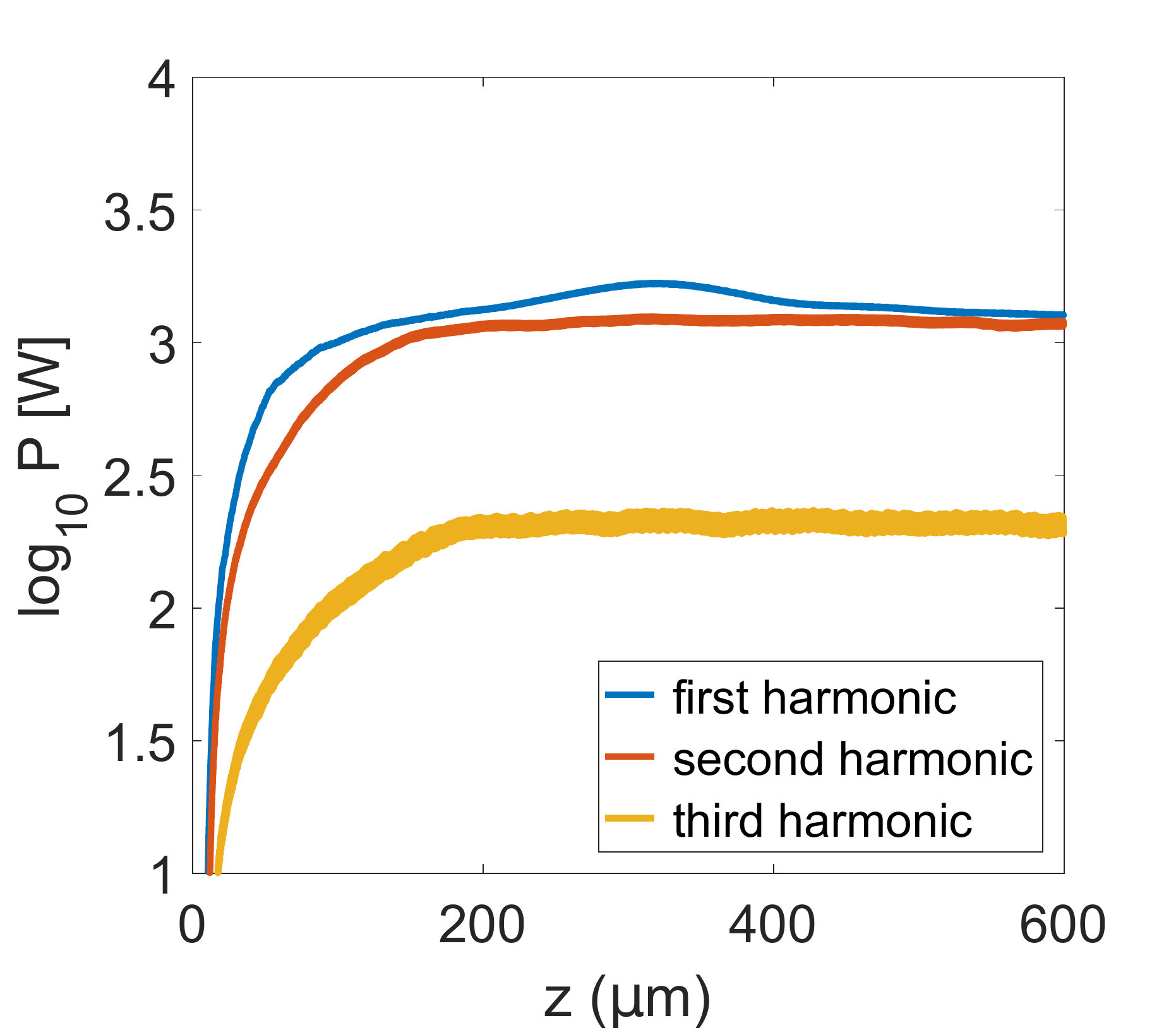} & \includegraphics[height=2.5in]{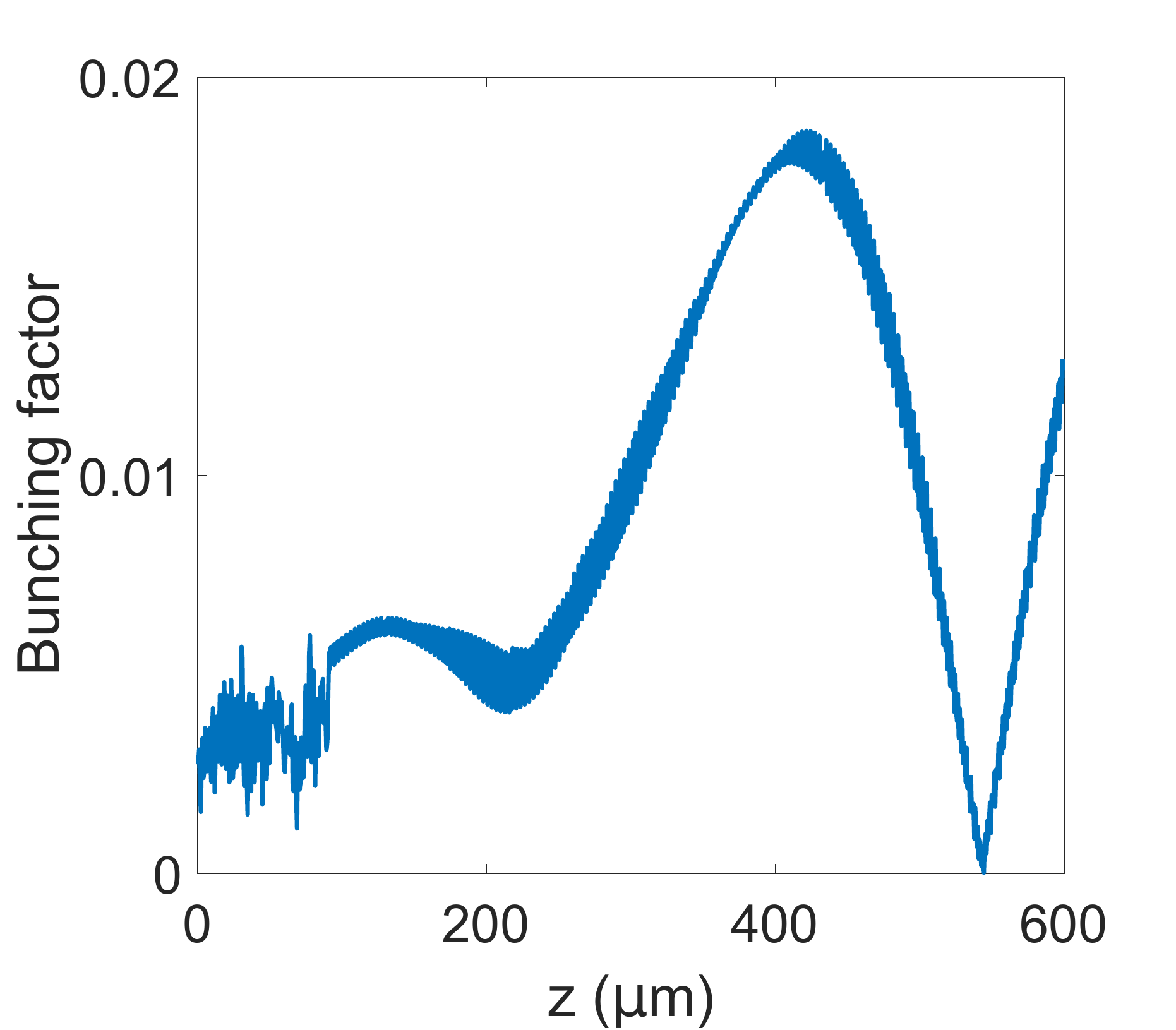} \\
(c) & (d)
\end{array}$
\caption{(a) Electric field of the generated radiation in front of the bunch, (b) the total radiated power measured at 82\,nm distance from the bunch center in terms of the traveled distance, (c) the same radiation power for various harmonic orders, and (d) bunching factor of the considered bunch during the ICS interaction.}
\label{power-example3}
\end{figure}

\begin{figure}[h]
\centering
\includegraphics[height=2.5in]{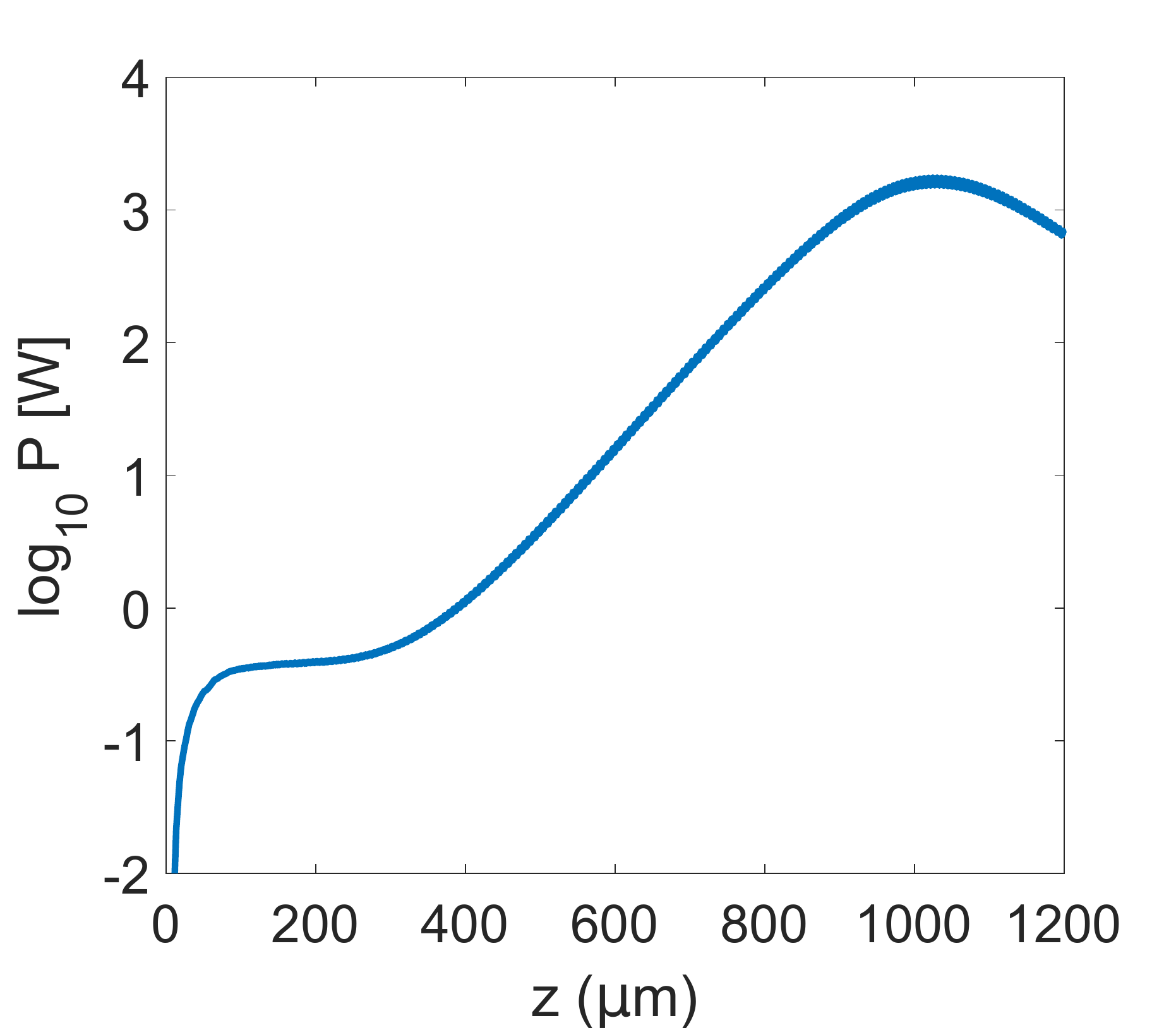}
\caption{The total radiated power measured at 82\,nm distance from the bunch center in terms of the traveled distance for an imaginary bunch where each electron is represented by a cloud of 1000 particles.}
\label{powerUnreal-example3}
\end{figure}

Fig.\,\ref{power-example3}a illustrates the radiation field 82\,nm away from the bunch center.
In addition, Fig.\,\ref{power-example3}b shows the radiated power in terms of travelled undulator distance computed using MITHRA, illustrating the effect of space charge and energy spread.
It is observed that the gain obtained in this regime is very small compared with a usual static undulators.
The reason for this effect is the very large shot noise in the bunch because of the low number of particles in each micro-bunch.
Note that in this simulation, each electron is modeled as one single particle.
The strong shot noise causes a strong initial radiation, which reaches the expected saturation power after a low gain.
As a matter of fact, the micro-bunching process increases the coherence of the output radiation rather than power amplification.
Another aspect in this regime of interaction is the generation of strong higher order harmonics, which are depicted up to the third harmonic in Fig.\,\ref{power-example3}c.
Note that the accuracy of the results decreases for higher harmonics due to the required resolution in the computational mesh.
To show that the micro-bunching effect takes place in this regime as well, the bunching factor of the electron beam is depicted in Fig.\,\ref{power-example3}d.
The bunching of the electrons due to the ICS interaction is clearly observed in the plot of bunching factor.
According to the depicted power and pulse shape, total number of emitted photons is approximately equal to $4.2\times10^3$.
To demonstrate the presented hypothesis related to the micro-bunching of bunches with low number of electrons per wavelength bucket, we perform an \emph{unreal} simulation, where each electron is presented by 1000 particles.
In this case, each particle represents a charge 1000 times smaller than the charge of one electron.
In addition, we assume an initial bunching factor equal to 0.001 for the input bunch to trigger the FEL gain.
In Fig.\,\ref{powerUnreal-example3}, the radiation of such a charge configuration is depicted.
The results clearly reveal the radiation start from much lower powers, possibility of achieving the FEL gain and saturating in the same power level as above, thereby confirming the above theory for radiation of low density electron bunches.

\printindex

\bibliographystyle{unsrtnat}


\end{document}